\PassOptionsToPackage{sort&compress}{natbib}
\documentclass[preprint,12pt,3p,onecolumn]{elsarticle}

\usepackage[utf8]{inputenc}
\usepackage[T1]{fontenc}

\usepackage{lineno}
\modulolinenumbers[5]

\usepackage{amsmath}
\usepackage{amssymb}
\usepackage{caption}
\usepackage{graphicx}
\usepackage{latexsym}
\usepackage{times}
\usepackage{rotating}
\usepackage{comment}
\usepackage{subcaption}
\usepackage{pdflscape}
\usepackage{booktabs}
\usepackage{multirow} 
\usepackage{threeparttable}
\usepackage{enumitem}
\usepackage{mathrsfs}
\usepackage{mathtools}
\usepackage{physics}
\usepackage{siunitx}
\usepackage{dimnum}
\newdimnum{M}{M}{Mach}{number}
\newdimnum{MM}{\ensuremath{\widetilde{M}}}{modified Mach}{number}
\newdimnum{N}{\ensuremath{N}}{non-equidiffusional}{coefficient}
\newdimnum{Np}{\ensuremath{\mathscr{N}}}{non-equidiffusional}{pressure-coefficient}
\usepackage{array}
\usepackage{colortbl}
\usepackage{xcolor}

\begin{document}

\begin{frontmatter}

\title{Numerical method for strongly variable-density flows at low Mach number: flame-sheet regularisation and a mass-flux immersed boundary method}

\author[USP]{Matheus P. Severino\corref{CA}}
\cortext[CA]{Corresponding author}
\ead{matheus.severino@usp.br}

\author[INPE]{Fernando F. Fachini}
\ead{fernando.fachini@inpe.br}

\author[UNESP]{Elmer M. Gennaro}
\ead{elmer.gennaro@unesp.br}

\author[UPM]{Daniel Rodríguez}
\ead{daniel.rodriguez@upm.es}

\author[USP]{Leandro F. Souza}
\ead{lefraso@icmc.usp.br}

\address[USP]{Instituto de Ciências Matemáticas e de Computação, Universidade de São Paulo \\
13566-590 São Carlos -- São Paulo, Brazil}

\address[INPE]{Grupo de Mecânica de Fluidos Reativos, Instituto Nacional de Pesquisas Espaciais \\
12630-000 Cachoeira Paulista -- São Paulo, Brazil}

\address[UNESP]{Departamento de Engenharia Aeronáutica, Universidade Estadual Paulista \\ 13876-750 São João da Boa Vista -- São Paulo, Brazil}

\address[UPM]{Escuela Técnica Superior de Ingeniería Aeronáutica y del Espacio, Universidad Politécnica de Madrid \\ 28040 Madrid -- Madrid, Spain}

\begin{abstract}
A low-Mach-number flow, in the laminar regime, has intrinsically two 
characteristic spatial scales for a given time scale, or 
two characteristic temporal scales for a given spatial scale, 
and these dual scales are very different due to the disparity 
between the flow and acoustic speed. 
Therefore low-Mach-number flows impose mathematical and computational 
challenges in their description.
Standard numerical methods for compressible flows, which are typically 
designed for problems with a single dominant spatial and temporal scale,
require alternative approaches such as preconditioning techniques or 
solvers tailored for low-Mach-number equations. 
The present work introduces a simplified fluid dynamics model for 
flows at low Mach number, based on the fractional time-step method. 
The proposed approach is suitable for handling strong temperature 
gradients and thermal diffusion, as encountered in combustion systems.
To address discontinuities at the flame front in reacting-flow 
cases, due to the hypothesis of infinitely fast chemistry, a 
regularisation procedure is employed. 
Additionally, the immersed boundary method (IBM) is extended 
to handle mass flux across the boundary surface, enabling  
simulations of fuel ejection from an arbitrary burner 
geometry, using a convenient Cartesian grid. 
The numerical method utilises a predictor-corrector scheme for 
time integration on a collocated grid, with flux interpolation 
to prevent numerical pressure oscillations (``odd-even decoupling'').
Relevant test cases are used to verify the methods and their 
implementations, demonstrating correctness and robustness.
\end{abstract}


\end{frontmatter}

\section{Introduction}
\label{Introduction}

Low Mach number flows, in which the characteristic flow speed 
is much smaller than the speed of sound, are prevalent in natural 
phenomena and engineering applications \cite{muller/1998, glegg/2017}.
They are commonly found in fields such as Combustion, Meteorology, and 
Geophysics \cite{klein/1995:JCP, happenhofer/2013:JCP}.
These flows are mathematically characterised by a small Mach number, 
which is a scaling factor for the pressure term, relating hydrodynamic 
and thermodynamic contributions.
The low Mach number limit represents a singular limit  
\cite{majda/1984,alazard/2006:ARMA,alazard/2006:JOMA,clausen/2013:PRE}, 
meaning that the mathematical nature of the equations undergoes a significant 
transformation as the Mach number approaches zero.
This characteristic presents challenges in both analytical 
\cite{alazard/2006:ARMA,alazard/2008:DCDS} and numerical treatments 
\cite{clausen/2013:PRE, lessani/2006:JCP}.
The singular nature introduces disparate scales into the problem -- 
a short time scale, associated with fast-propagating acoustic waves, 
and a large time scale, related to much slower fluid flow 
\cite{muller/1998} --, leading to a stiff system of equations. 
As a result, standard compressible flow solvers, become inaccurate 
and inefficient at low Mach numbers 
\cite{wall/2002:JCP,keshtiban/2004:IJNMF}.
Explicit methods require impractically small time steps, due to the 
Courant--Friedrichs--Lewy (CFL) condition, which is constrained by the 
speed of sound in the system. 
Although implicit methods allow for larger time steps, they face 
challenges with ill-conditioned matrices.

The most direct procedure to handling 
low Mach number flows involves applying preconditioning techniques to a compressible 
solver, improving the conditioning of the system \cite{turkel/1999:ARFM}. 
This method is ideal for flows with significant local variations in the 
Mach number. 
A second approach is to employ a solver specifically designed to solve the 
low-Mach-number asymptotic equations \cite{muller/1998}, 
making it more effective and accurate for cases involving purely 
low Mach number flows.
Within this latter category, the artificial compressibility
\cite{chorin/1967:JCP} and projection (or fractional time-step) methods 
\cite{chorin/1968:MC, temam/1969:ARMA} are notable. 
They were initially developed for the particular case of incompressible
flows (i.e, low Mach number flows at the quasi-isentropic or quasi-adiabatic 
thermodynamic limit \cite{alazard/2008:DCDS, clausen/2013:PRE}).

The strategy of the artificial compressibility method is to introduce artificial 
subsonic acoustic waves that propagate without influencing the fluid velocity
\cite{dupuy/2021:PRE}. 
This approach recuperates the physical-mathematical structure of a compressible system, explicit in time and local in space.
However, for accurate unsteady simulations, it usually requires a dual-time stepping 
approach with sub-iterations in artificial time (towards the steady-state 
solution), for each physical time step \cite{madsen/2006:CE}.
Additionally, the selection of the artificial wave propagation speed is not 
a trivial task.
A comprehensive analysis of the artificial compressibility method, including 
its physical significance, is presented in \cite{clausen/2013:PRE}, and an 
extension for non-isothermal flows can be found in \cite{dupuy/2021:PRE}.

The projection method, also known as the fractional time-step method, is a type 
of operator-splitting technique \cite{macnamara/2016}. 
It decomposes the original equation (operator) into independent sub-equations
(operators). 
Each sub-equation is solved separately, and the solution to the original 
equation is, then, constructed by combining them. 
The primary advantage of this approach is that solving each sub-equation 
individually is, in general, much more straightforward and computationally 
efficient than solving the original problem directly, although it does introduce 
splitting errors.
An interesting perspective is to interpret this method as a block LU decomposition 
\cite{perot/1993:JCP, perot/1995:JCP}.
This view clarifies the implied approximations and helps avoid the traditional 
confusion regarding the order of accuracy and boundary conditions for intermediate 
(temporary) velocity and pressure.
A key advantage of the projection method over the artificial compressibility method 
is its ability to ensure mass conservation strictly, without relying on an artificial 
compressibility parameter or requiring dual-time stepping.
On the other hand, the global (elliptic) nature of the hydrodynamic pressure makes 
its integration (by inverting a Poisson equation) the most computationally 
intensive part of the method.

The immersed boundary method (IBM) is a numerical technique designed to simulate 
fluid flows involving complex geometries, using convenient orthogonal grids. 
Initially introduced in the 1970s to model blood flow through heart valves 
\cite{peskin/1972:JCP}, IBM has been widely recognised as a versatile and effective 
approach for addressing fluid-structure interaction problems \cite{peskin/2002:AN}.
Fundamentally, the method incorporates a volumetric forcing term into the 
conservation equations \cite{verzicco/2023:ARFM}.
The specific definition and treatment of this forcing differentiate IBM into 
distinct methodological categories, such as continuous and discrete forcing 
approaches \cite{mittal/2005:ARFM}.
As the names suggest, a continuous forcing method incorporates the forcing 
term before discretising the conservation equations, making it independent 
of the numerical schemes. 
Conversely, a discrete forcing method applies the forcing term after 
discretisation, rendering it dependent on the specific numerical 
implementation.
Within the continuous forcing category, penalisation (or penalty) methods 
model the immersed body as a porous medium 
\cite{iwakami/2014:CCP,specklin/2018:EJMBF,mittal/2005:ARFM,verzicco/2023:ARFM}, 
in which the forcing term acts as a damping force \cite{iwakami/2014:CCP}.
It was originally proposed to model natural convection
in porous medium \cite{arquis/1984}.
As the permeability tends to zero, the porous structure approaches 
an impermeable solid.
Penalisation methods can be viewed as a subset of feedback methods 
\cite{verzicco/2023:ARFM}, which were originally introduced 
in \cite{goldstein/1993:JCP}. 
Physically, forcing terms of feedback methods are damped oscillators 
\cite{iaccarino/2003:AMR,mittal/2005:ARFM,verzicco/2023:ARFM}.

A substantial challenge revolves around modelling diverse phenomena properly. 
The model should adeptly encapsulate the desired physics while maintaining 
simplicity to allow insightful interpretation. 
Given this challenge, the current study endeavours to propose a model for fluid 
dynamics at low Mach numbers -- with strong temperature gradients and thermal 
energy diffusion --, using a simplified formulation. 
To address this, a model for reacting flows at low Mach numbers is employed, 
assuming that the chemical reaction is much faster than the characteristic 
flow speed, or, equivalently, the adoption of the Burke--Schumann limit 
(or kinetics) \cite{burke/1948}. 
Under this assumption, reactant concentrations and the system energy can be described 
using coupling functions -- mixture fraction and excess enthalpy --, which are 
transported across the domain without chemical reaction 
\cite{zeldovich/1985, linan/91, fachini/1999:CF, fachini/1999:AIAA, Fachini/2007:IJHMT}.
This approach permits the numerical investigations of diffusion flames, 
and will be detailed later.

Despite the formal simplifications, an infinitely narrow flame presents 
numerical challenges. 
Namely, certain coefficients and derivatives exhibit a discontinuity at the 
flame, due to distinct behaviours within the fuel and oxidiser subdomains.
To avoid numerical oscillations, associated with discontinuities, 
this work proposes a regularisation procedure for these properties.
In addition, it introduces a novel application of the immersed boundary 
method by incorporating mass flux and, thus, expanding its traditional use.
Specifically, the fuel ejection from a cylindrical burner, as well as the 
conditions of all variables at the burner surface, are prescribed using a 
penalisation IBM.
In addition to this clear physical interpretation, as previously described, the penalisation method is chosen because 
of its formal convergence results 
\cite{angot/1999:NM,carbou/2003:ADE}.

Temporal integration is performed using a predictor-corrector scheme, combined 
with a projection method, which has proven efficient in handling strong 
temperature gradients \cite{kooshkbaghi/2013:IJNMF}.
To prevent odd-even decoupling -- between velocity and pressure --, a flux 
interpolation approach is applied on a collocated grid, used for spatial 
discretisation.

In what follows, the text is organised as indicated: 
Section 2 provides a detailed explanation of the physical assumptions 
and the related mathematical model;
Section 3 outlines the numerical method and discusses related details;
Section 4 presents relevant test cases, verifying the numerical model and
its computational implementation;
Finally, Section 5 concludes the study with a summary of findings.

\section{Physical and Mathematical Modelling}
\label{Formulation}
\addvspace{10pt}

This section outlines the physical hypotheses, and the 
corresponding mathematical description, for modelling 
reacting flows at low Mach number.

\subsection{A general model}

Consider a fluid with a density $\hat{\rho}$, flowing at 
a velocity $\hat{\vb*{v}} \coloneqq [\hat{v}_1,\hat{v}_2,\hat{v}_3]^T$, under a 
pressure $\hat{p}$ and with temperature $\hat{T}$. 
This fluid comprises a mixture of $N$ chemical species, with the mass fraction of the $n$-th species given by $\hat{Y}_{n} \coloneqq \hat{\rho}_{n} / \hat{\rho}$, in which $\rho_n$ is the density of the $n$-th species for $n = 1, \dots, N$.
These variables depend, in 
general, both on time $\hat{t}$ and position 
$\hat{\vb*{x}} \coloneqq [\hat{x}_1,\hat{x}_2,\hat{x}_3]^T$. 
Note that the caret symbol ($\hat{\cdot}$)  denotes a dimensional quantity.

The dynamics of this system, subject to buoyancy, can be described by the following set 
of (dimensional) partial differential equations, representing physical laws of conservation \cite{poinsot/2022}:
\begin{equation}\label{Eq2.1}
    \partial_{\hat{t}} 
    \hat{\rho}
    + 
    \hat{\grad} \dotproduct
    \left(
        \hat{\rho} \hat{\vb*{v}}
    \right)
    = 
    0,
\end{equation}
\begin{equation}\label{Eq2.2}
    \partial_{\hat{t}} 
    \left(
        \hat{\rho} \hat{\vb*{v}}
    \right) 
    +
    \hat{\grad} \dotproduct 
    \left(
        \hat{\rho} \hat{\vb*{v}} \hat{\vb*{v}}
    \right) 
    = 
    -\hat{\grad} \hat{p}
    + 
    \hat{\grad} \dotproduct
    \hat{\vb*{\tau}} 
    +
    \left(
        \hat{\rho}
        -
        \hat{\rho}_{c}        
    \right)
    \hat{\vb*{g}},
\end{equation}
\begin{equation}\label{Eq2.3}
    \partial_{\hat{t}} 
    \left(
        \hat{\rho} \hat{Y}_{n}
    \right) 
    +
    \hat{\grad} \dotproduct 
    \left(
        \hat{\rho} 
        \hat{\vb*{v}} 
        \hat{Y}_{n}
    \right)
    = 
    \hat{\grad}
    \dotproduct
    \left(
         \hat{\rho} \hat{D}_{n} 
         \hat{\grad} 
         \hat{Y}_{n}
    \right)
    +
    \hat{\omega}_{n},
\end{equation}
\begin{equation}\label{Eq2.4}
    \partial_{\hat{t}} 
    \left(
    \hat{\rho} \hat{h}
    \right)
    +
    \hat{\grad} \dotproduct
    \left(
        \hat{\rho} 
        \hat{\vb*{v}}
        \hat{h}
    \right)
    =
    \partial_{\hat{t}}
    \hat{p}
    +
    \left(
        \hat{\vb*{v}}
        \dotproduct
        \hat{\grad}
    \right) 
    \hat{p}
    -
    \hat{\grad} \dotproduct
    \hat{\vb*{q}}
    +
    \hat{\grad} \hat{\vb*{v}}
    \boldsymbol{:}
    \hat{\vb*{\tau}}
    -
    \sum_{n=1}^N
    \Delta\hat{h}^0_{n}
    \hat{\omega}_{n},
\end{equation}
supplemented by an equation of estate.

Conventional notation is employed, with 
$\partial_{\hat{t}}$, $\hat{\grad}$ and 
$\hat{\grad} \dotproduct$ denoting 
(dimensional) partial temporal derivative, gradient and divergence 
operators, respectively.
In addition, $\hat{\vb*{v}} \hat{\vb*{v}}$ is the juxtaposition representation of the dyadic product $\hat{\vb*{v}} \otimes \hat{\vb*{v}}$, $\hat{\vb*{\tau}}$ 
is the viscous stress tensor, $\hat{\rho}_{c}$ is a characteristic density for buoyancy, $\hat{\vb*{g}} \coloneqq -\hat{g} 
\vb*{e}_{x_2}$ (with $\vb*{e}_{x_2} \coloneqq [0,1,0]^T$) is the gravity acceleration, 
$\hat{D}_{n}$ is the mass diffusivity of the $n$-th species in the mixture, 
$\hat{\omega}_{n} \coloneqq 
\sum_m \hat{\omega}_{n, m}$ is the chemical 
source term (in which $\hat{\omega}_{n, m}$ 
is the specific production rate of the $n$-th species due to the $m$-th reaction), $\hat{h} 
\coloneqq \int_{\hat{T}_{ref}}^{\hat{T}} \hat{c}_p d\hat{T}$ (for a reference temperature $\hat{T}_{ref}$)
is the sensible enthalpy and $\Delta\hat{h}^0_{n}$ is the standard enthalpy of formation for the $n$-th species.

Note that specific heats of the mixture 
($\hat{c}_v$ and $\hat{c}_p$) are represented by the 
weighted average of the specific heats of each 
species, i.e., $\hat{c}_v 
\coloneqq \sum_{n=1}^N \hat{c}_{v,n} \hat{Y}_{n}$
and $\hat{c}_p
\coloneqq \sum_{n=1}^N \hat{c}_{p,n} \hat{Y}_{n}$.
In addition, while the mass fraction ($\hat{Y}_{n}$) is inherently dimensionless, it is also marked with a caret symbol to distinguish it from $Y_{n}$, which will represent its version normalised by a characteristic value.

The viscous stress tensor ($\hat{\vb*{\tau}}$) is given by the Navier--Poisson law, i.e.,
\begin{equation}\label{Eq2.5}
    \hat{\vb*{\tau}}
    \coloneqq
    \hat{\mu}
    \left[
        \hat{\grad} \hat{\vb*{v}} 
        + 
        \left(
            \hat{\grad} \hat{\vb*{v}}
        \right)^T
        -
        \dfrac{2}{3}
        \left(
            \hat{\grad} \dotproduct \hat{\vb*{v}} 
        \right)
        \vb*{I}
    \right]
    +
    \hat{\zeta}
    \left(
            \hat{\grad} \dotproduct \hat{\vb*{v}} 
        \right)
        \vb*{I},
\end{equation}
in which $\hat{\mu}$ is the shear viscosity, and $\hat{\zeta}$ is the bulk (or volumetric) viscosity.
Moreover, the energy flux ($\hat{\vb*{q}}$) is determined by
\begin{equation}\label{Eq2.6}
    -
    \hat{\vb*{q}}
    \coloneqq
    \hat{\kappa}
    \hat{\grad}
    \hat{T}
    -
    \hat{\rho}
    \sum_{n=1}^N
    \hat{h}_{n}
    \hat{D}_{n}
    \hat{\grad}
    \hat{Y}_{n}
\end{equation}
The first term on the right-hand side corresponds to the Fourier's law for thermal conduction, and the second one, to the diffusion of species with distinct enthalpies (thermal capacity).
In this definition, $\hat{\kappa}$ is the thermal 
conductivity of the mixture and $\hat{h}_{n}$ is the sensible enthalpy for the $n$-th species.
Moreover, the mass diffusion flux is given by the Fick's law, and the thermal energy flux due to radiation is not being explicitly accounted for.

\subsection{Additional assumptions}

It is assumed that the gas mixture behaves 
like a perfect gas, leading to the adoption 
of the ``ideal gas law'':
\begin{equation}\label{Eq2.7}
   \hat{p} 
   = 
   \hat{\rho} 
   \hat{T}
   \hat{R} / \hat{W},
\end{equation}
in which $\hat{R}$ is the universal
gas constant and $\hat{W} \coloneqq 
1/( \sum_{n=1}^N \hat{Y}_{n} / \hat{W}_{n})$ is the (mean) molar mass of 
the mixture, denoting the molar mass of 
the $n$-th species as $\hat{W}_{n}$.

A power law is employed to represent the temperature dependence of transport coefficients \cite{smoke/1991}.
Concretely, considering definitions for the numbers of \Pr[s], $\Pr \coloneqq \hat{\mu} / (\hat{\kappa} / \hat{c}_{p})$, and \Le[s] for the $n$-th species, $\Le_n \coloneqq [ \hat{\kappa} / (\hat{\rho} \hat{c}_{p}) ] / \hat{D}_n $,  the relations $\hat{\mu}/\Pr = \hat{\rho} \hat{D}_n \Le_n = \hat{\kappa} / \hat{c}_p$ hold, in which
\begin{equation}\label{Eq2.8}
     \dfrac{\hat{\kappa}}{\hat{c}_p}
     = \hat{A}
     \left(
        \dfrac{\hat{T}}{\hat{T}^*}
    \right)^\sigma
\end{equation}
with $A \coloneqq \SI{2.58e-5}{\kilogram \per(\meter \cdot \second)} $ and $\sigma \coloneqq 0.7$, for a reference temperature $\hat{T}^* = \SI{298}{\kelvin}$.
Alternatively, the Sutherland's law could be used \cite{Sutherland/1893:TLDPMJS}.

Take into account identical specific heats for all species, i.e., $\hat{c}_{v} = \hat{c}_{v,i} = \hat{c}_{v,j}$
and
$\hat{c}_{p} = \hat{c}_{p,i} = \hat{c}_{p,j}$
for each $i,j \in \{1, \dots, N \}$.
Hence, the second term of energy flux, in Eq. (\ref{Eq2.6}), is zero \cite{smoke/1991, poinsot/2022}.
In addition, for a calorifically
perfect gas (constant specific heats), the sensible enthalpy can be expressed as $\hat{h} = \hat{c}_p \hat{T}$.

Combustion is modelled through an irreversible, one-step global reaction between a fuel and an oxidant, as expressed symbolically by:
\begin{equation}\label{Eq2.9}
    \text{F} + \hat{s} \text{O} \rightarrow (1+\hat{s}) \text{P} \quad \quad - \Delta \hat{h}^0,
\end{equation}
$\hat{s}$ represents the stoichiometric coefficient in terms of mass, i.e., to react with one unit of fuel (`F') mass -- producing $(1+\hat{s})$ units of product (`P') mass and releasing heat $\Delta \hat{h}^0$ --, $\hat{s}$ units of oxidiser (`O') mass are required.
Similarly to the case of the mass fraction ($\hat{Y}_{n}$), it is emphasised that the parameter $\hat{s}$ is dimensionless. 
However, it is introduced with a hat to distinguish it from its normalised version ($s$), found during the nondimensionalisation procedure.
Therefore, considering $n=1$ for fuel, $n=2$ for oxidiser and $n=3$ for products, 
\begin{equation}\label{Eq2.10}
    \hat{\omega}_{n}
    \coloneqq
    c_n \ \hat{\omega}_{1}
\end{equation}
in which $c_n = 0$ for an inert $n$-th species (i.e., for $n \ne 1, 2, 3$), and $c_1 = 1$, $ c_2 = \hat{s}$ and  $c_3 = -(1+\hat{s})$, since $\hat{\omega}_{1} = \hat{\omega}_{2}/\hat{s} = -\hat{\omega}_{3}/(1+\hat{s})$.
Also, $\sum_{n=1}^N \Delta\hat{h}^0_{n}\hat{\omega}_{n} = \hat{\omega}_1 \sum_{n=1}^N c_n \Delta\hat{h}^0_{n} = \Delta \hat{h}^0 \hat{\omega}_1$, by definition.
Moreover, an Arrhenius equation models the reaction rate, viz.,
\begin{equation}\label{Eq2.11}
    \hat{\omega}_1 
    =
    \hat{\rho}
    \hat{B}
    \hat{Y}_1 \hat{Y}_2
    e^{-\hat{E}/(\hat{R}\hat{T})}
\end{equation}
in which $\hat{B}$ is a pre-exponential factor, and
$\hat{E}$ is the activation energy for the chemical reaction.

These assumptions on chemical reaction, combined with the supposition of constant mean molar mass ($\hat{W} = \text{const.}$) -- i.e., the fluid is a homogeneous mixture in whole domain --, permits the description of the system dynamics in terms of only two species, fuel ($\hat{Y}_1$) and oxidiser ($\hat{Y}_2$) \cite{linan/91}.

Considering the previous hypotheses, numerical indices will be replaced by $F$ (for fuel) and $O$ (for oxidiser), to imbue them with more suggestive notation, for subsequent discussions.

\subsection{Nondimensionalisation}

Denoting the characteristic values with a subscript `$c$', dimensionless quantities are obtained as:
\begin{equation}\label{Eq2.12}
     \begin{array}{lcl}
        t \coloneqq \hat{t} / \hat{t}_c &, & \vb*{x} \coloneqq \hat{\vb*{x}}/ \hat{l}_c \\
        \rho \coloneqq \hat{\rho}/ \hat{\rho}_c &, & \vb*{v} \coloneqq \hat{\vb*{v}}/ \hat{v}_c \\
        T \coloneqq \hat{T}/ \hat{T}_c &, & p \coloneqq \hat{p}/ \hat{p}_c \\
        Y_{F} \coloneqq \hat{Y}_{F}/ \hat{Y}_{{F},c} &, & Y_{O} \coloneqq \hat{Y}_{O}/ \hat{Y}_{{O},c} \\
        D_{F} \coloneqq \hat{D}_{F}/ \hat{D}_{{F},c} &, & D_{O} \coloneqq \hat{D}_{O}/ \hat{D}_{{O},c} \\
        \mu \coloneqq \hat{\mu}/ \hat{\mu}_c &, & \zeta \coloneqq \hat{\zeta}/ \hat{\mu}_c \\
        \kappa \coloneqq \hat{\kappa}/ \hat{\kappa}_c &, & c_p \coloneqq \hat{c}_p/ \hat{c}_{p,c} = 1 \\
        s \coloneqq \hat{s} \hat{Y}_{F,c} / \hat{Y}_{O,c} &, & \Delta h^0 \coloneqq \Delta \hat{h}^0 \hat{Y}_{F,c} / (\hat{c}_{p,c} \hat{T}_c) \\
        \vb*{g} \coloneqq \hat{\vb*{g}} / \hat{g}_{c} = - \vb*{e}_{x_2}
    \end{array}
\end{equation}
which induce dimensionless operators:
\begin{equation}\label{Eq2.13}
     \begin{array}{ccccc}
        \partial_t \coloneqq \hat{t}_c \partial_{\hat{t}} 
        &, & 
        \grad \coloneqq \hat{l}_c \hat{\grad}
        &, & 
        \div \coloneqq \hat{l}_c \hat{\grad} \boldsymbol{\cdot} 
    \end{array}
\end{equation}

The system, under these additional hypotheses and definitions, is described by the following set of dimensionless equations:
\begin{equation}\label{Eq2.14}
    \St 
    \partial_{t} 
    \rho
    + 
    \div
    \left(
        \rho \vb*{v}
    \right)
    = 
    0,
\end{equation}
\begin{equation}\label{Eq2.15}
    \St
    \partial_{t} 
    \left(
        \rho \vb*{v}
    \right) 
    +
    \div 
    \left(
        \rho \vb*{v} \vb*{v}
    \right) 
    = 
    -
    \dfrac{1}{ \MM^2 }
    \grad p
    + 
    \dfrac{\Pr}{\Pe}
    \div
    \tau 
    +
    \dfrac{1}{\Fr^2}
    \left(
        1
        -
        \rho 
    \right)
    \vb*{e}_{x_2},
\end{equation}
\begin{equation}\label{Eq2.16}
    \St
    \partial_{t} 
    \left(
        \rho Y_{F}
    \right) 
    +
    \div
    \left(
        \rho 
        \vb*{v}
        Y_{F}
    \right)
    = 
    \dfrac{1}{\Pe \Le_F}
    \div
    \left(
         \kappa 
         \grad 
         Y_{F}
    \right)
    +
    \Dah
    \omega_{F},
\end{equation}
\begin{equation}\label{Eq2.17}
    \St
    \partial_{t} 
    \left(
        \rho Y_{O}
    \right) 
    +
    \div
    \left(
        \rho 
        \vb*{v}
        Y_{O}
    \right)
    = 
    \dfrac{1}{\Pe \Le_O}
    \div
    \left(
         \kappa 
         \grad 
         Y_{O}
    \right)
    +
    s 
    \Dah
    \omega_{F},
\end{equation}
\begin{multline}\label{Eq2.18}
    \St
    \partial_{t} 
    \left(
        \rho T
    \right)
    +
    \div
    \left(
        \rho \vb*{v} T 
    \right)
    =
    \dfrac{\gamma - 1}{\gamma}
    \left[
        \partial_{t}
        p
        +
        \left(
            \vb*{v}
            \dotproduct
            \grad
        \right)
        p
        +
        \MM^2
        \dfrac{\Pr}{\Pe}
        \left(
            \grad \vb*{v}
            \boldsymbol{:}
            \vb*{\tau}
        \right)
    \right]
    + \\ +
    \dfrac{1}{\Pe}
    \div
    \left(
        \kappa
        \grad T
    \right)
    -
    \Delta h^0
    \Dah
    \omega_{F},
\end{multline}
together with the equation of state:
\begin{equation}\label{Eq2.19}
   p 
   =
   \rho 
   T,
\end{equation}
in which 
$\St \coloneqq (\hat{l}_c/\hat{v}_c)/ \hat{t}_c$ is the \Sr[l],  
$\Pr \coloneqq \hat{\mu}_c / (\hat{\kappa}_c / \hat{c}_{p,c})$ is the \Pr[l], 
$\Pe \coloneqq \hat{l}_c \hat{v}_c / [\hat{\kappa}_c / (\hat{\rho}_c \hat{c}_{p,c})]$ is the \Pe[l], 
$\Fr \coloneqq \hat{v}_c / (\hat{g}_c \hat{l}_c)^{1/2}$ is the \Fr[l], 
$\Le_\eta \coloneqq [ \hat{\kappa}_c / (\hat{\rho}_c \hat{c}_{p,c}) ] / \hat{D}_\eta $ is the \Le[l] for the species $\eta \in \{ F, O \}$, 
$\Dah \coloneqq (\hat{l}_c / \hat{v}_c) / [\hat{Y}_{O,c} \hat{B} e^{-\hat{E}/(\hat{R}\hat{T})}]^{-1}$ is the \Dah[l], 
$\gamma \coloneqq \hat{c}_{p,c} / \hat{c}_{v,c}$ is the adiabatic index, 
$\MM \coloneqq \gamma^{1/2} \M$ is proportional to the \M[l] $\M \coloneqq \hat{v}_c / (\gamma \hat{p}_c / \hat{\rho}_c)^{1/2}$.

Moreover, the viscous stress tensor ($\vb*{\tau}$) is now expressed as
\begin{equation}\label{Eq2.20}
    \vb*{\tau} 
    =
    \kappa
    \left[
        \grad \vb*{v} 
        + 
        (\grad \vb*{v})^T
        -
        \dfrac{2}{3} 
        \left(
            \div \vb*{v} 
        \right)
        \vb*{I}
    \right]
    +
    \zeta
    \left(
        \div \vb*{v} 
    \right)
    \vb*{I},
\end{equation}

In this context, it is emphasised that the relative intensity between convection and diffusion is expressed as a proportion of the \Pe[l] ($\Pe$), in each equation. 
Furthermore, the dimensionless transport coefficients reads
\begin{equation}\label{Eq2.21}
     \kappa = \mu = \rho D_{\eta} = T^\sigma
\end{equation}
with $\sigma = 0.7$ and $\eta \in \{ F, O \}$.
%

Henceforth, $\St \coloneqq 1$, indicating that $\hat{t}_c$ is the residence time (i.e., $\hat{t}_c \coloneqq \hat{l}_c / \hat{v}_c$).

\subsection{Low-Mach-number limit}\label{lowMach}

Simulating flows at low Mach numbers presents a significant challenge 
in computational fluid dynamics (CFD) due to the large disparity 
between acoustic and convective time scales.
In such regimes, the speed of sound exceeds the characteristic flow 
velocity by orders of magnitude, creating a multi-scale problem that 
complicates numerical modelling. 
This phenomenon is ubiquitous in natural systems -- such as atmospheric 
or biological flows -- and engineering applications -- such as cooling and 
reacting flows --, for which resolving acoustic interactions alongside bulk 
fluid motion demands computationally intensive techniques to ensure accuracy 
and stability.

To address this challenge, the present section derives the 
low-Mach-number asymptotic limit of the conservation equations. 
By systematically isolating the dominant physical processes through 
asymptotic analysis, the conservation equations are formally simplified 
to eliminate (instantaneous) acoustic effects while preserving convective and diffusive 
dynamics. 
This approach decouples the disparate scales inherent to low-Mach-number 
flows, yielding a simplified system that retains fidelity to the 
underlying physics, while enabling computationally viable simulations.

In deriving the low-Mach-number equations, 
a multi-scale expansion technique is employed
for each dependent variable, expressed in 
terms of powers of the \M[l] 
\cite{muller/1998, schochet/2005:ESAIM}.

Defining the fast time scale, 
\begin{equation}\label{Eq2.22}
    \xi 
    \coloneqq
    \dfrac{\hat{t}}
    {\hat{l}_c / (\hat{p}_c / \hat{\rho}_c)^{1/2}}
    =
    \dfrac{t}
    {\MM}
\end{equation}
related to the propagation of acoustic waves, dependent variables are 
expanded in power series of $\MM$, as functions of two temporal 
scales ($t$ and $\xi$) and one spatial scale 
($\vb*{x}$):
\begin{equation}\label{Eq2.23}
    \vb*{\mathcal{A}}(\vb*{x},t;\MM)
    = 
    \vb*{\mathcal{A}}_0(\vb*{x},t,\xi)
    +
    \MM
    \vb*{\mathcal{A}}_1(\vb*{x},t,\xi)
    +
    \MM^2
    \vb*{\mathcal{A}}_2(\vb*{x},t,\xi)
    +
    \order{\MM^3},
\end{equation}
in which $\vb*{\mathcal{A}} \coloneqq [\rho, \vb*{v}, p, Y_F, Y_O, T]^T$ 
is the state vector.

Hence, the temporal derivative consists of a flow-changing 
contribution ($\partial / \partial t$) and an acoustic-propagation 
contribution ($\MM^{-1} \partial / \partial \xi$), for a given 
position ($\vb*{x}$) and \MM[l] ($\MM$).
Mathematically,
\begin{equation}\label{Eq2.24}
    \left.
        \partial_t \vb*{\mathcal{A}}
    \right|_{\vb*{x}, \MM}
    = 
    \left(
        \dfrac{\partial}{\partial t}
        +
        \dfrac{1}{\MM}
        \dfrac{\partial}{\partial \xi}
    \right)
    \left[
        \vb*{\mathcal{A}}_0
        +
        \MM
        \vb*{\mathcal{A}}_1
        +
        \MM^2
        \vb*{\mathcal{A}}_2
        +
        \order{\MM^3}
    \right]
\end{equation}

The derivation of the low-Mach-number equations 
involves substituting this expansion into the equations 
and integrating the first-order equations of mass, 
species, and energy, as well as the second-order 
equation of momentum, over the acoustic wave period 
($\mathrm{C_a}$).
Note that it only captures the net acoustic effect on
the average velocity field ($\overline{\vb*{v}_0}$),
within the average velocity tensor 
($\overline{\vb*{v}_0 \vb*{v}_0} 
\coloneqq 
\mathrm{C_a} ^{-1}\int_0^{\mathrm{C_a}} 
\vb*{v}_0 \vb*{v}_0 d \xi$), as detailed in 
\cite{muller/1998, schochet/2005:ESAIM}. 
Alternatively, one could have considered the 
acoustic spatial scale instead of the acoustic 
temporal scale \cite{klein/1995:JCP}.

The influence of acoustics can be completely 
eliminated by approximating the average velocity 
tensor ($\overline{\vb*{v}_0 \vb*{v}_0}$) as the 
tensor of average velocity 
($\overline{\vb*{v}}_0 \overline{\vb*{v}}_0$), culminating 
in the zero-Mach-number limit \cite{majda/1985:CST},
which is equivalent to consider solely the flow 
time scale within the expansion 
\cite{sivashinsky/1979:AA, majda/1984, majda/1985:CST, muller/1998},
viz.,
\begin{equation}\label{Eq2.25}
    \vb*{\mathcal{A}}(\vb*{x},t;\MM)
    = 
    \vb*{\mathcal{A}}_0(\vb*{x},t)
    +
    \MM
    \vb*{\mathcal{A}}_1(\vb*{x},t)
    +
    \MM^2
    \vb*{\mathcal{A}}_2(\vb*{x},t)
    +
    \order{\MM^3}.
\end{equation}
The zero-Mach-number equations will be considered in this work.
Introducing Eq. (\ref{Eq2.25}) into Eqs. (\ref{Eq2.14}) - (\ref{Eq2.21}), 
yields the following zeroth-order equations:
\begin{equation}\label{Eq2.26}
    \partial_{t} 
    \rho_0
    + 
    \div
    \left(
        \rho_0 \vb*{v}_0
    \right)
    = 
    0,
\end{equation}
\begin{equation}\label{Eq2.27}
    \partial_{t} 
    \left(
        \rho_0 \vb*{v}_0
    \right) 
    +
    \div 
    \left(
        \rho_0 \vb*{v}_0 \vb*{v}_0
    \right) 
    = 
    -
    \grad p_2
    + 
    \dfrac{\Pr}{\Pe}
    \div
    \tau_0 
    +
    \dfrac{1}{\Fr^2}
    \left(
        1
        -
        \rho_0 
    \right)
    \vb*{e}_{x_2},
\end{equation}
\begin{equation}\label{Eq2.28}
    \partial_{t} 
    \left(
        \rho_0 Y_{F,0}
    \right) 
    +
    \div
    \left(
        \rho_0 \vb*{v}_0 Y_{F,0}
    \right)
    = 
    \dfrac{1}{\Pe \Le_F}
    \div
    \left(
         \kappa_0
         \grad 
         Y_{F,0}
    \right)
    +
    \Dah
    \omega_{F,0},
\end{equation}
\begin{equation}\label{Eq2.29}
    \partial_{t} 
    \left(
        \rho_0 Y_{O,0}
    \right) 
    +
    \div
    \left(
        \rho_0 \vb*{v}_0 Y_{O,0} 
    \right)
    = 
    \dfrac{1}{\Pe \Le_O}
    \div
    \left(
         \kappa_0
         \grad 
         Y_{O,0}
    \right)
    +
    \Dah
    s
    \omega_{F,0},
\end{equation}
\begin{equation}\label{Eq2.30}
    \partial_{t} 
    \left(
        \rho_0 T_0
    \right)
    +
    \div
    \left(
        \rho_0 \vb*{v}_0 T_0
    \right)
    =
    \dfrac{\gamma - 1}{\gamma}
    \dfrac{d p_0}{dt} 
    +
    \dfrac{1}{\Pe}
    \div
    \left(
        \kappa_0
        \grad T_0
    \right)
    -
    \Dah
    \Delta h^0
    \omega_{F,0},
\end{equation}
\begin{equation}\label{Eq2.31}
   p_0 
   = 
   \rho_0 
   T_0,
\end{equation}
in which $\kappa_0 \coloneqq \kappa(T_0)$, and
\begin{equation}\label{Eq2.32}
    \vb*{\tau}_0
    =
    \kappa_0
    \left[
        \grad \vb*{v}_0 
        + 
        (\grad \vb*{v}_0)^T
        -
        \dfrac{2}{3} 
        \left(
            \div \vb*{v}_0 
        \right)
        \vb*{I}
    \right]
\end{equation}

Observe that, the term corresponding to the bulk viscous pressure 
-- the second term on the right-hand side (RHS) of Eq. (\ref{Eq2.20}) -- 
is not explicitly included into Eq. (\ref{Eq2.32}). 
In the low-Mach-number limit, the hydrodynamic pressure and viscous 
bulk pressure can be integrated, because both contribute to stresses 
as volumetric (isotropic) dilatations, but do not affect the energy balance 
or thermodynamic state.
A detailed discussion can be found in \cite{papalexandris/2020:CMT}.

The zeroth-order pressure ($p_0$) can be associated with the 
thermodynamic pressure, while the second-order one ($p_2$) can be 
associated with the hydrodynamic pressure.
It is justified because they share the same order of magnitude in the 
momentum and state equations, respectively, as evidenced in their 
representation as order-unit terms.

Additionally, a constraint for $p_0$ emerges from the momentum equation,
\begin{equation}\label{Eq2.33}
   \grad p_0 
   = 
   0,
\end{equation}
which implies $p_0 = p_0(t)$, i.e., the thermodynamic pressure spreads 
along the system instantaneously.

In the interest of clarity, only the zeroth-order index for the 
thermodynamic pressure (i.e., $p_0$) will be retained henceforward.

\subsection{Flame sheet approximation}

For scenarios in which the reaction rate significantly exceeds the 
flow-rate variation (i.e., for $\Dah \gg 1$), coexistence of reactants 
within the domain is not possible.
Under the premise of comparable magnitudes between the contributions of 
convective and reactive terms (i.e., $\Dah Y_F Y_O \sim 1$), it is 
established that $Y_F Y_O \sim \Dah^{-1}$.
Therefore, as $\Dah$ approaches infinity, $Y_F Y_O$ tends to zero, implying 
that $Y_F = 0$ or $Y_O = 0$ at every point in the domain.
Thus, an infinitely thin reaction zone (known as ``flame sheet'' or 
``flame surface'') separates the fuel subdomain (where $Y_F \ne 0$ and 
$Y_O = 0$) from the oxidiser subdomain (where $Y_F = 0$ and $Y_O \ne 0$). 
Meanwhile, at the flame surface, both reactants have zero concentration 
($Y_F = 0$ and $Y_O = 0$).

The flame sheet approximation was initially proposed by Burke \& Schumann, 
in a seminal paper that formalised diffusion (non-premixed) flames theory 
for the first time \cite{burke/1948}.
Under this Burke--Schumann limit (or kinetics), a Schvab--Zeldovich--Liñán 
formulation \cite{zeldovich/1985, linan/91, fachini/1999:CF, fachini/1999:AIAA, Fachini/2007:IJHMT} 
can be employed to describe a non-equidiffusional reacting flow using 
coupling functions, which are transported without reaction.
These functions are known as mixture fraction ($Z$) and excess enthalpy ($H$), 
and their equations replace  species and energy equations.

By subtracting Eq. (\ref{Eq2.29}) multiplied by $\Le_O / \Le_F$, from 
Eq. (\ref{Eq2.28}) multiplied by $S$, an equation for $Z \coloneqq S Y_F - Y_O + 1$ 
is obtained as
\begin{equation}\label{Eq2.34}
    \partial_{t} 
    \left(
        \rho
        Z
    \right)
    +
    \div
    \left(
        \rho
        \vb*{v}
        Z        
    \right)
    = 
    \dfrac{1}{\Le \Pe}
    \div
    \left(
        \kappa \grad Z
    \right)
\end{equation}
in which $S \coloneqq s\Le_O / \Le_F$ is a generalised stoichiometric 
coefficient, and
\begin{equation}\label{Eq2.35}
    \Le(Z) 
    \coloneqq
    (\Le_F - \Le_O)  \theta(Z-1) + \Le_O,
\end{equation}
with $\theta$ being the Heaviside step function.
Note that under this definition of mixture fraction 
(a rescaled version), the flame is always located on the level 
curve $Z = 1$.

Analogously, by adding Eq. (\ref{Eq2.28}) multiplied by $\Le_F$, 
Eq. (\ref{Eq2.29}) multiplied by $\Le_O$, and Eq. (\ref{Eq2.30}) 
multiplied by $(S+1)/Q$, an equation for $H \coloneqq (S+1)T/Q + Y_F + Y_O$ 
is derived as
\begin{equation}\label{Eq2.36}
    \partial_{t} 
    \left(
        \rho
        H
    \right)
    +
    \div
    \left(
        \rho
        \vb*{v}
        H
    \right)
    =
    \dfrac{S+1}{Q}
    \dfrac{\gamma - 1}{\gamma}
    \dfrac{d p_0}{dt}
    +
    \dfrac{1}{\Pe}
    \left[
        \div
        \left(
            \kappa
            \grad H
        \right)
        +
        \N
        \div
        \left(
            \kappa
            \grad Z
        \right)
    \right]
\end{equation}
in which $Q \coloneqq \Delta h^0 / \Le_{F}$ is a generalised heat 
released, and
\begin{equation}\label{Eq2.37}
    \N(Z) 
    \coloneqq 
    (\N_F - \N_O)\theta(Z-1) + \N_O,
\end{equation}
with $\N_F \coloneqq (1 - \Le_F)/(S \Le_F)$ and $\N_O \coloneqq (\Le_O - 1)/\Le_O$.

To close the system, it is missing an equation for the thermodynamic 
pressure ($p_0$) \cite{nicoud/2000:JCP}.
Returning the left-hand side (LHS) of Eq. (\ref{Eq2.36}) to primitive variables, using the equation of state (\ref{Eq2.31}) to identify the pressure term in LHS, taking into account the mixture fraction equation (\ref{Eq2.34}) and the definition of the function $\N$ in Eq. (\ref{Eq2.37}), one finds

\begin{equation}\label{Eq2.38}
    \dfrac{d p_0}{dt}
    =
    \dfrac{\gamma}{V}
    \left\{
        \dfrac{Q}{S+1}
        \dfrac{1}{\Pe}
        \int\limits_V
        \bigg[
            \div
            \left(
                \kappa
                \grad 
                H
            \right)
            +
            \Np
            \div
            \left(
                \kappa
                \grad 
                Z
            \right)
        \bigg]
        dV
        -
        p_0
        \int\limits_V
        \left(
            \div
            \vb*{v}
        \right)
        dV
    \right\}
\end{equation}
i.e., $p_0$ is obtained as solution of an ordinary differential equation, 
over a domain with volume $V$.
Evidently, $\Np$ has the same behaviour as $\N$,
\begin{equation}\label{Eq2.39}
    \Np(Z) 
    \coloneqq 
    (\Np_F - \Np_O)\theta(Z-1) + \Np_O,
\end{equation}
but with $\Np_F \coloneqq -1/S$ and $\Np_O \coloneqq 1$.

Two simplifications are possible depending on the system boundary:

(i) for enclosed systems, 
\begin{equation}\label{Eq2.40}
   p_0
   \int\limits_V \dfrac{1}{T} dV
   = 
   \int\limits_V \rho dV
   =
   m,
\end{equation}
i.e., $p_0$ can be obtained by integrating the equation of state over 
the domain volume ($V$), since the total mass ($m$) remains constant 
over time; and

(ii) for open systems,
\begin{equation}\label{Eq2.41}
   p_0
   = 
   \dfrac{\hat{p}_0}{\hat{p}_{c}}
   = 
   1,
\end{equation}
i.e., the thermodynamic pressure is determined by atmospheric pressure, 
as its influence is instantaneously transmitted from the boundary to the 
entire domain.

Therefore, the direct use of Eq. (\ref{Eq2.38}) is only required for 
semi-enclosed systems, i.e., for which the degree of openness varies over 
time (e.g., internal combustion engine cylinders).

\subsection{Final model}

Summarising, a chemically reacting flow at low Mach number, considering 
buoyancy effects, and under the flame surface approximation 
(infinitely fast chemistry), can be
mathematically described by the following system of (dimensionless) 
equations:
\begin{equation}\label{Eq2.42}
    \partial_{t} 
    \rho  
    + 
    \div
    (\rho \vb*{v})
    = 
    0,
\end{equation}
\begin{equation}\label{Eq2.43}
    \partial_{t} 
    (\rho \vb*{v}) 
    +
    \div (
    \rho \vb*{v} \vb*{v}) 
    = 
    -
    \grad   p
    + 
    \dfrac{\Pr}{\Pe} 
    \div
    \vb*{\tau} 
    +
    \dfrac{1}{\Fr^2}
    (1 - \rho)
    \vb*{e}_{x_2},
\end{equation}
\begin{equation}\label{Eq2.44}
        \rho
        \partial_{t} 
        Z 
        +
        \rho
        \left(
            \vb*{v}
            \dotproduct
            \grad
        \right)
        Z
     = 
     \dfrac{1}{\Le \Pe}
     \div
     \left(
        \kappa \grad Z
     \right),
\end{equation}
\begin{equation}\label{Eq2.45}
        \rho
        \partial_{t} 
        H
        +
        \rho
        \left(
            \vb*{v}
            \dotproduct
            \grad
        \right)
        H
        =
        \zeta
        d_t p_0
        +
        \dfrac{1}{\Pe}
        \left[
            \div
            \left(
                \kappa
                \grad H
            \right)
            +
            \N
            \div
            \left(
                \kappa
                \grad Z
            \right)
        \right],
\end{equation}
with $\zeta \coloneqq (S+1)(\gamma - 1)/(\gamma Q)$, $d_t p_0 \coloneqq d p_0 / dt$, 
and the viscous stress tensor ($\vb*{\tau}$) modelled by the (simplified) Navier--Poisson law
\begin{equation}\label{Eq2.46}
    \vb*{\tau} 
    =
    \kappa
    \left[
        \grad \vb*{v} 
        + 
        (\grad \vb*{v})^T
        -
        \dfrac{2}{3} 
        \left(\grad \dotproduct \vb*{v} \right)
        \vb*{I}
    \right],
\end{equation}
In addition, the following piecewise definitions for the Lewis number ($\Le$) and $\N$ account for differential diffusion:
\begin{equation}\label{Eq2.47}
    \Le(Z) 
    \coloneqq
    (\Le_F - \Le_O)  \theta(Z-1) + \Le_O,
\end{equation}
and
\begin{equation}\label{Eq2.48}
    \N(Z) 
    \coloneqq 
    (\N_F - \N_O)\theta(Z-1) + \N_O.
\end{equation}

The system is complemented by the ideal gas law
\begin{equation}\label{Eq2.49}
   p_0 
   = 
   \rho T,
\end{equation}
to describe the thermodynamic state, and a power law to capture the 
temperature dependence of transport coefficients
\begin{equation}\label{Eq2.50}
   \kappa
   =
   T^\sigma
\end{equation}
with $\sigma = 0.7$.

The equations for mixture fraction ($Z$) and excess enthalpy ($H$) variables 
have been expressed in the non-divergent (or non-conservative) form for 
convenience.
Specially, for the calculus of the regularised temperature derivative, as 
described in Sec. \ref{flameReg}.
The modelled system is, naturally, subject to specific initial and boundary 
conditions, which will be described on a case-by-case basis, in 
Sec. \ref{simulations}.

Compared to related works, the proposed model additionally accounts for the 
effects of buoyancy and preferential or differential diffusion, in a more 
simplified combustion formulation. \\



\section{Numerical Method}
\label{nMethod}
\addvspace{10pt}

Hydrodynamic variables evolution relies on a fractional time-step 
(or projection) method \cite{Denaro/2003:IJNMF}, as proposed 
in \cite{lessani/2006:JCP, najm/1998:JCP}.
This method was introduced by Chorin \cite{chorin/1968:MC} and 
Témam \cite{temam/1969:ARMA}, whose the former proposes an elegant
interpretation based on the decomposition theorem of Ladyzhenskaya 
(or Helmholtz–Hodge decomposition) \cite{Denaro/2003:IJNMF, bladel/1958}.
Specifically, the fractional time-step method involves a formal separation 
of the pressure term from the other momentum terms, consisting, 
therefore, an operator splitting technique \cite{macnamara/2016}. 
An insightful approach is to interpret this method as a block LU decomposition, 
as proposed by Perot \cite{perot/1993:JCP, perot/1995:JCP}. 
This perspective clarifies the underlying assumptions, addressing persistent 
questions about boundary conditions for intermediate (temporary) velocity 
and pressure fields, as well as about the degradation 
of the temporal accuracy.
An elliptic equation for pressure enforces the velocity divergence constraint 
(i.e., pressure acts as a Lagrange multiplier) \cite{perot/1993:JCP}.
In general, an inhomogeneous restriction on the velocity divergence is imposed (related to thermal compressibility), 
which is simplified to a homogeneous one in the incompressible limit 
\cite{alazard/2008:DCDS}.

Partial differential equations are converted to a system of ordinary differential 
equations via spatial discretisation (method of lines) \cite{hairer/1993, leveque/2007}, 
enabling the modular use of temporal integrators. 
Therefore, a predictor-corrector method is employed for time marching.
The predictor stage uses a second-order Adams–Bashforth scheme 
\cite{ferziger/2002}, and incorporates a pressure correction step to ensure
mass conservation.
Similarly, the corrector stage is constructed under a second-order 
Adams--Moulton scheme \cite{ferziger/2002}, also involving a pressure 
correction step. 
The combination of multi-step methods, explicit for the predictor 
(Adams–Bashforth method) and implicit for the corrector 
(Adams--Moulton method), is well-established and has already been used 
in related studies \cite{najm/1998:JCP, tyliszczak/2014:IJNMHFF}.
In both stages, the pressure calculation arises from the solution of a 
Poisson equation, as previously mentioned, representing the most significant computational demand 
of the method.

The physical domain is discretised into a collocated grid. 
The spatial discretisation of operators is based on centred approximations, 
along with the introduction of auxiliary fluxes to prevent
non-physical pressure oscillations, known as ``odd-even decoupling''  
\cite{lessani/2006:JCP, morinishi/1998:JCP} (or ``checkerboarding''), 
which arises from the inability of the system to conserve mass and 
momentum with the equivalent accuracy, when all variables are calculated 
at the same points.

\subsection{Temporal discretisation}\label{tempDis}

Superscripts are used for time indication.
Known values from the past ($t_{n-m}$) and at the present ($t_n$) 
are indicated by indices `$n-m$' and `$n$', respectively, 
in which $t_{n-m} \coloneqq t_n - m\Delta t$, for an integer $m$ and the
time step $\Delta t$.
Unknown, predicted and corrected values at the future 
$t_{n+1} \coloneqq t_n + \Delta t$ are represented by indices
`$*$' and `$n+1$', in this order.

\subsubsection{Predictor stage}

\begin{enumerate}[label=(\roman*)]

\item The estimated value of the mixture fraction ($Z^*$) is computed using 
Eq. (\ref{Eq2.44}), from known values of instants $t_n$ and $t_{n-1}$, namely

\begin{equation}\label{Eq3.1}
     \dfrac{Z^* - Z^{n}}{\Delta t}
     =
     \dfrac{3}{2}
     \text{Res}_Z(\rho^{n},\vb*{v}^{n},Z^{n})
     -
     \dfrac{1}{2}
     \text{Res}_Z(\rho^{n-1},\vb*{v}^{n-1},Z^{n-1}), 
\end{equation}
in which
\begin{equation}\label{Eq3.2}
    \text{Res}_Z
    \coloneqq
    \dfrac{1}{\rho}
    \left[
        -
        \rho
        \left(\vb*{v} \dotproduct \grad \right)Z
        +    
        \dfrac{1}{\Le \Pe}
        \div \left(\kappa \grad Z\right)
    \right]
\end{equation}

    \item Similarly, the predicted value for the excess enthalpy 
    ($H^*$) is derived from Eq. (\ref{Eq2.45}) as

\begin{multline}\label{Eq3.3}
     \dfrac{H^* - H^{n}}{\Delta t}
     =
    \dfrac{3}{2}
    \left[
         \text{Res}_H(\rho^{n},\vb*{v}^{n},H^{n})
         +
         f_H(\rho^{n},d_t p_0|^{n}, Z^n)
    \right]
    - \\ -
    \dfrac{1}{2}
    \left[
         \text{Res}_H(\rho^{n-1},\vb*{v}^{n},H^{n-1})
         +
         f_H(\rho^{n-1},d_t p_0|^{n-1}, Z^{n-1})
    \right],
\end{multline}
in which
\begin{subequations}\label{Eq3.4}
    \begin{gather}
        \text{Res}_H
        \coloneqq
        \dfrac{1}{\rho}
        \left[
            -
            \rho
            \left(
                \vb*{v}
                \dotproduct
                \grad
            \right)
            H
            +
            \dfrac{1}{\Pe}
            \div
            \left(
                \kappa
                \grad H
            \right)
        \right],
       \label{Eq3.4a} \\ 
        f_H
        \coloneqq
        \underbrace{
            \dfrac{1}{\rho}
            \left[
                \zeta      
                d_t p_0
                +
                \frac{\N}{\Pe}
                \div
                \left(
                    \kappa
                    \grad Z
                \right)
            \right]
        }_{source \ term \ for \ H-equation},
    \label{Eq3.4b}
    \end{gather}
\end{subequations}
with
\begin{equation}\label{Eq3.5}
     \begin{array}{ccc}
        \left.
        d_t p_0
    \right|^{n}
    \approx
    \dfrac{3 p_0^{n} - 4 p_0^{n-1} + p_0^{n-2} }
    {2 \Delta t}, & \text{and} & \left.
        d_t p_0
    \right|^{n-1}
    \approx
    \dfrac{3 p_0^{n-1} - 4 p_0^{n-2} + p_0^{n-3} }
    {2 \Delta t}.
    \end{array}
\end{equation}

\item The predicted value for temperature, $T^* = T(Z^*,H^*)$, is 
determined using the definitions of $Z$ and $H$ 
(see Sec. \ref{flameReg} for details). 
Consequently, density ($\rho^*$) is predicted from 
Eq. (\ref{Eq2.49}). 
Effectively,
\begin{equation}\label{Eq3.6}
 \rho^{*} = \dfrac{p_0^*}{T^*}.
\end{equation}
The thermodynamic pressure ($p_0^*$) is computed using 
Eqs. (\ref{Eq2.38}), (\ref{Eq2.40}) or (\ref{Eq2.41}), depending on the system boundary.

\item At this step, the predicted value for the hydrodynamic pressure ($p^*$) 
can be computed.
    
The momentum equation, i.e., Eq. (\ref{Eq2.43}), can be semi-discretised as
\begin{equation}\label{Eq3.7}
    \dfrac{
    \rho^* \vb*{v}^* 
    - 
    \rho^n \vb*{v}^n    
    }
    {\Delta t}
    =
    \dfrac{3}{2}
    \left[
        \vb*{\text{Res}_{v}}(\rho^n, \vb*{v}^{n}) + \vb*{f_v}(\rho^{n})
    \right]
    -
    \dfrac{1}{2}
    \left[
        \vb*{\text{Res}_{v}}(\rho^{n-1}, \vb*{v}^{n-1}) + \vb*{f_v}(\rho^{n-1})
    \right]
    -  
    \dfrac{1}{\Delta t}
    \grad P^*,
\end{equation}
in which
\begin{subequations}\label{Eq3.8}
    \begin{gather}
    \vb*{\text{Res}_{v}}
    \coloneqq
    -
    \div (
    \rho \vb*{v} \vb*{v})
    + 
    \dfrac{\Pr}{\Pe} 
    \div
    \vb*{\tau},     
    \label{Eq3.8a} \\
    \vb*{f_v}
    \coloneqq
    \underbrace{
        \dfrac{1}{\Fr^2}
        (1 - \rho)
        \vb*{e}_y.
    }_{source \ term \ for \ \vb*{v}-equation}
    \label{Eq3.8b}\\
    P^* \coloneqq  p^* \Delta t
    \label{Eq3.8c}
    \end{gather}
\end{subequations}

Taking the divergence of Eq. (\ref{Eq3.7}), 
one can get a Poisson equation for pressure, i.e.,
\begin{equation}\label{Eq3.9}
    -
    \laplacian 
    P^*
    =
    \div
    \left(
        \rho^* \vb*{v}^*
    \right)
    -
    \div
    \left(
        \rho^* \Tilde{\vb*{v}}
    \right),
\end{equation}
in which
\begin{equation}\label{Eq3.10}
    \dfrac{\rho^* \Tilde{\vb*{v}}}
    {\Delta t}
    \coloneqq    
    \dfrac{\rho^n \vb*{v}^n}
    {\Delta t}
    +
    \dfrac{3}{2}
    \left[
        \vb*{\text{Res}_{v}}
        (\rho^{n}, \vb*{v}^{n})
        +
        \vb*{f_{v}}
        (\rho^{n})
    \right]
    -
    \dfrac{1}{2}
    \left[
        \vb*{\text{Res}_{v}}
        (\rho^{n-1}, \vb*{v}^{n-1})
        +
        \vb*{f_{v}}
        (\rho^{n-1})
    \right]
\end{equation}

Finally, from the continuity equation, i.e., 
Eq. (\ref{Eq2.42}), 
\begin{equation}\label{Eq3.11}
    \div
    \left(
        \rho^{*} \vb*{v}^{*}
    \right)
    =
    -
    \left.        \partial_t \rho
    \right|^*
    \approx
    -
    \dfrac{3 \rho^* - 4 \rho^n + \rho^{n-1} }
    {2 \Delta t},
\end{equation}
In essence, this approach estimates the temporal derivative of 
density using backward second-order finite differences.
Therefore, mass conservation is enforced via pressure. 

Employing Eqs. (\ref{Eq3.10}) and (\ref{Eq3.11}), the 
predicted pressure value ($p^*$) can be determined using 
Eq. (\ref{Eq3.9}).

\item Finally, returning to Eq. (\ref{Eq3.7}), a prediction
for velocity ($\vb*{v}^*$) can be obtained.
     
\end{enumerate}

\subsubsection{Corrector stage}

\begin{enumerate}[label=(\roman*)]

\item The mixture fraction ($Z^{n+1}$) at the next time step ($t_{n+1}$) 
is computed from Eq. (\ref{Eq2.44}), incorporating values from the 
predictor step (future) and the previous time step (present), namely
\begin{equation}\label{Eq3.12}
     \dfrac{Z^{n+1} - Z^{n}}{\Delta t}
     =
     \dfrac{1}{2}
     \text{Res}_Z(\rho^{*},\vb*{v}^{*},Z^{*})
     +
     \dfrac{1}{2}
     \text{Res}_Z(\rho^{n},\vb*{v}^{n},Z^{n}).
\end{equation}

    \item Analogously, the corrected value of excess enthalpy ($H^{n+1}$) 
     can be derived from Eq. (\ref{Eq2.45}) as
\begin{multline}\label{Eq3.13}
     \dfrac{H^{n+1} - H^{n}}{\Delta t}
     =
     \dfrac{1}{2}
     \left[
         \text{Res}_H(\rho^{*},\vb*{v}^{*},H^{*})
         +
         f_H(\rho^{*},d_t p_0|^{*}, Z^{*})
     \right]
     + \\ +
     \dfrac{1}{2}
     \left[
         \text{Res}_H(\rho^{n},\vb*{v}^{n},H^{n})
         +
         f_H(\rho^{n},d_t p_0|^{n},Z^{n})
     \right],
\end{multline}
in which
\begin{equation}\label{Eq3.14}
    \left.
        d_t p_0
    \right|^*
    \approx
    \dfrac{3 p_0^* - 4 p_0^{n} + p_0^{n-1} }
    {2 \Delta t}.
\end{equation}
    
    \item The temperature, $T^{n+1} = T(Z^{n+1},H^{n+1})$, 
    is computed using the definitions of $Z$ and $H$
    From Eq. (\ref{Eq2.49}), the value for 
    $\rho^{n+1}$ is, then, obtained as
\begin{equation}\label{Eq3.15}
 \rho^{n+1}
 =
 \dfrac{p_0^{n+1}}{T^{n+1}},
\end{equation}
in which $p_0^{n+1}$ is derived from Eqs. (\ref{Eq2.38}), (\ref{Eq2.40}) or (\ref{Eq2.41}), 
as appropriated.

\item The corrected pressure ($p^{n+1}$) is determined 
    analogously to the predicted 
    pressure ($p^*$), namely
\begin{equation}\label{Eq3.16}
    \dfrac{
    \rho^{n+1} \vb*{v}^{n+1} 
    - 
    \rho^n \vb*{v}^n    
    }
    {\Delta t}
    =
    \dfrac{1}{2}
    \left[
        \vb*{\text{Res}_{v}}(\rho^*, \vb*{v}^*) + \vb*{f_v}(\rho^*)
    \right]
    +
    \dfrac{1}{2}
    \left[
        \vb*{\text{Res}_{v}}(\rho^{n}, \vb*{v}^{n}) + \vb*{f_v}(\rho^{n})
    \right]
    -
    \dfrac{1}{\Delta t}
    \grad P^{n+1},
\end{equation}
in which $P^{n+1}$ is the corrector-stage analogous of Eq. (\ref{Eq3.8c}).

Taking the divergence of Eq. (\ref{Eq3.16}), 
a Poisson equation for pressure emerges, i.e.,
\begin{equation}\label{Eq3.17}
    -
    \laplacian 
    P^{n+1}
    =
    \div
    \left(
        \rho^{n+1} \vb*{v}^{n+1}
    \right)
    -
    \div
    \left(
        \rho^{n+1} \Breve{\vb*{v}}
    \right)
\end{equation}
in which
\begin{equation}\label{Eq3.18}
    \dfrac{\rho^{n+1} \Breve{\vb*{v}}}
    {\Delta t}
    \coloneqq    
    \dfrac{\rho^n \vb*{v}^n}
    {\Delta t}
    +
    \dfrac{1}{2}
    \left[
        \vb*{\text{Res}_{v}}(\rho^*, \vb*{v}^*) + \vb*{f_v}(\rho^*)
    \right]
    +
    \dfrac{1}{2}
    \left[
        \vb*{\text{Res}_{v}}(\rho^{n}, \vb*{v}^{n}) + \vb*{f_v}(\rho^{n})
    \right]
\end{equation}

Finally, from the mass conservation equation (\ref{Eq2.42}), 
\begin{equation}\label{Eq3.19}
    \div
    \left(
        \rho^{n+1} \vb*{v}^{n+1}
    \right)
    =
    -
    \left.
        \partial_t \rho
    \right|^{n+1}
    \approx
    -
    \dfrac{3 \rho^{n+1} - 4 \rho^n + \rho^{n-1} }
    {2 \Delta t},
\end{equation}
utilising the same approximation as the predictor stage.

Hence, using Eqs. (\ref{Eq3.18}) and (\ref{Eq3.19}), 
the pressure value at time $t_{n+1} $(i.e, $p^{n+1}$) 
can be computed from Eq. (\ref{Eq3.17}).

\item The corrected velocity ($\vb*{v}^{n+1}$) is calculated 
from Eq. (\ref{Eq3.16}). \\

Note that multi-step (or multi-point) methods require the 
prescription of values for various (past) time points, which 
are not always available. 
Therefore, it is recommended to start the simulation with 
lower-order versions and gradually transitioning to 
their higher-order analogous, as the necessary data becomes 
available.
Concretely, for this case, one starts with a first order
Adams–Bashforth method (i.e., the Euler method) for the predictor.
While the corrector can continue using the same second-order
method.
This combination is known as the Heun's method. 

\end{enumerate}

\subsection{Spatial Discretisation}

\begin{figure}[t]
\centering
\includegraphics[width=210pt]{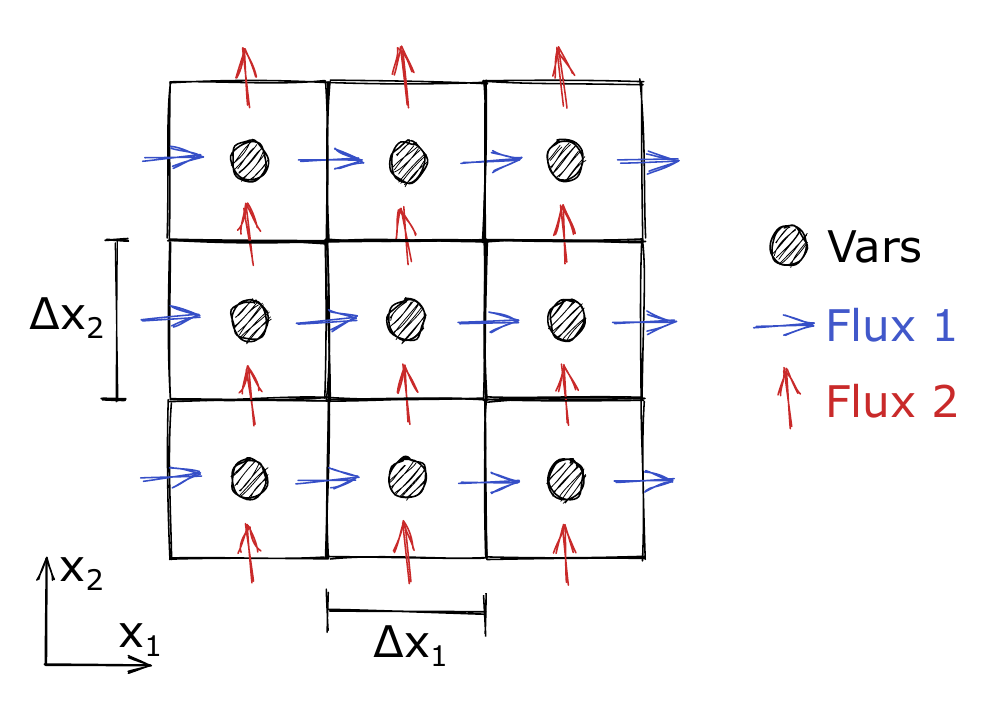}
\caption{
A collocated bidimensional grid in Cartesian coordinates, with spacing 
$\Delta x_1$ and $\Delta x_2$ for $x_1$- and $x_2$-directions, respectively. 
Each variable is stored at the centre of grid cells (circles), while 
auxiliary fluxes are defined at cell interfaces of the corresponding 
direction (arrows), i.e, flux 1 ($F_1$) for the horizontal direction ($x_1$), 
and flux 2 ($F_2$) for the vertical direction ($x_2$).
}
\label{fig-grid}
\end{figure}

Basically, it is possible to discretise the physical domain into regular, 
staggered, or collocated grids \cite{morinishi/1998:JCP}. 
Regular grids -- where all variables are placed at the same position 
-- provide advantages in terms of easier computational implementation, 
parallelisation, adaptability to curvilinear coordinates, and extension 
to high-order schemes.
Staggered grids -- where velocity components are strategically dislocated 
in relation to scalar variables -- avoid numerical pressure oscillations 
due to the resulting strong ``odd-even coupling'' 
(i.e., fluxes at cell-interfaces are appropriated to enforces  mass 
conservation locally).
Collocated grids -- where all variables are located at the same point, 
similar to regular grids, while auxiliary fluxes are conveniently defined 
around that point, similar to staggered grids -- combining the good features 
of both regular and staggered grids, in principle.

This study follows the approach of \cite{morinishi/1998:JCP}, developed for 
discrete conservation of kinetics energy in incompressible flows, and 
extended to variable-density low-Mach-number flows, encompassing both 
non-reacting \cite{nicoud/2000:JCP} and reacting \cite{lessani/2006:JCP} 
scenarios.
Concretely, it employs a collocated grid in Cartesian coordinates, as 
illustrated in Fig. \ref{fig-grid}.
The operators are discretised using a centred finite-differences scheme, 
with flux interpolation to avoid numerical pressure oscillations.

\subsubsection{Interpolation and finite-difference operators}\label{oper}

The interpolation and finite-difference operators are defined below.
It should be noted that odd-index operators (i.e., for odd $n$) transfer data 
between central points and interfaces, while even-index operators 
(i.e., for even $n$) keep the data where it is originally defined. 
For example, when information is to be passed from a centroid to an 
interface, or from an interface to a centroid, the respective odd-index 
operator needs to be used. 
When information corresponding to a centroid or an interface must be 
processed in the same place, the respective even-index operator should be invoked.

Viz., the adopted interpolation operator is given by
\begin{equation}\label{Eq3.20}
    \left.
        \Bar{\phi}^{n{x_i}}
    \right|_{x_{i,I}}
    \coloneqq
    \dfrac{
    (x_{i,I} - x_{i,I-n/2})
    \phi(x_{i,I+n/2}) 
    + 
    (x_{i,I+n/2} - x_{i,I})
    \phi(x_{i,I-n/2})}
    {x_{i,I+n/2} - x_{i,I-n/2}},
\end{equation}
and the finite-difference operator is given by
\begin{equation}\label{Eq3.21}
    \left.
        \dfrac{\delta_n \phi}{\delta_n {x_i}}
    \right|_{x_{i,I}}
    \coloneqq
    \dfrac{
    \phi(x_{i,I+n/2})
    - 
    \phi(x_{i,I-n/2})
    }
    {x_{i,I+n/2} - x_{i,I-n/2}}
\end{equation}
both are derived from Taylor series expansion for a 
non-uniform grid.
Note that the first index (`$i$') refers to the direction 
(i.e., $x_1$, $x_2$ or $x_3$) and the second index (`$I$'), 
to the domain partition (or grid) index, omitting constant 
directions.
Thus, $I$ could indicate either a cell centre or interface.
They are (locally) second-order accurate, and can 
be combined (varying `$n$') to produce higher-order 
approximations. 

\subsubsection{Integration operator}

A collocated grid (see Fig. \ref{fig-grid}) naturally 
induces numerical integration via a middle Riemann sum 
(or midpoint rule), as the value at the centroid 
represents the value across the entire cell.

Therefore, integrals are discretised as
\begin{equation}\label{Eq3.22}
    \int\limits_V
    \phi
    (x_1,x_2,x_3)
    dV
    \approx
    \sum_{I,J,K}
    \phi
    (x_{1,I}\ ,\ x_{2,J}\ ,\ x_{3,K})
    \Delta V_{(I,J,K)}
\end{equation}
in which $\phi(x_{1,I}\ , \ x_{2,J}\ , \ x_{3,K})$ is $\phi$ 
evaluated at the point $(x_{1,I}\ , \ x_{2,J}\ ,\ x_{3,K})$,
whose volume of the corresponding cell is $\Delta V_{(I,J,K)}$.
The same idea applies to bidimensional and 
unidimensional integrals, with the overall 
spatial error of second order \cite{moukalled/2016}.

\subsubsection{Discretisation strategy}

The discretisation is based on the operators previously 
delineated, as proposed by 
\cite{lessani/2006:JCP, nicoud/2000:JCP, morinishi/1998:JCP}. 
Only the corrector-stage schemes will be presented herein. 
The schemes for the predictor stage are entirely analogous 
and, for the sake of brevity, will be omitted.

Auxiliary fluxes are defined at the midpoint of interfaces, 
necessitating the transfer of data from centroids to 
interfaces. 
Thus, odd-index operators should be utilised. 
Hence, from Eqs. (\ref{Eq3.16}) and (\ref{Eq3.18}), and 
discrete operators of Eqs. (\ref{Eq3.20}) and (\ref{Eq3.21}),
\begin{equation}\label{Eq3.23}
        F^{n+1}_i
        \coloneqq
        \overline{ {\rho^{n+1} \Breve{v}_i} }^{1{x_i}}
        -
        \dfrac{\delta_1 P^{n+1}}{\delta_1 x_i}
\end{equation}

Alternatively to the derivation in Sec. \ref{tempDis}, 
a discretisation of the pressure Poisson equation can be 
accomplished, directly, from the continuity equation. 
Starting from Eq. (\ref{Eq3.19}) and employing auxiliary 
fluxes ($F^{n+1}_i$) of Eq. (\ref{Eq3.23}), one gets
\begin{equation}\label{Eq3.24}
        \left.
        \partial_t \rho
        \right|^{n+1}
        +
        \dfrac{\delta_1 F^{n+1}_i}{\delta_1 x_i}
        = 0
\end{equation}
In addition, velocity components ($v_i$) can be computed 
from Eq. (\ref{Eq3.16}), using $P^{n+1}$ calculated 
from Eq. (\ref{Eq3.24}), as
\begin{equation}\label{Eq3.25}
        v_i^{n+1}
        =
        \dfrac{1}{\rho^{n+1}}
        \left(
            \rho^{n+1} \Breve{v}_i
            -
            \dfrac{\delta_2 P^{n+1}}{\delta_2 x_i}
        \right)  
\end{equation}
Note that even-index operator were used, 
since $v_i$ and $P$ are defined at the same 
grid.
Convective terms can be derived from auxiliary 
fluxes, given by Eq. (\ref{Eq3.23}), as
\begin{subequations}\label{Eq3.26}
    \begin{gather}
    \dfrac{\partial }{\partial x_j}
        (\rho v_i v_j )
        \approx
        \dfrac{\delta_1 }{\delta_1 x_j}
        (F_j\overline{ {v_i} }^{1{x_i}}),     
    \label{Eq3.26a} \\
    \rho v_j
        \dfrac{\partial Z}{\partial x_j}
        \approx
        \overline{
            F_j \dfrac{\delta_1 Z}{\delta_1 x_j}
        }^{1{x_j}}
    \label{Eq3.26b}
    \end{gather}
\end{subequations}
with the same strategy of mixture fraction ($Z$) being 
employed for excess enthalpy ($H$).
Odd-index operators were used for convective terms, since 
auxiliary fluxes and primitive variables are defined at 
distinct grids.
This procedure culminates in strong ``odd-even coupling'',
between pressure (mass conservation) and velocity 
(momentum conservation), avoiding numerical 
oscillations in the pressure field, common in grids where all variables are 
calculated at the same points.

Lastly, it is pertinent to note that operators pertaining 
to diffusion phenomena are discretised using standard 
centred-finite differences, yielding the same order of 
accuracy as operators associated with convection.
It is equivalent to use a composition of 
odd-index operators.

\subsection{Poisson Equation}\label{PoissonEq}

Physically, there are no explicit boundary conditions for 
pressure in the low Mach number regime, which, naturally, 
includes the particular case of incompressibility.
In this context, pressure acts as a restriction 
(Lagrange multiplier) on the velocity divergence to enforce 
the conservation of mass \cite{rajagopal/2015:IJONLM}.

Given the lack of consensus in the literature on this issue 
(see, for instance, \cite{gresho/1987:IJNMF,rempfer/2006:AMR,sani/2006:IJNMF,rempfer/2008:IJNMF}), 
it is recommended to avoid explicitly defining boundary 
conditions for pressure. 
When working with staggered or collocated grids with 
auxiliary fluxes, boundary conditions for mass fluxes 
are known \cite{kim/1985:JCP, zang/1994:JCP}.
In other words, auxiliary fluxes do not need to be 
split, as proposed in Eq. (\ref{Eq3.23}).
Thus, boundary conditions for pressure or intermediate velocity are not required
\cite{perot/1993:JCP, kim/1985:JCP, zang/1994:JCP}.
Another consequence is that the compatibility 
condition \cite{yoon/2016:JSC} is satisfied at 
discrete level \cite{abdallah/1987:JCP, henshaw/1994:JCP} 
-- under consistent boundary conditions --,
and the existence of a solution is guaranteed.
However, it results in an indeterminate system, as
it implies a problem subjected to pure Neumann boundary 
conditions \cite{yoon/2016:JSC}: an infinite 
set of solutions is possible, as adding any constant 
to a solution results in another solution.
As the low-Mach-number flows only perceive pressure 
gradients, any solution is sufficient, and this study 
employs the augmented matrix method to obtain the 
zero-mean solution \cite{yoon/2016:JSC, henshaw/1994:JCP}.
This approach also regularises the solvability condition 
for the linear system associated with the discretisation 
of the Poisson equation \cite{porzridis/2001:JCP}, 
compensating any eventual residue in the discrete 
compatibility condition.

The pressure solution within the domain can be 
extrapolated to ghost cells, enabling the computation 
of derivatives at points adjacent to the boundary, 
consistently to the procedure for interior points. 
This also allows for the determination of pressure 
on the boundary.
More specifically, a quadratic extrapolation is 
applied at a ghost-cell point (e.g., $I = 0$) as
\begin{equation}\label{Eq3.27}
    \phi(x_{i,0}) 
    =
    A \phi(x_{i,1}) 
    + 
    B \phi(x_{i,2}) 
    + 
    C \phi(x_{i,3})
\end{equation}
in which
\begin{equation}\label{Eq3.28}
    \begin{aligned}
        \begin{aligned}
        A &\coloneqq \frac{(x_{i,2} - x_{i,0})(x_{i,3} - x_{i,0})}{(x_{i,2} - x_{i,1})(x_{i,3} - x_{i,1})}, \\[2mm]
        B &\coloneqq \frac{(x_{i,1} - x_{i,0})(x_{i,3} - x_{i,0})}{(x_{i,1} - x_{i,2})(x_{i,3} - x_{i,2})}, \\[2mm]
        C &\coloneqq \frac{(x_{i,1} - x_{i,0})(x_{i,2} - x_{i,0})}{(x_{i,1} - x_{i,3})(x_{i,2} - x_{i,3})}.
        \end{aligned}
    \end{aligned}
\end{equation}
Note that this approach is equivalent to consider 
one-sided (second-order) finite differences for 
pressure derivatives at first-interior points.

\subsection{Immersed Boundary Method}\label{sub-IBM}

The immersed boundary method (IBM), firstly presented by 
Peskin \cite{peskin/1972:JCP}, provides an ingenious 
technique for modelling intricate fluid dynamics, initially applied to circulatory system modelling of flows around heart valves. 
This method represents a notable advancement in 
computational fluid dynamics, enabling the description 
of complex flow-structure interactions using orthogonal 
coordinates on non-conforming grids.

The original proposal of Peskin involves augmenting the 
(incompressible) momentum equation with a forcing term 
to account for the presence of an immersed body within 
the flow, which is, precisely, the essence of this method 
\cite{peskin/2002:AN}.
In fact, a volumetric forcing term ($\vb*{F}_{\vb*{v}}$) can be added 
to Eq. (\ref{Eq2.43}), resulting in
\begin{equation}\label{Eq3.29}
    \partial_{t} 
    (\rho \vb*{v}) 
    +
    \div (
    \rho \vb*{v} \vb*{v}) 
    = 
    -
    \grad   p
    + 
    \dfrac{\Pr}{\Pe} 
    \div
    \vb*{\tau} 
    +
    \dfrac{1}{\Fr^2}
    (1 - \rho)
    \vb*{e}_{x_2}
    +
    \vb*{F}_{\vb*{v}},
\end{equation}
with $\vb*{F}_{\vb*{v}} = \vb*{F}_{\vb*{v}}(\vb*{x},t)$ adjusted dynamically 
to represent the response forces, related to the presence 
of an immersed body.
The specific form of $\vb*{F}_{\vb*{v}}$ depends on the particular method, 
with several possibilities \cite{mittal/2005:ARFM,verzicco/2023:ARFM}.

An elegant idea is to consider a solid obstacle as a porous 
medium of negligible permeability ($\hat{K}$ [length$^2$]),
i.e., $0 < \hat{K} \ll 1$.
This approach was proposed originally by Arquis \cite{arquis/1984},
subsequently known as the penalisation (or penalty) method.
Because of the physical motivation, as well as the 
mathematical robustness \cite{angot/1999:NM,carbou/2003:ADE}, 
the present study adopts a penalisation method, which 
implies the following forcing term definition
\begin{equation}\label{Eq3.30}
    \vb*{F}_{\vb*{v}}
    \coloneqq
    -
    \chi
    \dfrac{\Pr}{\Pe \Dar_{ib}}
    \kappa
    \left(
        \vb*{v}
        -
        \vb*{v}_{ib}
    \right)
\end{equation}
in which $\chi = \chi(\vb*{x},t)$ is a characteristic (or mask) function, 
$\Dar_{ib} \coloneqq \hat{K}_{ib} / \hat{l}^2_c \ll 1$ is
the immersed-boundary (IB) \Dar[l] -- associated with the imposed 
permeability $\hat{K}_{ib}$ -- and the velocity $\vb*{v}_{ib}$.
Physically, $\vb*{F}_{\vb*{v}}$ represents a linear momentum sink 
(or a linear damping force), inversely proportional to $\Dar_{ib}$, 
acting against the difference $\vb*{v} - \vb*{v}_{ib}$.
In the context of IBM, it is known as the penalisation (or penalty) term.
A relevant observation is that, the work done by this damping force
scales with $\MM^2$, and does not contribute to the enthalpy 
in the low-Mach-number equations (see Sec. \ref{lowMach}).

The parameter $\Dar_{ib}$ is critical, as it controls the proper convergence to the actual solution, 
on the one hand, and the integration stability, on the other, requiring a 
compromise between correct imposition ($\Dar_{ib} \rightarrow 0$), 
and relaxation of the time-step restriction ($\Dar_{ib} \rightarrow 1$).
Indeed, it imposed the stability condition $\Delta t \leq c (\Rey \Dar_{ib})$ -- for a real, 
scheme-dependent constant $c$ --, when using an explicit method.
The characteristic function ($\chi$) is commonly related to a discrete version of
the Heaviside step function -- defined as $0$, for a fluid point, and as 
$1$, for an IB point.
The present formulation considers the shifted characteristic function derived by Iwakami \cite{iwakami/2014:CCP}.
The basic idea is to extend the body over the fluid by 
$(\Dar_{ib})^{1/2}$, which corresponds to the (dimensionless)
Brinkman layer thickness (or Brinkman screening length),  
a transition zone connecting the fluid and porous medium 
behaviours.
This approach categorically reduces the error in the fluid region,
after the transition layer.
The physical justification of this method efficacy appears to stem from the 
appropriate representation of the solid dimensions -- i.e.,
the ``effective length'' is a proper representation of the body length -- 
because the transition layer is transferred from the porous-medium interior to the fluid region.
Mathematically, the wavenumbers associated with the penalised-numerical 
solution coincide with those related to the actual 
solution in the fluid region (except into the Brinkman layer), as
shown in \cite{iwakami/2014:CCP}.
Naturally, the grid resolution should be suffice to capture the transition zone (i.e., $\Delta x_i \lesssim (\Dar_{ib})^{1/2}$ for $i=1,2,3$).

Analogously to the momentum forcing, the imposition of boundary conditions
on the immersed body can be consistently accomplished -- for mixture fraction 
($F_{Z}$) and excess enthalpy ($F_{H}$), respectively -- 
defining the penalisation terms

\begin{equation}\label{Eq3.31}
    F_{Z}
    \coloneqq
    -
    \chi
    \dfrac{1}{\Le \Pe \Dar_{ib}}
    \kappa
    \left(
        Z
        -
        Z_{ib}
    \right)
\end{equation}
and
\begin{equation}\label{Eq3.32}
    F_{H}
    \coloneqq
    -
    \chi
    \dfrac{1}{\Pe \Dar_{ib}}
    \kappa
    \left(
        H
        -
        H_{ib}
    \right),
\end{equation}
in which $Z_{ib}$ and $H_{ib}$ are the respective conditions 
for $Z$ and $H$ on the body surface. 

Penalisation terms are discretised using the same second-order 
scheme as described for other forcing terms in Sec. \ref{nMethod}.
Additionally, for the case of mass ejection from the body, a 
mass source term, $S_m = S_m(t,\vb*{x})$, must be added to 
the continuity equation, in consonance with the imposed velocity. 
Namely,
\begin{equation}\label{Eq3.33}
    \partial_{t} 
    \rho  
    + 
    \div
    (\rho \vb*{v})
    = 
    S_m
\end{equation}
which will be, actually, accounted for into the forcing term of the pressure Poisson equation.

This penalisation method follows a dimensionally consistent derivations of forcing terms, because
\textit{a mathematical model for physical systems should always incorporate physically 
consistent terms. 
Even for purely numerical purposes, terms inherently carry physical meaning}.
Neglecting this principle makes interpreting results difficult or, worse, may 
lead to simulation divergence or non-physical solutions.

\subsection{Flame Regularisation}\label{flameReg}

Although formally simplified, an infinitely thin 
flame implies a discontinuity of properties between the 
two sides (fuel and oxidiser subdomains) of the flame 
surface. 
Especially critical for finite differences (strong form), 
discontinuities should lead to numerical oscillations 
during the time-march process and may also hinder 
post-processing or utilisation as a base flow for 
additional analyses (e.g., stability analyses).
This is because, the basis of finite difference 
methods is the Taylor series expansion, which requires 
functions to be infinitely differentiable and, 
therefore, continuous at the point of expansion.

To circumvent this issue, a regularisation technique 
is applied to the Heaviside function ($\theta$), which 
originally models the behaviour of these properties.
This method entails utilising a regularised variant of 
the Heaviside function ($\theta_{\varepsilon}$) to achieve 
smooth property transitions over a thickness $\varepsilon$, 
between the fuel and oxidant subdomains, adjacent to the 
flame.
Concretely, it involves the using of \cite{gutierrez/2022:IJNMF}
\begin{equation}\label{Eq3.34}
    \theta_\varepsilon
    (Z-Z_{st})
    \coloneqq
    \dfrac{
        1 
        + 
        \tanh[ k (Z - Z_{st}) ]
    }
    {2}
\end{equation}
which acts around $Z_{st} = 1$ 
(i.e. around the stoichiometric-mixture level, 
representing the flame), continuously linking 
properties from either flame side.

The regularisation parameter $k$ 
controls the effective support.
By determining a proper value for 
$k \coloneqq [1 / (2 \varepsilon)] \, \ln [(1 - \text{tol}) / \text{tol} ]$, 
such that it converges to $\theta$, within a specified tolerance ($tol$), 
when $Z = \pm \varepsilon$.
An appropriate choice for $tol$ is 
based on the spatial truncation errors.

Hence, the regularised coefficients $\Le_\varepsilon$, $\N_\varepsilon$ and $\Np_\varepsilon$ read
\begin{subequations}\label{Eq3.35}
    \begin{gather}
        \Le_\varepsilon(Z) 
        =
        (\Le_F - \Le_O)  \theta_\varepsilon(Z-1) + \Le_O
       \label{Eq3.35a} \\ 
        \N_\varepsilon(Z) 
        =
        (\N_F - \N_O)  \theta_\varepsilon(Z-1) + \N_O
         \label{Eq3.35b} \\ 
        \Np_\varepsilon(Z) 
        =
        (\Np_F - \Np_O)  \theta_\varepsilon(Z-1) + \Np_O
         \label{Eq3.35c}
    \end{gather}
\end{subequations}

Besides these three coefficients, temperature 
derivatives are also discontinuous at the flame. 
The following procedure may be adopted to circumvent 
this issue:
\begin{enumerate}[label=(\roman*)]

\item Use the definition of $H$ and $Z$ to write $T = T(Z,H)$, i.e,

\begin{equation}\label{Eq3.36}
    T = \dfrac{Q}{S+1} \left[ F + H \right]
\end{equation}
in which $F = F(Z)$ is defined by
\begin{equation}\label{Eq3.37}
    F(Z)
    \coloneqq
     \Bigg\{
     \begin{array}{lr}
        (1-Z)/S , & \text{ for } Z > 1\\
        (Z-1),    & \text{ for } Z \leq 1
    \end{array}
\end{equation}

\item Identify that the discontinuity is in the derivative of $F$
\begin{equation}\label{Eq3.38}
    F'(Z)
    =
    \Bigg\{
    \begin{array}{lr}
       -1/S , & \text{ for } Z > 1\\
       1,     & \text{ for } Z \leq 1
    \end{array}
\end{equation}
and, thus, apply the regularisation to it:
\begin{equation}\label{Eq3.39}
    F'(Z)
    =
    \left(
        -1/S
        -
        1
    \right)
    \theta_\varepsilon(Z-1)
    +
    1
\end{equation}

So, the regularised $F$ can be recovered by integration, i.e.,
\begin{equation}\label{Eq3.40}
    F_\varepsilon(Z)
    =
    \int\limits_{1}^{Z}
    \left[
        -
        \left(
            1/S
            +
            1
        \right)
        \theta_\varepsilon(z-1)
        +
        1
    \right]
    dz
\end{equation}
or, explicitly,
\begin{equation}\label{Eq3.41}
    F_\varepsilon(Z) 
    = 
    \frac{1}{2S} 
    \left[ 
        - 
        (S + 1) 
        \dfrac{ \ln \bigl\{ \cosh[ k (Z - 1) ]\bigr\}}{k} 
        + 
        (S - 1)
        (Z - 1) 
      \right].
\end{equation}

Note that it imposes the correct slopes for $F_\varepsilon$ from the flame position, but does not guarantee the right level, which varies with the thickness $\varepsilon$. 
To adjust the level, $F_\varepsilon$ must be translated to satisfy the values at the boundary, which implies a smaller maximum temperature, naturally.

\begin{figure}[t]
\centering
     \begin{subfigure}[t]{230pt}
         \centering
         \includegraphics[width=\textwidth]{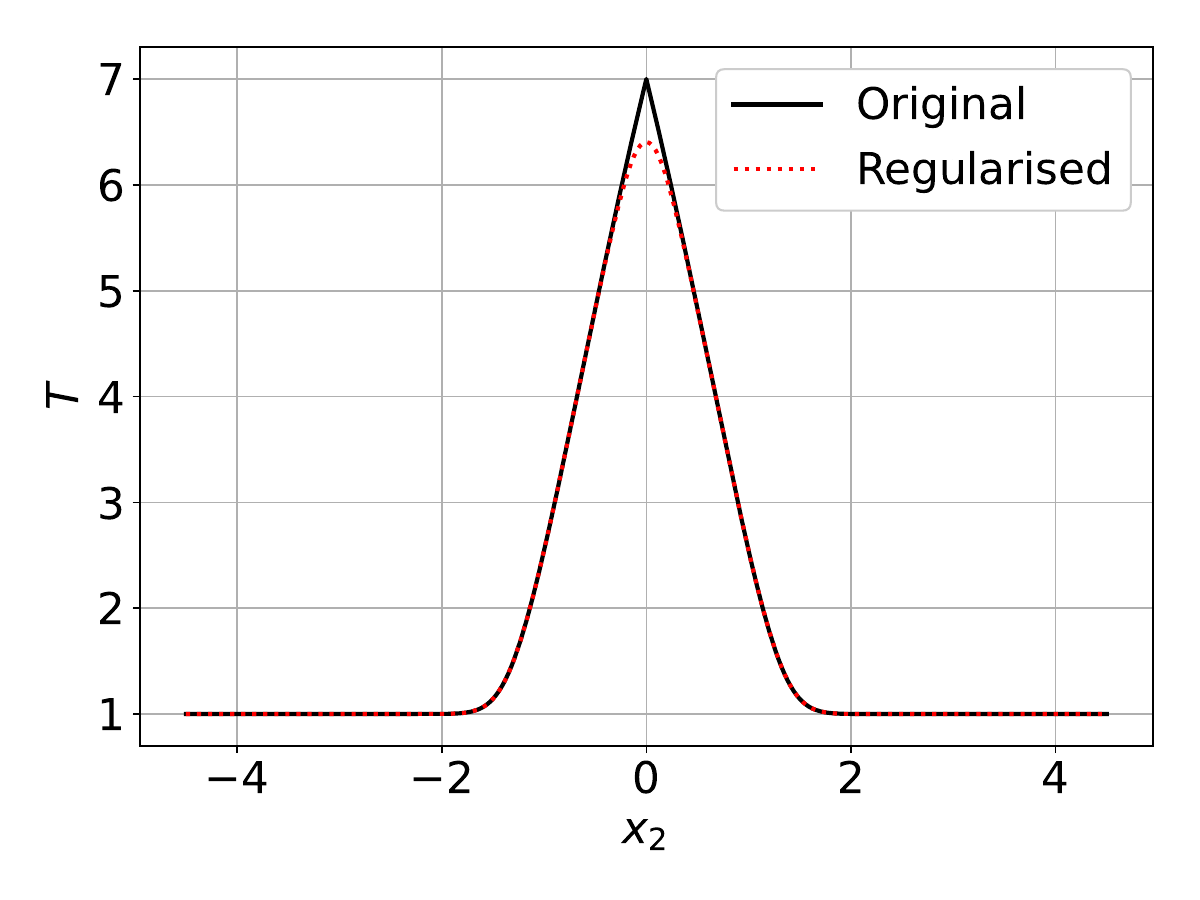}
         \caption{}
         \label{fig-Rega}
     \end{subfigure}
     \begin{subfigure}[t]{230pt}
         \centering
         \includegraphics[width=\textwidth]{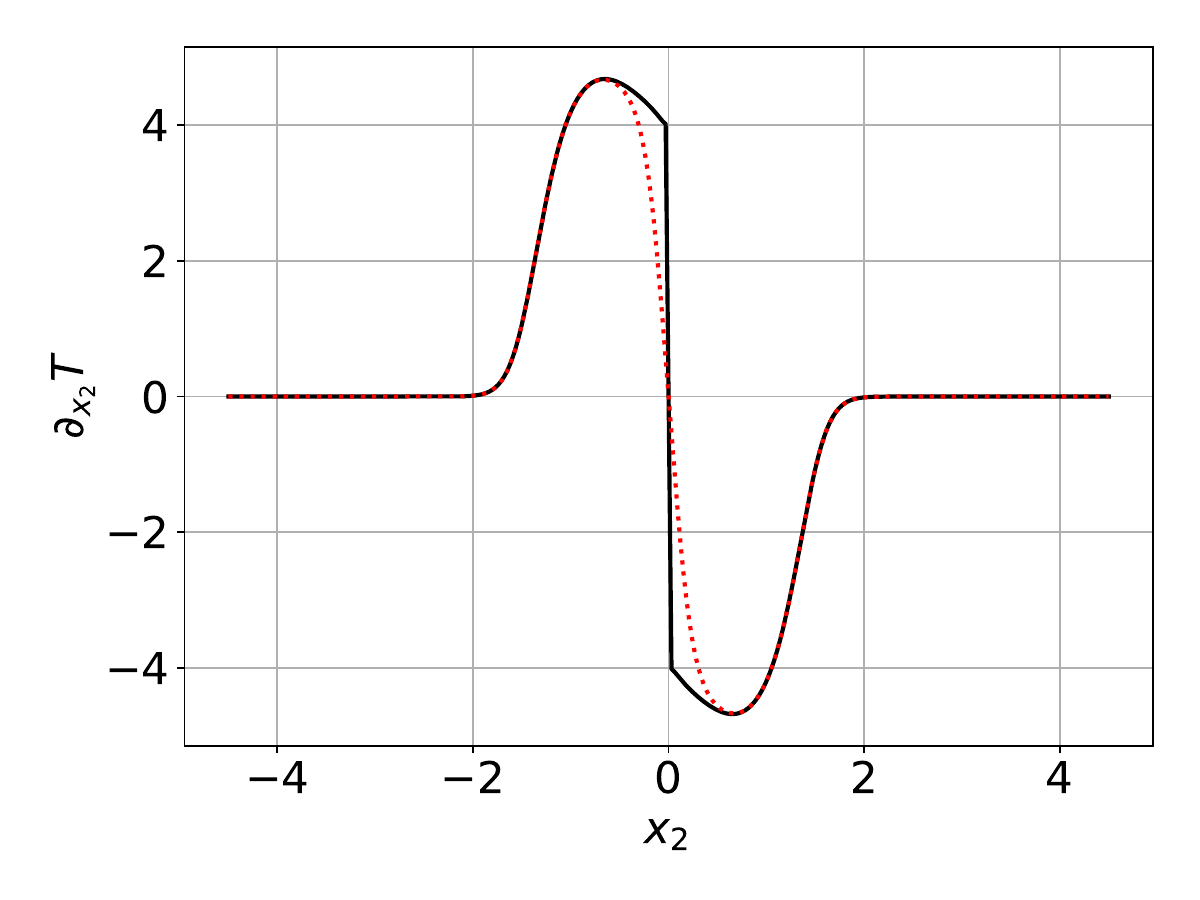}
         \caption{}
         \label{fig-Regb}
     \end{subfigure}
\caption{
    Comparison of original and regularised temperature (a) and temperature derivative flame-crossed profiles for a planar counterflow diffusion 
    flame.
}
\label{fig-Reg}
\end{figure}

Therefore, the regularised temperature ($T_\varepsilon$) can be expressed as
\begin{equation}\label{Eq3.42}
    T_\varepsilon
    = 
    \dfrac{Q}{S+1} 
    \left[ 
        F_\varepsilon
        + 
        H 
    \right]
\end{equation}

Figure \ref{fig-Reg} shows a comparison between original and regularised profiles
of temperature and its variations, along the normal direction to a planar counterflow diffusion flame.
In what follows, the index `$\varepsilon$' in the regularised properties 
shall be omitted, for the sake of notation simplicity.

\end{enumerate}

\subsection{Boundary conditions}

Boundary conditions are implemented using a 
ghost-cell method, derived from operators in 
Sec. \ref{oper}.
For instance, to impose a condition 
$\Phi = \Phi(x_i, t)$ 
at the left boundary ($I = 1/2$), the proper value for the 
ghost-cell $\phi(x_{i,0})$ 
can be set as follows:

\paragraph{For a Dirichlet condition}
\begin{equation}\label{Eq3.43}
    \phi(x_{i,0})
    =
    2\Phi
    -
    \phi(x_{i,1}),
\end{equation}
which admits the same spacing among the ghost and first interior points (i.e., a mirrored ghost-cell).

\paragraph{For a Neumann condition}
\begin{equation}\label{Eq3.44}
    \phi(x_{i,0})
    =
    -
    ({x_{i,1} - x_{i,0}})
    \Phi    
    +
    \phi(x_{i,1})
\end{equation}

\paragraph{For a periodic condition}
\begin{equation}\label{Eq3.45}
    \phi(x_{i,0})
    =
    \phi(x_{i,N_{x_1}-1})
\end{equation}
in which $N_{x_i}$ is the last index in 
the $i$-direction partition, also representing
a ghost-cell.

\paragraph{For an extrapolation condition}
\begin{equation}\label{Eq3.46}
    \phi(x_{i,0}) 
    =
    A \phi(x_{i,1}) 
    + 
    B \phi(x_{i,2}) 
    + 
    C \phi(x_{i,3})
\end{equation}
with the same coefficients such as defined in Eq. (\ref{Eq3.28}).

\subsection{Computational Implementation}

The numerical model is computationally implemented 
in Fortran 90, procedurally, for sequential execution.
The pressure Poisson equation is solved via MUMPS 
\cite{amestoy/2000}, for robustness and efficiency, 
a public domain parallel sparse direct solver, which 
implements a multifrontal method to solve large 
linear systems \cite{amestoy/2000:CMAME}.
The post-processing is carried out in Python 3.


\section{Numerical Experiments}
\label{simulations}
\addvspace{10pt}

For the purposes of this work, 3D cases should only 
add unnecessary complexity.
Therefore, this section concentrates on 2D test-case 
simulations for constant (incompressible) and 
variable density flows, at low Mach numbers, 
with and without chemical reactions (combustion).
These numerical experiments were carefully selected to 
check different aspects of the proposed numerical 
method, highlighting both its capacities and 
limitations.
For consistency, post-processing employs discrete 
operators (for differentiation, interpolation, 
integration, etc.) with an accuracy equivalent 
(or greater) to that of the proposed method.
This approach mitigates the introduction of 
additional errors.


\subsection{Preliminaries}
\label{pre}

The order of convergence ($b$) is given by \cite{roache/1998}
\begin{equation}\label{Eq4.1}
    b = \dfrac{\ln{(E_m/E_{m+1})}}{\ln{(\delta)}}
\end{equation}
in which $E_m$ is the absolute (scalar) or a norm (vector) error 
for the grid with index $m$, and $\delta$ is the refinement ratio from 
the coarser ($m$) to the finer ($m+1$) grid.
Alternatively, for a constant refinement ratio, the order of convergence
can be directly estimated using solutions ($f$) for three levels ($m, m+1$ and $m+2$) of refinement,
without the necessity of an exact solution.
In fact,
\begin{equation}\label{Eq4.2}
    b = \ln \left( \dfrac{f_{m}-f_{m+1}}{f_{m+1}-f_{m+2}} \right)/\ln{(\delta)}
\end{equation}
from Richardson extrapolations.

The grid convergence index (GCI), for the finer grid, has the form
\begin{equation}\label{Eq4.3}
    \text{GCI} \coloneqq F_s \dfrac{\abs{\epsilon}}{\delta^a-1}
\end{equation}
in which $\epsilon \coloneqq (f_{m} - f_{m+1}) / f_{m+1}$ and $F_s$ 
is a ``safe factor'', considered here as $1.25$.

For the first two, incompressible cases, the diffusion-term coefficient is
rewritten explicitly as the reciprocal of the \Rey[l], i.e., $\Rey \coloneqq \Pe/\Pr$
is used.
This reformulation permits a straightforward physical interpretation, since
there is no thermal energy flux under constant-density scenario.

In tables and other mentions, values are presented with few decimal places, by space limitation; however, the 
calculations were always performed in double precision.


\subsection{Taylor--Green vortex}
\label{TGvortex}

This configuration is proposed to verify the accuracy of the 
hydrodynamical-variables integration.
It also enables a discussion on the fractional-time step (or projection)
method, with concepts commonly misunderstood.

Consider a transient, incompressible flow defined in a dimensional 
periodic-square domain $[0,2\pi]^2$.
Taylor--Green vortices were proposed
as an analytical tool to study the laminar-turbulent 
transition \cite{taylor/1937:PRSL}.
Indeed, a closed-form solution exists for the laminar regime:
\begin{align}
    v_1(x_1,x_2,t) &= \sin(2\pi x_1) \cos(2\pi x_2) \exp\left(-\frac{8\pi^2}{\Rey}t\right), \label{Eq4.4}  \\
    v_2(x_1,x_2,t) &= -\cos(2\pi x_1) \sin(2\pi x_2) \exp\left(-\frac{8\pi^2}{\Rey}t\right), \label{Eq4.5} \\
    p(x_1,x_2,t) &= \frac{1}{4} \left[ \cos(4\pi x_1) + \cos(4\pi x_2) \right] \exp\left(-\frac{16\pi^2}{\Rey} t \right), \label{Eq4.6}
\end{align}
given in terms of dimensionless velocity components and pressure.
The characteristic values are based on the initial 
state, i.e., $\hat{L}_c \coloneqq 2\pi \ \SI{}{\meter}$ 
(the vortex wave-length), $\hat{v}_c \coloneqq \SI{1}{\meter / \second}$ 
(maximum initial velocity component), $\hat{t}_c \coloneqq \hat{L}_c / \hat{v}_c$ and 
$\hat{p}_c \coloneqq \hat{\rho}_c \hat{v}_c^2$, for a given $\hat{\rho}_c$.

\begin{figure}[t]
\centering
\includegraphics[width=270pt]{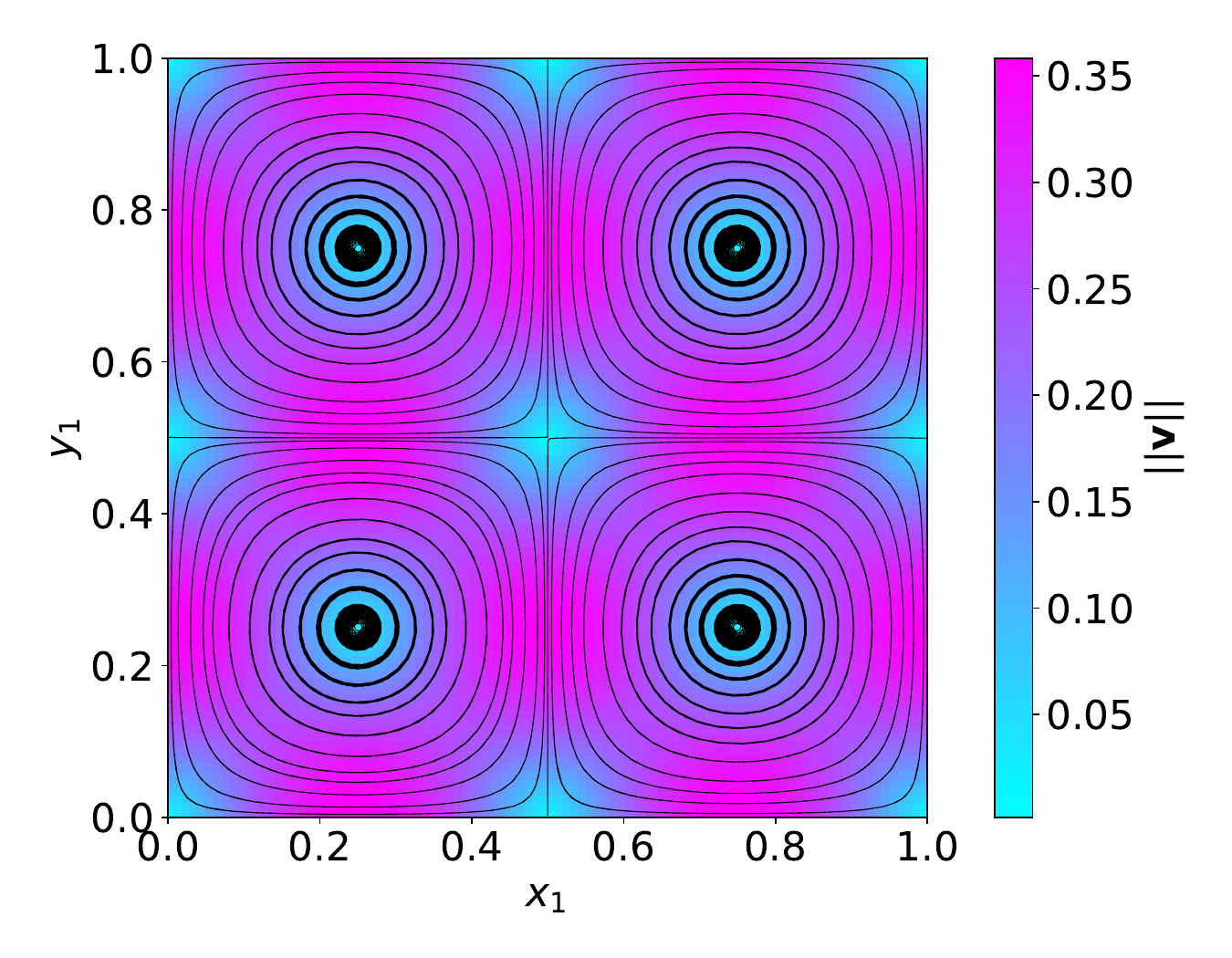}
\caption{
Streamlines of Taylor--Green vortices at $t \approx 1.3$.
Light (dark) tones indicate low (high) local speeds. 
}
\label{fig-TGvortex}
\end{figure}

In what follows, numerical and analytical results are compared at 
$t_{ref} = \Rey / (8 \pi^2) \approx 1.3$ for $\Rey = 100$, which is the 
characteristic decay-time for velocity, related to viscous effects.
After that time interval, the velocity magnitude is reduced by a 
factor of $e^{-1}$, as depicted in Fig. \ref{fig-TGvortex}.
For the pressure field, the viscous decaying is twice faster.

\paragraph{Boundary conditions}
Periodic for the complete boundary.

\paragraph{Initial condition}
Given by the exact solution at $t=0$.

\subsubsection{Spatial Accuracy}

\begin{sidewaystable}
  \centering
  \caption{Spatial convergence analysis.}
  \label{tab:spatialConvergence}
  
  \resizebox{\linewidth}{!}{%
    \begin{tabular}{@{} lll *{14}{S[table-format=2.5]} @{}}
      \toprule
      \multirow{3}{*}{Var.} & \multirow{3}{*}{Step} & \multirow{3}{*}{Type} & \multicolumn{14}{c}{Grid Resolution (Error $\times 10^m$ \& Order)} \\
      \cmidrule(lr){4-17}
      & & & \multicolumn{2}{c}{10$\times$10} & \multicolumn{2}{c}{20$\times$20} & \multicolumn{2}{c}{40$\times$40} & \multicolumn{2}{c}{80$\times$80} & \multicolumn{2}{c}{160$\times$160} & \multicolumn{2}{c}{320$\times$320} & \multicolumn{2}{c}{640$\times$640} \\
      \cmidrule(lr){4-5} \cmidrule(lr){6-7} \cmidrule(lr){8-9} \cmidrule(lr){10-11} \cmidrule(lr){12-13} \cmidrule(lr){14-15} \cmidrule(lr){16-17}
      & & & {L2} & {L$\infty$} & {L2} & {L$\infty$} & {L2} & {L$\infty$} & {L2} & {L$\infty$} & {L2} & {L$\infty$} & {L2} & {L$\infty$} & {L2} & {L$\infty$} \\
      \midrule
      \multicolumn{17}{@{}l}{$v_1$} \\
      \cmidrule(lr){1-17}
      & $\Delta t_1$ & Error & 4.88549 & 8.14144 & 7.51857 & 1.53352 & 1.87103 & 3.73776 & 4.66922 & 9.34800 & 1.16668 & 2.33834 & 2.91627 & 5.84660 & 7.29024 & 1.46168 \\
      & & Order & {--} & {--} & 2.70 & 2.41 & 2.01 & 2.04 & 2.00 & 2.00 & 2.00 & 2.00 & 2.00 & 2.00 & 2.00 & 2.00 \\
      \cmidrule(lr){2-17}
      & $\Delta t_2$ & Error & 4.90125 & 8.19313 & 7.61725 & 1.55024 & 1.89338 & 3.77411 & 4.72348 & 9.43587 & 1.18022 & 2.36012 & 2.95013 & 5.90127 & 7.37507 & 1.47557 \\
      & & Order & {--} & {--} & 2.69 & 2.40 & 2.01 & 2.04 & 2.00 & 2.00 & 2.00 & 2.00 & 2.00 & 2.00 & 2.00 & 2.00 \\
      \cmidrule(lr){2-17}
      & $\Delta t_3$ & Error & 4.90285 & 8.19831 & 7.62717 & 1.55192 & 1.89565 & 3.77775 & 4.72893 & 9.44496 & 1.18157 & 2.36231 & 2.95352 & 5.90687 & 7.38356 & 1.47802 \\
      & & Order & {--} & {--} & 2.68 & 2.40 & 2.01 & 2.04 & 2.00 & 2.00 & 2.00 & 2.00 & 2.00 & 2.00 & 2.00 & 2.00 \\
      \midrule
      \multicolumn{17}{@{}l}{$v_2$} \\
      \cmidrule(lr){1-17}
      & $\Delta t_1$ & Error & 4.74295 & 8.59583 & 7.51852 & 1.52749 & 1.87103 & 3.73597 & 4.66922 & 9.34774 & 1.16668 & 2.33834 & 2.91627 & 5.84660 & 7.29024 & 1.46168 \\
      & & Order & {--} & {--} & 2.66 & 2.49 & 2.01 & 2.03 & 2.00 & 2.00 & 2.00 & 2.00 & 2.00 & 2.00 & 2.00 & 2.00 \\
      \cmidrule(lr){2-17}
      & $\Delta t_2$ & Error & 4.75961 & 8.65399 & 7.61722 & 1.54562 & 1.89338 & 3.77373 & 4.72348 & 9.43577 & 1.18022 & 2.36012 & 2.95013 & 5.90127 & 7.37507 & 1.47557 \\
      & & Order & {--} & {--} & 2.64 & 2.49 & 2.01 & 2.03 & 2.00 & 2.00 & 2.00 & 2.00 & 2.00 & 2.00 & 2.00 & 2.00 \\
       \cmidrule(lr){2-17}
      & $\Delta t_3$ & Error & 4.76130 & 8.65981 & 7.62714 & 1.54744 & 1.89565 & 3.77756 & 4.72893 & 9.44494 & 1.18157 & 2.36231 & 2.95352 & 5.90687 & 7.38356 & 1.47802 \\
      & & Order & {--} & {--} & 2.64 & 2.48 & 2.01 & 2.03 & 2.00 & 2.00 & 2.00 & 2.00 & 2.00 & 2.00 & 2.00 & 2.00 \\
      \midrule
      \multicolumn{17}{@{}l}{$p$} \\
      \cmidrule(lr){1-17}
      & $\Delta t_1$ & Error & 7.24861 & 16.7021 & 1.15242 & 2.71154 & 2.67762 & 5.52558 & 6.75480 & 1.35842 & 1.78206 & 3.56474 & 5.40399 & 1.08125 & 2.30111 & 4.60359 \\
      & & Order & {--} & {--} & 2.65 & 2.62 & 2.11 & 2.29 & 1.99 & 2.02 & 1.92 & 1.93 & 1.72 & 1.72 & 1.23 & 1.23 \\
      \cmidrule(lr){2-17}
      & $\Delta t_2$ & Error & 7.26272 & 16.7617 & 1.15869 & 2.74032 & 2.68247 & 5.63697 & 6.66419 & 1.34427 & 1.67344 & 3.35241 & 4.27711 & 8.55603 & 1.16419 & 2.32869 \\
      & & Order & {--} & {--} & 2.65 & 2.61 & 2.11 & 2.28 & 1.99 & 2.07 & 1.97 & 2.00 & 1.88 & 1.97 & 1.88 & 1.88 \\
       \cmidrule(lr){2-17}
      & $\Delta t_3$ & Error & 7.26413 & 16.7676 & 1.15933 & 2.74327 & 2.68355 & 5.66160 & 6.65646 & 1.34871 & 1.66261 & 3.33243 & 4.16443 & 8.33318 & 1.05050 & 2.10369 \\
      & & Order & {--} & {--} & 2.65 & 2.61 & 2.11 & 2.28 & 2.00 & 2.07 & 1.99 & 2.02 & 2.00 & 2.00 & 1.99 & 2.00 \\
      \midrule
      \multicolumn{17}{@{}l}{$\bar{p}$} \\
      \cmidrule(lr){1-17}
      & $\Delta t_1$ & Error & 7.24816 & 16.6996 & 1.15128 & 2.70913 & 2.66501 & 5.50300 & 6.62812 & 1.33378 & 1.65536 & 3.31240 & 4.13704 & 8.27912 & 1.03417 & 2.06982 \\
      & & Order & {--} & {--} & 2.65 & 2.62 & 2.11 & 2.30 & 2.01 & 2.04 & 2.00 & 2.01 & 2.00 & 2.00 & 2.00 & 2.00 \\
      \cmidrule(lr){2-17}
      & $\Delta t_2$ & Error & 7.26267 & 16.7614 & 1.15857 & 2.74008 & 2.68121 & 5.63446 & 6.65153 & 1.34181 & 1.66077 & 3.32724 & 4.15042 & 8.30309 & 1.03750 & 2.07541 \\
      & & Order & {--} & {--} & 2.65 & 2.61 & 2.11 & 2.28 & 2.01 & 2.07 & 2.00 & 2.01 & 2.00 & 2.00 & 2.00 & 2.00 \\
       \cmidrule(lr){2-17}
      & $\Delta t_3$ & Error & 7.26413 & 16.7676 & 1.15932 & 2.74324 & 2.68342 & 5.66135 & 6.65519 & 1.34846 & 1.66134 & 3.32991 & 4.15175 & 8.30786 & 1.03782 & 2.07834 \\
      & & Order & {--} & {--} & 2.65 & 2.61 & 2.11 & 2.28 & 2.00 & 2.07 & 2.00 & 2.02 & 2.00 & 2.00 & 2.00 & 2.00 \\
      \bottomrule
    \end{tabular}%
  } 

  \vspace{0.2cm}
  \raggedright
  \textbf{Notes:}
  \begin{itemize}[leftmargin=*,nosep]
     \item Errors scaled by $10^{-m}$, in which $m$ corresponds to grid sizes: \\
     10$\times$10: $m=3$; 20$\times$20: $m=4$; 40$\times$40: $m=5$; 80$\times$80: $m=6$; \\
     160$\times$160: $m=7$; 320$\times$320: $m=8$; 640$\times$640: $m=9$.
     \item For $p$ and $\bar{p}$, $m=2$ for L$_\infty$ error at 10x10.
     \item Temporal steps: $\Delta t_1 = 5\times10^{-5}$, $\Delta t_2 = 5\times10^{-6}$, $\Delta t_3 = 5\times10^{-7}$.
  \end{itemize}
\end{sidewaystable}

In consonance with the previous discussion, the theoretical convergence rate 
(second order) are achieved for all variables, as 
the solutions enter the asymptotic convergence regime, starting from 
the 80 $\times$ 80 grid resolution, as shown in Tab. \ref{tab:spatialConvergence}.
In addition, the symmetries of the flow are clearly present in the errors of 
horizontal ($v_1$) and vertical ($v_2$) components of the velocity, as expected, 
culminating in very close errors.

Furthermore, the most interesting aspect of this analysis is the pressure 
($p$) behaviour, as previously underlined. 
To show that, values of $\Delta t$ were carefully selected. 
Notably, for the most refined grids, the pressure convergence order 
deteriorates for the largest time-step, because temporal errors accumulates with 
spatial errors, insofar as
$\Delta t \sim (\Delta x)^2$.
The theoretical order is restored with $\Delta t$ reduction, as
spatial errors become dominant.
To emphasize this distinction, the time-averaged pressure 
($\bar{p}$), over the interval 
$[t - \Delta t, t]$, is also analysed:
\begin{equation}\label{Eq4.7}
  \bar{p}(x_1,x_2,t) = \frac{\exp(\alpha \Delta t)-1}{\alpha \Delta t} p
\end{equation}
in which $\alpha \coloneqq 16 \pi^2/\Rey$.
Unlike the instantaneous pressure, $\bar{p}$ consistently 
adheres to the theoretical convergence order, even for
the largest time-step, as exhibit in Tab. 
\ref{tab:spatialConvergence}.

A conclusion is that the discrete pressure is unconditionally 
first-order time accurate, requiring some clarification: the 
pressure in the projection method is not an ``instantaneous 
variable''. 
Instead, as stated before, it arises as a constraint 
(Lagrangian multiplier), enforcing consistency with the 
continuity equation over an interval 
of size $\Delta t$, 
effectively behaving as a time-averaged pressure over this 
interval.
By the mean value theorem, this introduces an error of 
$\order{\Delta t}$, when it is interpreted as an 
instantaneous value, for a specific time.

If the discrete pressure should be instantaneously described 
with accuracy of second order, one could refers to the 
modifications proposed in \cite{brown/2001:JCP}.

\subsubsection{Temporal Accuracy}

Isolating temporal errors in explicit (or semi-explicit) 
methods is generally unviable. 
Spatial errors dominate, and are coupled to temporal errors 
through stability conditions, such as the CFL constraint:
$
\Delta t \leq C (\Delta x)^2,
$
for a real, method-specific constant $C$.

A potential workaround involves statistical error modelling:
parameterising the error ($E$) -- 
e.g., $E = C_t (\Delta t)^{p_t} + C_x (\Delta x)^{p_x} + C_y (\Delta y)^{p_y}$,
for real coefficients ($C_t ,C_x, C_y$) and powers ($p_t,p_x,p_y$) --, 
constructing a balanced set of simulations across spatial 
and temporal resolutions, and performing non-linear regression 
to disentangle temporal and spatial contributions.
However, this requires extensive simulations.

The present approach circumvents this challenge by 
leveraging predictor-corrector temporal integration.
The local truncation error is quantified using the 
$L^2$-norm of the instantaneous difference between 
predicted and corrected solutions. 
The $L^2$-norm acts as a high-order spatial error 
filter, contributing to the isolation of temporal 
errors. 
While temporal averaging (a low-pass filter in time) 
could suppress high-frequency temporal artifacts, linked 
to spatial discretisation errors, it was unnecessary 
here.

\begin{table}[ht]
  \centering
  \caption{Temporal convergence analysis.}
  \label{tab:tempConvergence}
  
  \resizebox{\linewidth}{!}{%
    \begin{tabular}{@{} ll *{4}{S[table-format=1.3e-2]} *{3}{S[table-format=1.2]} @{}}
      \toprule
      \multirow{2}{*}{Grid} & \multirow{2}{*}{Var.} & \multicolumn{4}{c}{Error} & \multicolumn{3}{c}{Local Order} \\
      \cmidrule(lr){3-6} \cmidrule(lr){7-9}
      & & {$\Delta t_1$} & {$\Delta t_2$} & {$\Delta t_3$} & {$\Delta t_4$} & {$\Delta t_{1\to2}$} & {$\Delta t_{2\to3}$} & {$\Delta t_{3\to4}$} \\
      \midrule
      \multirow{3}{*}{50$\times$50} & $p$ & 2.546e-09 & 3.173e-10 & 3.960e-11 & 4.946e-12 & 3.00 & 3.00 & 3.00 \\
                                   & $v_1$ & 2.858e-09 & 3.562e-10 & 4.445e-11 & 5.552e-12 & 3.00 & 3.00 & 3.00 \\
                                   & $v_2$ & 2.858e-09 & 3.562e-10 & 4.445e-11 & 5.552e-12 & 3.00 & 3.00 & 3.00 \\
      \midrule
      \multirow{3}{*}{100$\times$100} & $p$ & 3.202e-10 & 3.996e-11 & 4.991e-12 & 6.236e-13 & 3.00 & 3.00 & 3.00 \\
                                     & $v_1$ & 3.537e-10 & 4.415e-11 & 5.516e-12 & 6.892e-13 & 3.00 & 3.00 & 3.00 \\
                                     & $v_2$ & 3.537e-10 & 4.415e-11 & 5.516e-12 & 6.892e-13 & 3.00 & 3.00 & 3.00 \\
      \midrule
      \multirow{3}{*}{200$\times$200} & $p$ & 2.561e-12 & 3.200e-13 & 3.999e-14 & 4.998e-15 & 3.00 & 3.00 & 3.00 \\
                                     & $v_1$ & 2.823e-12 & 3.528e-13 & 4.409e-14 & 5.511e-15 & 3.00 & 3.00 & 3.00 \\
                                     & $v_2$ & 2.823e-12 & 3.528e-13 & 4.409e-14 & 5.511e-15 & 3.00 & 3.00 & 3.00 \\
      \midrule
      \multirow{3}{*}{400$\times$400} & $p$ & 4.001e-14 & 5.001e-15 & 6.251e-16 & 7.831e-17 & 3.00 & 3.00 & 3.00 \\
                                     & $v_1$ & 4.409e-14 & 5.512e-15 & 6.894e-16 & 8.683e-17 & 3.00 & 3.00 & 2.99 \\
                                     & $v_2$ & 4.410e-14 & 5.511e-15 & 6.894e-16 & 9.126e-17 & 3.00 & 3.00 & 2.92 \\
      \bottomrule
    \end{tabular}%
  } 

  \vspace{0.15cm}
  \small
  \raggedright
  \textbf{Notes:} The error is based on L$_2$ norm of functions at the instant $t = 1.3$; \textit{Local order} refers to the local truncation order of the temporal discretisation method.
\end{table}

Table \ref{tab:tempConvergence} synthesises the results of temporal 
analysis.
Time steps are selected based on grid resolution as
$(\Delta t_1, \Delta t_2, \Delta t_3, \Delta t_4) 
= 
10^{-3} \cdot (4,2,1,0.5);
10^{-3} \cdot (2,1,0.5,0.25);
10^{-4} \cdot (4,2,1,0.5);
10^{-4} \cdot (1,0.5,0.25,0.125)$, from the coarser
to finer grids, respectively.
Note, once again, the expected equivalence of the errors for the 
vertical and horizontal velocity components, originated in the
flow symmetry.
All variables exhibit a third-order local truncation error, 
consistent with the expected (globally) second-order temporal 
accuracy of the method \cite{hairer/1993}. 
The global accuracy is degraded by one order, due to error 
accumulation during temporal marching, which scales as 
$(t_{total}/\Delta t) (\Delta t)^3 \sim (\Delta t)^2$, 
in which $t_{total}$ is the simulation final time.

Observe that, this analysis does not rely on access to an 
exact reference solution.
Instead, it estimates local error by calculating the 
discrepancy between predicted and corrected variable 
values -- quantities that are inherently computed at 
each time step.
This error estimation strategy can also be used to 
dynamically adjust the time-step size in adaptive 
time-stepping methods.

The error order should be greater than the precision of the 
floating-point system (double-precision herein).
The grid with higher resolution (400 $\times$ 400) is selected
to show this limitation.
For the case $\Delta t_4$, discretisation errors exceeds 
the machine precision, producing an artificially
degraded convergence order.



\subsection{Taylor--Couette flow}
\label{TCflow}

\begin{figure}[t]
\centering
\includegraphics[width=210pt]{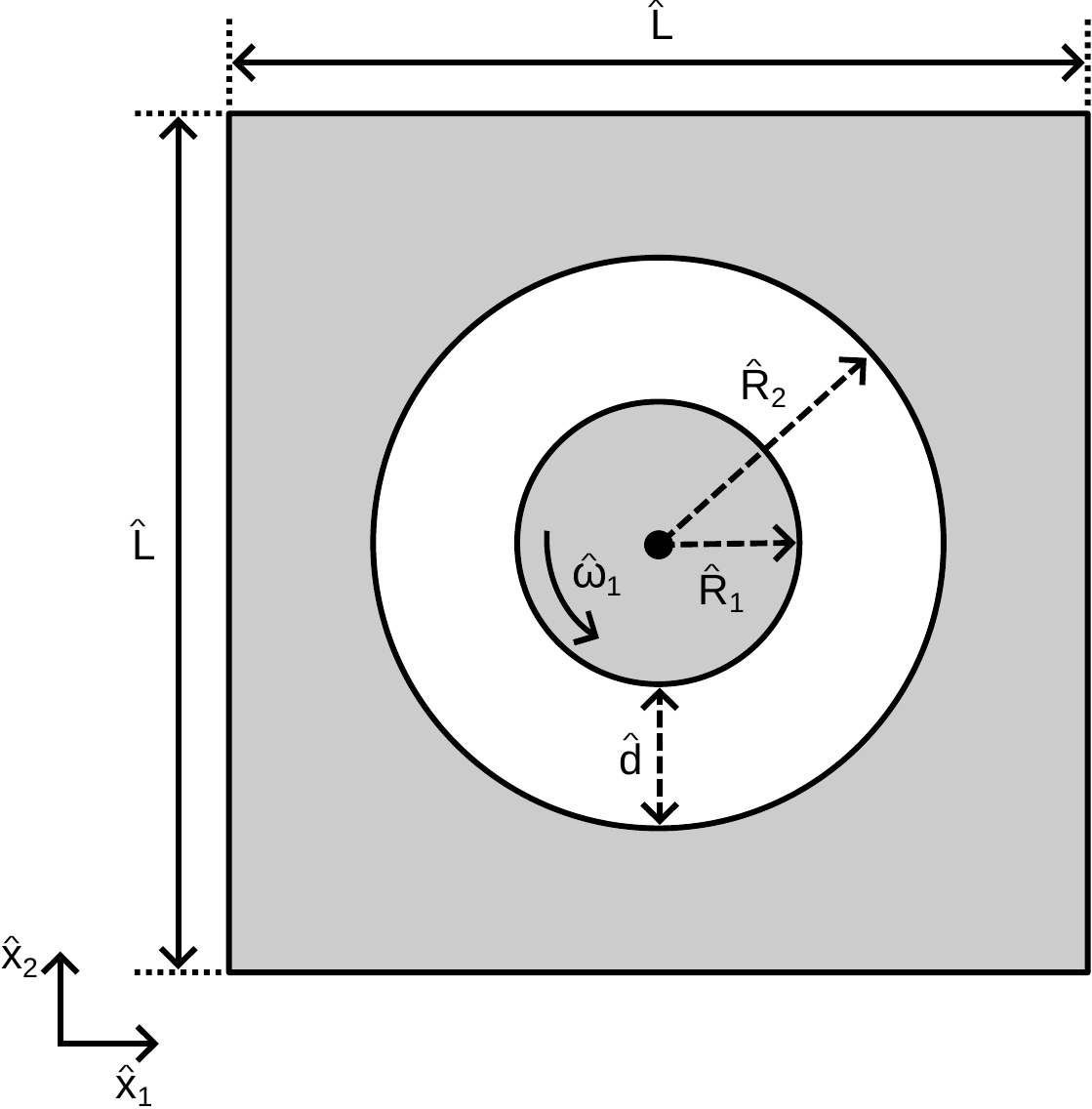}
\caption{
Axisymmetric representation of a Taylor--Couette flow.
}
\label{fig-TCflow}
\end{figure}

This problem is selected as a virtual laboratory to test 
the immersed boundary method (IBM).
Since, it also admits an exact solution for comparison.

Consider the flow between two concentric cylinders, with inner radius 
$\hat{R}_1$ (rotating at angular velocity $\omega_1$) and outer radius 
$\hat{R}_2$ (stationary, $\omega_2 = 0$).
An illustration is provided in Fig. \ref{fig-TCflow}.

The dimensionless variables are given by:
\begin{equation}
    r \coloneqq \dfrac{\hat{r}}{\hat{d}} \quad \text{and} \quad v_\theta 
    \coloneqq \dfrac{\hat{v}_\theta}{\hat{\omega_1} \hat{R}_1}, 
    \quad 
    \text{for} 
    \quad 
    \hat{R}_1 \leq \hat{r} \leq \hat{R}_2
\end{equation}
in which $\hat{d} \coloneqq \hat{R}_2 - \hat{R}_1$ is the annular gap, 
and $\hat{v}_\theta$ is the azimuthal velocity.

The general solution for the dimensionless azimuthal velocity ($v_\theta$) 
is given by:
\begin{equation}\label{Eq:FinalProfile}
    v_\theta(r) = \dfrac{\eta}{1 - \eta^2} \left[ \dfrac{1}{r(1 - \eta)} 
    - 
    r(1 - \eta) \right],
\end{equation}
expressed in terms of the radii ratio, $\eta \coloneqq \hat{R}_1/\hat{R}_2$.

Taking $\eta = \frac{1}{2}$ ($\hat{R}_2 = 2\hat{R}_1$ and $\hat{d} = \hat{R}_1$),
\begin{equation*}
    v_\theta(r) = \dfrac{4 - r^2}{3r} \quad \text{for} \quad 1 \leq r \leq 2
\end{equation*}
which, in Cartesian coordinates, reads
\begin{equation*}
    \vb*{v}(x_1,x_2) = \dfrac{1}{3}\left(\dfrac{4}{x_1^2+x_2^2} - 1 \right) 
     \begin{bmatrix}
         -x_2 \\
          x_1
     \end{bmatrix}
     \quad \text{for} \quad 1 \leq (x_1^2+x_2^2) \leq 4
\end{equation*}

\paragraph{Boundary conditions}
Periodic for the complete boundary, to avoid artificial 
interferences.

\paragraph{Initial condition}
Given by the exact solution.

\subsubsection{Accuracy of penalised solutions}

\begin{figure}[t]
\centering
     \begin{subfigure}[t]{230pt}
         \centering
         \includegraphics[width=\textwidth]{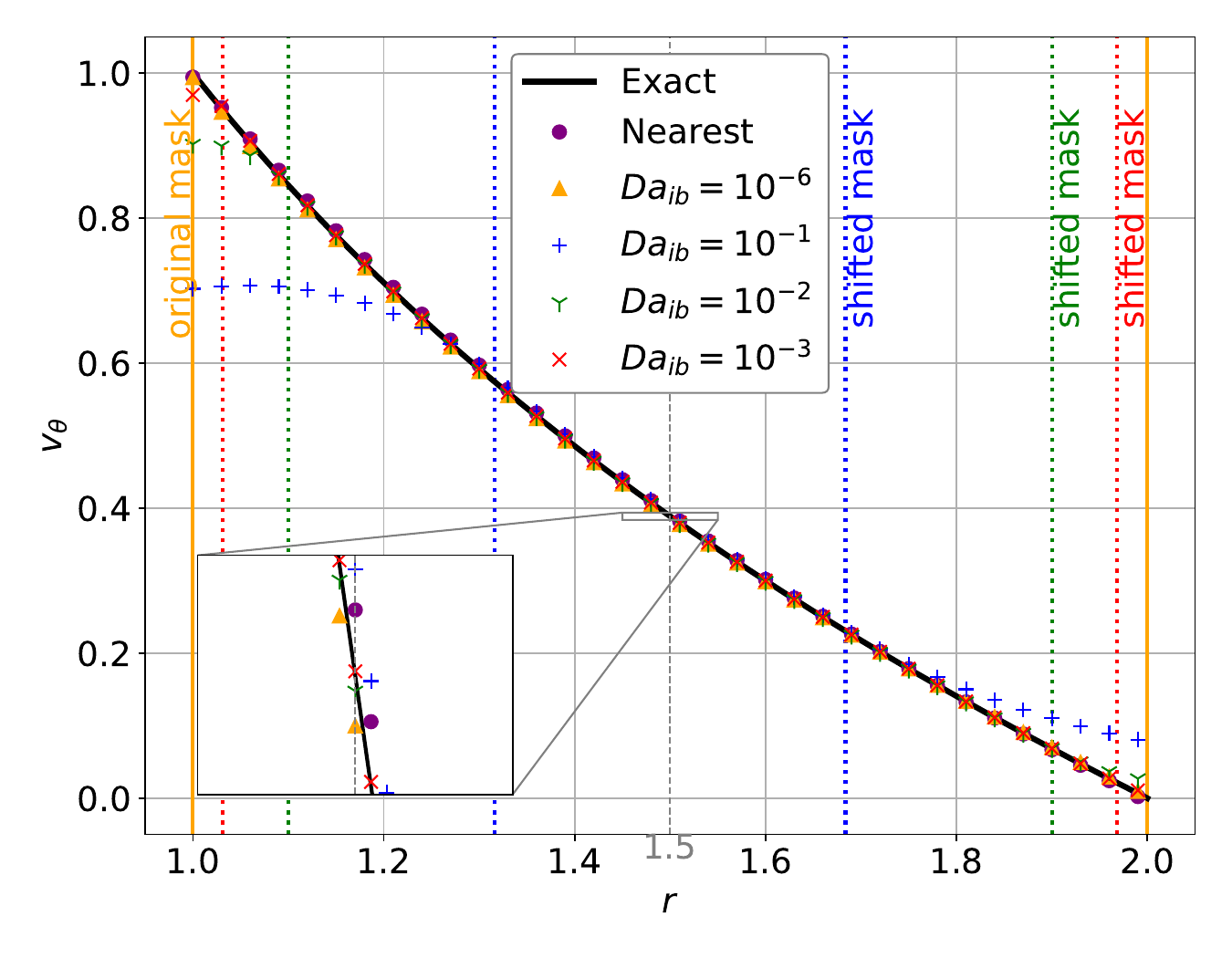}
         \caption{}
         \label{fig-TCa}
     \end{subfigure}
     \begin{subfigure}[t]{230pt}
         \centering
         \includegraphics[width=\textwidth]{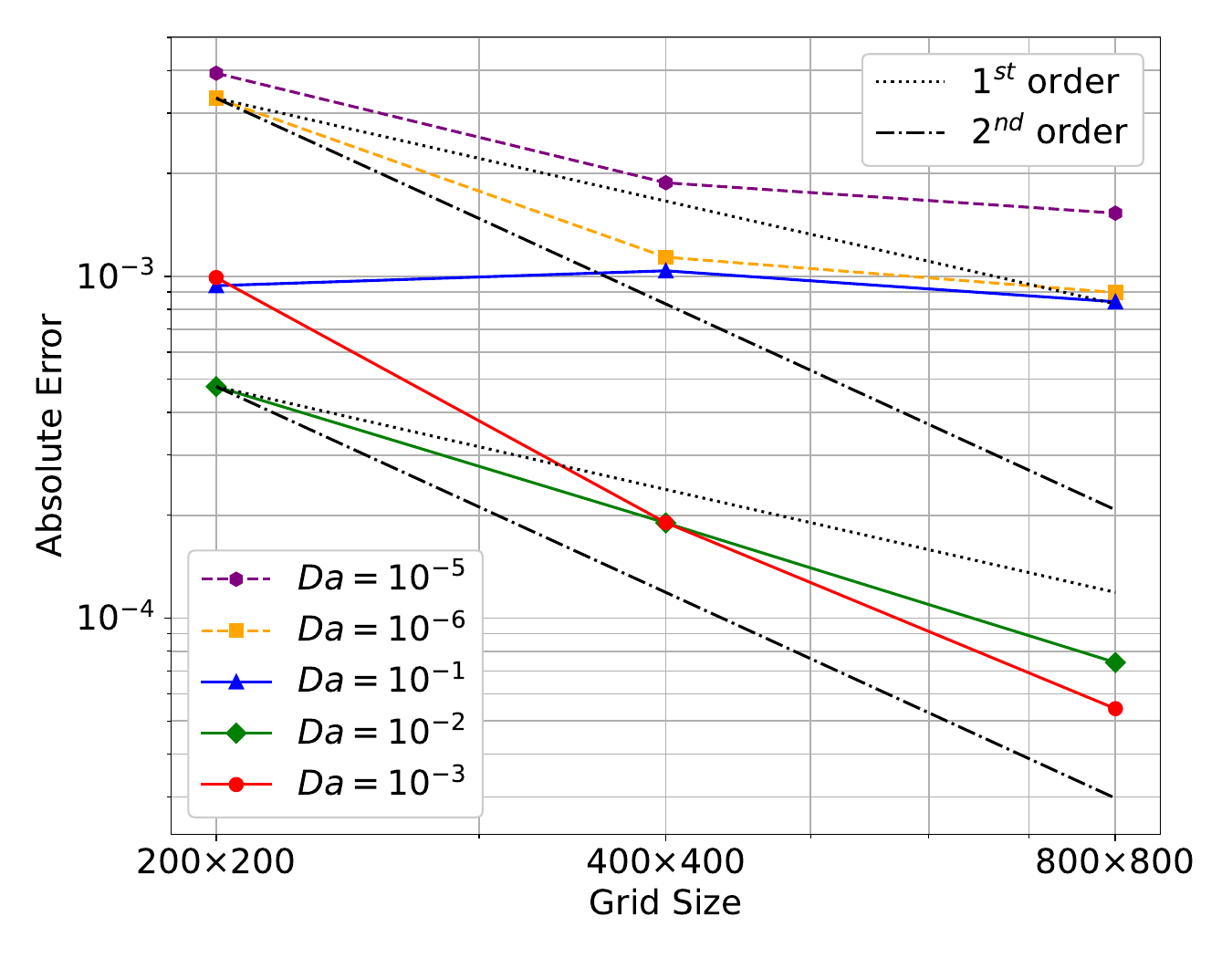}
         \caption{}
         \label{fig-TCb}
     \end{subfigure}
\caption{
    Comparison of penalised solutions accuracy (a) and order of convergence (b) 
    for $\Rey = 1$.
    The behaviours under $\Rey = 10$ are practically indistinguishable from those 
    presented, and have therefore been omitted for clarity.
}
\label{fig-TC}
\end{figure}

Penalised solutions for different immersed-boundary (IB) Darcy numbers 
($\Dar_{ib} = 10^{-1}, 10^{-2}$ and $10^{-3}$), using the shifted 
characteristic (or mask) function, are compared with the exact solution, 
a nearest-point implementation of the direct-forcing method \cite{fadlun/2000:JCP}, 
and a penalised solution for much smaller IB Darcy number ($\Dar_{ib} = 10^{-6}$), 
using the original mask function, for reference.
In addition, a grid convergence study is also explored for cases using 
shifted and original mask functions.

The finds are concisely presented in Fig. \ref{fig-TC}, for $\Rey = 1$:
Figure \ref{fig-TCa} depicts a comparison for the azimuthal 
velocity at $\theta = \pi/4$, to test both velocity components 
($v_i$ for $i = 1,2$);
Figure \ref{fig-TCb}, in turn, synthesises a grid convergence study for 
penalised solutions.
Concretely, the absolute error is based on the L$^2$-norm of differences
between exact and numerical solutions for the velocity component $v_2$,
in the complete annular section (i.e., for each $\theta \in [0,2\pi]$ ).
Errors for the $v_1$-component are near identical (due to the symmetry).
The results for $\Rey = 10$ are visually indistinguishable from those for 
$Re = 1$, making their presentation redundant.

The penalised solutions using the shifted mask function (i.e., the extension of 
the solid-to-fluid frontier) are the best descriptions of the exact solution -- 
for $\Dar_{ib} = 10^{-2}$ and $\Dar_{ib} = 10^{-3}$ --, except into the respective 
transition zones, delimited in Fig. \ref{fig-TCa} by dotted vertical lines.
On the other hand, the pensalised solution using the original mask function 
(solid vertical lines in Fig. \ref{fig-TCa}), even for a several orders of 
magnitude smaller IB Darcy number (i.e., $\Dar_{ib} = 10^{-6}$), was not able to 
overcome the accuracy of the shifted-mask-function solutions.
This fact is due to its penalisation error \cite{angot/1999:NM,carbou/2003:ADE}, 
evidenced by the saturation of dashed curves in Fig. \ref{fig-TCb}, as the 
penalisation error becomes dominant over the spatial discretisation error, due 
to the grid refinement.

The penalised solutions using the shifted characteristic function exhibit 
error reduction and improved order of convergence, as $\Dar_{ib}$ increases, 
according to Fig. \ref{fig-TCb}.
While $\Dar_{ib} = 10^{-3}$ attains, asymptotically, second-order convergence 
under grid refinement (the first grid resolution is not sufficient to capture 
the transition zone properly), $\Dar_{ib} = 10^{-2}$ displays intermediate 
convergence (between first and second order).
For $\Dar_{ib} = 10^{-1}$, errors saturate at $\order{10^{-3}}$, 
revealing forcing-term rigidity deficiencies -- i.e., the porous medium
becomes too permissive.
This behaviour suggests an optimal $\Dar_{ib}$: the smallest one 
recovering the theoretical spatial-convergence rate, while minimally constraining 
grid refinement, since the condition $\Delta x_i < \Dar_{ib}^{1/2}$ should be
satisfied along the extended solid region.
Nevertheless, a formal consolidation of this optimum is deferred to future work.

It is also pertinent to mention, within the context of IBMs, 
that the direct forcing technique described in \cite{uhlmann/2005:JCP} -- with 
transfer of properties between Lagrangian and Eulerian locations --, 
demonstrated pronounced numerical instability when integrated with the present 
method. 
This instability manifested, apparently, acute sensitivity to outflow conditions in the body region.
For a flow around a cylinder, an outlet sponge layer offered only partial 
remediation.
Dedicated research is needed to complete understand this issue.


\subsection{Thermally driven cavity}
\label{tdCavity}

\begin{figure}[t]
\centering
\includegraphics[width=210pt]{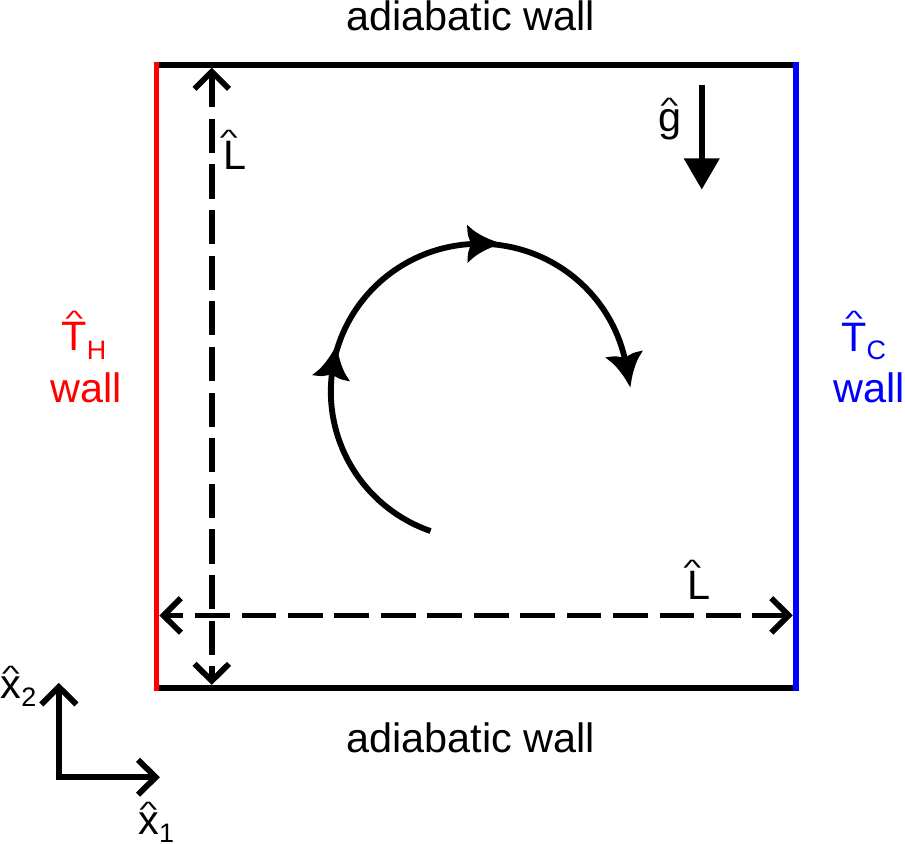}
\caption{
Scheme of enclosed square cavity of area $\hat{L}^2$, under thermally induced flow.
The left wall is maintained at temperature $\hat {T}_H$, and the right wall, at $\hat {T}_C$, 
with $\hat {T}_H > \hat {T}_C$.
Horizontal walls are considered adiabatic.
}
\label{fig-cavity}
\end{figure}

Thermally driven cavity is a standard test-case for variable-density 
solvers \cite{gutierrez/2022:IJNMF}.
Specially, in the context of natural convection \cite{tyliszczak/2014:IJNMHFF, paillere/2000:HMT}.
Benchmark solutions of \cite{vierendeels/2003:IJNMHFF} -- using a fully compressible Navier-Stokes solver over a high-resolution mesh -- will be used as reference.

Consider a differentially heated square cavity with side length $\hat{L}$, as presented in Fig. \ref{fig-cavity}.
Left and right walls are held at distinct temperatures, $\hat{T}_H$ and $\hat{T}_C$, 
respectively, with $\hat{T}_H > \hat{T}_C$.
Horizontal walls are perfectly insulated (adiabatic), i.e., with null thermal energy flux by them.

For a buoyancy-driven flow, it is relevant to define the \Ra[l] (\Ra).
Namely,
\begin{equation}\label{Eq4.8}
    \Ra
    \coloneqq
    \Pr \dfrac{\hat{g}_c \hat{\rho}_c^2 \left( \hat{T}_H - \hat{T}_C \right) \hat{l}_c^3 }{ \hat{T}_c  \hat{\mu}_c^2 },
\end{equation}
which represents the relative importance between convection and thermal diffusion (or conduction).
In fact, a $\Ra \gg 1$ ($\Ra \ll 1$) condition indicates a flow predominantly dictated by convection (conduction).
Hence, it is the analogous of the \Pe[l] ($\Pe$) for the free convection case, in which the flow is induced by density gradients.

The characteristic temperature is the mean temperature, i.e., $\hat{T}_c \coloneqq (\hat{T}_H + \hat{T}_C) / 2$, which permits the definition of the dimensionless temperature difference as
\begin{equation}\label{Eq4.9}
    \epsilon
    \coloneqq
    \dfrac{\hat{T}_H - \hat{T}_C}{2\hat{T}_c}
\end{equation}
Note that it quantifies the influence of the temperature on the flow field not directly related to buoyancy effects \cite{vierendeels/2003:IJNMHFF}.
Because of this, as $\epsilon$ increases, the Boussinesq approximation becomes invalid, while the present formulation can handle arbitrary temperature differences (respecting the limits of numerical stability).
To demonstrate it, the greatest value of temperature difference (i.e., $\epsilon = 0.6$) will be considered here.

A characteristic velocity can be induced from the solution for natural convection along a vertical wall \cite{white/1974}: $\hat{v}_c \coloneqq \Ra^{1/2} \hat{\mu}_c / (\hat{\rho}_c \hat{l}_c)$, in which $\hat{\mu}_c \coloneqq \hat{\mu}(\hat{T}_c)$, $\hat{\rho}_c \coloneqq \hat{\rho}(\hat{T}_c)$ and $\hat{l}_c \coloneqq \hat{L}$ \cite{vierendeels/2003:IJNMHFF}.
A reference hydrodynamic pressure is also derived from this scales as $\hat{p}_c \coloneqq \hat{\rho}_c \hat{v}_c^2$.
Therefore, it is possible to relate the \Ra[l] (\Ra) to the \Pe[l] (\Pe) as \cite{gutierrez/2022:IJNMF}
\begin{equation}\label{Eq4.10}
    \Pe
    =
    \Pr \Ra^{1/2}
\end{equation}
and, also, the \Fr[l] (\Fr) is obtained as \cite{gutierrez/2022:IJNMF}
\begin{equation}\label{Eq4.11}
    \Fr
    =
    (\Pr 2 \epsilon )^{1/2}
\end{equation}

For a consistent comparison, $\Pr = 0.71$, $\gamma = 1.4$, and the  Sutherland's law \cite{Sutherland/1893:TLDPMJS} is considered to model transport coefficients, i.e.,
\begin{equation}\label{Eq4.12}
   \kappa(T)
   =
   T^{3/2}
   \dfrac{1 + C}{T + C} \ ,
\end{equation}
in which $C \coloneqq \hat{C} / \hat{T}_c$, with $\hat{C} = \SI{110.5}{\kelvin}$ denoting a substance-dependent constant, in this case, air.

An initial (characteristic) thermodynamic state is specified by $\hat{p}_{0,c} = \SI{101325}{\pascal}$, $\hat{T}_{c} = \SI{600}{\kelvin}$ and $\hat{\rho}_{c} = \hat{\rho}(\hat{p}_{0,c},\hat{T}_{c})$, which, in conjunction with the pair $(\Ra, \epsilon)$, is sufficient to describe the complete state of the system.
In fact, hydrodynamic conditions are imposed from \Ra, and dimensionless temperatures $T_H = (1 + \epsilon) = 1.6$ and $T_C = (1 - \epsilon) = 0.4$ are implicit in $\epsilon$.

To recapitulate, for an enclosed domain, the thermodynamic pressure ($p_0$) evolutes as (see Sec. \ref{Formulation} )
\begin{equation}\label{Eq4.13}
   p_0
   =
    \dfrac{m}{\int\limits_{V} T^{-1} dV}
\end{equation}
in which $m$ is the initial (dimensionless) mass, a constant.

Finally, in a problem of ``heat transfer'', it is convenient and representative to work in terms of the \Nus[l] ($\Nus$), which quantifies the enhancement of thermal energy transfer due to fluid motion
relative to conduction alone.
Its local version is defined by
\begin{equation}\label{Eq4.14}
    \Nus(x_2)
    \coloneqq
    \dfrac{1}{2 \epsilon}
    \left(
        \Pe
        \rho
        v_1
        T        
        -
        \kappa
        \dfrac{\partial T}{\partial x_1}
    \right)
\end{equation}
and, its spatially average (over a vertical profile), as
\begin{equation}\label{Eq4.15}
   \overline{\Nus}
   \coloneqq
   \int\limits_{x_2=0}^{x_2=1}
   \Nus
   dx_2
\end{equation}
\paragraph{Boundary conditions}

For the convenient choice of $Q = S+1$,
\begin{subequations}\label{Eq4.16}
    \begin{gather}
        v_1 = v_2 = Z = H - (T_H + 1) = 0
        \quad
        ,
        \quad
        \text{on the left wall ($x_1 = 0$)};
       \label{Eq4.16a} \\ 
        v_1 = v_2 = Z = H - (T_C + 1) = 0
        \quad
        ,
        \quad
        \text{on the right wall ($x_1 = 1$)};
        \label{Eq4.16b} \\
        v_1 = v_2 = Z = \frac{\partial H}{\partial x_2} 
        = 0
        \quad
        ,
        \quad
        \text{on the horizontal walls ($x_2 = 0$ and $x_2 = 1$)}.
        \label{Eq4.16c}
    \end{gather}
\end{subequations}

\paragraph{Initial condition}

A quiescent system, under the mean temperature 
($T = 1$), composed by a uniform, non-reacting 
gaseous mixture (air),
\begin{equation}\label{Eq4.17}
    v_1 = v_2 = Z = H - 2 = 0
    \quad
    ,
    \quad
    \text{at $t = 0$}
\end{equation}

The verification is performed by comparing the results, 
for $(\Ra, \varepsilon) = (10^n, 0.6)$ with $n \in 
\{2,7\}$, against those from 
\cite{vierendeels/2003:IJNMHFF}.
These two cases were chosen as they represent the two extremes,
with the highest relative contributions from diffusion and 
convection, respectively.
In this sense, intermediate cases, i.e., $n \in \{3,4,5,6\}$, 
do not provide additional information and are, therefore, omitted.
Note that for both cases, the flow is in the steady state,
considered here when temporal derivative of the thermodynamic 
pressure ($p_0$) is, at least, as small as $10^{-6}$.
In fact, due to its ``global nature'', $p_0$ is an appropriate 
choice for monitoring global convergence, including for the 
steady state, grid resolution, and other related factors. \\

\subsubsection{Grid convergence study}

\begin{figure}[t]
\centering
\includegraphics[width=270pt]{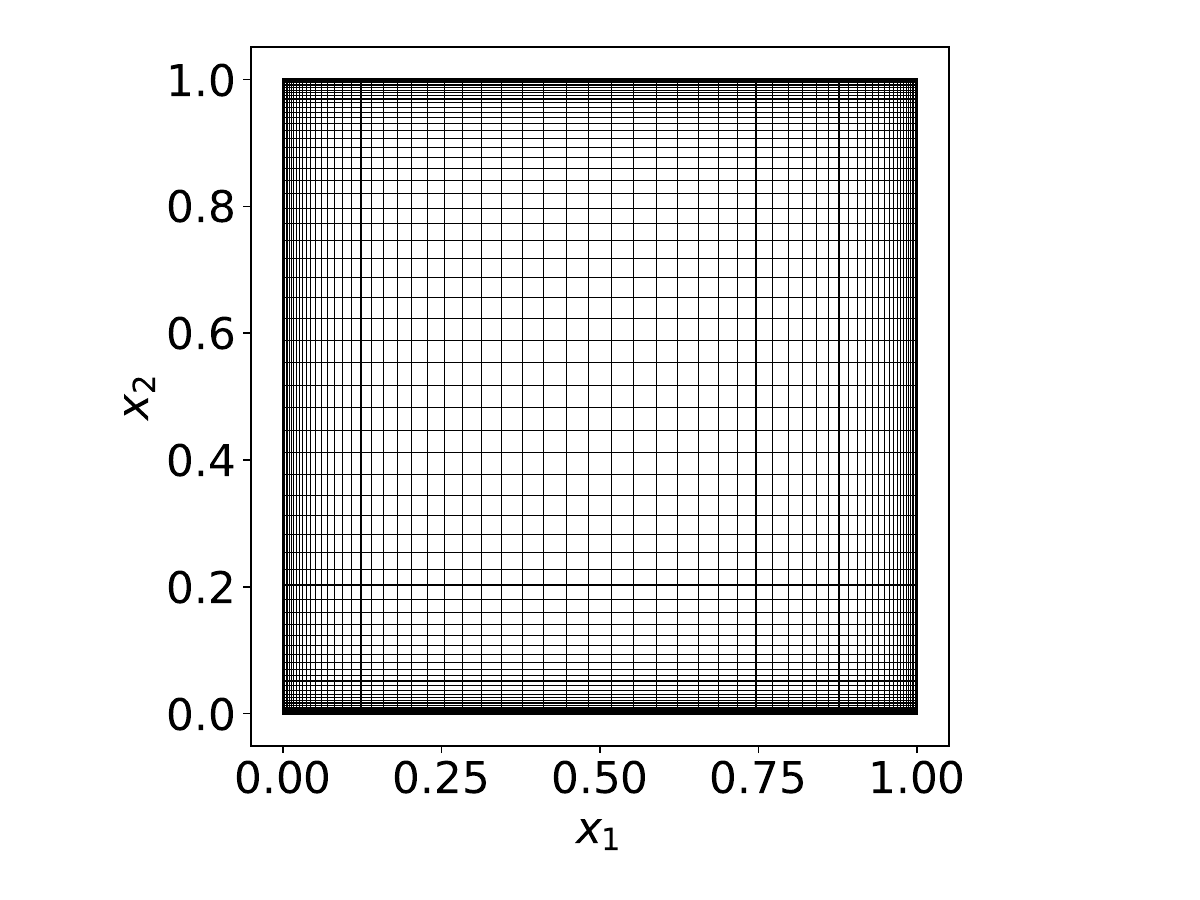}
\caption{
Non-uniformly spaced grid for a unity square, with increased resolution 
of twenty-one ($21$) times near boundary.
}
\label{fig-gridCav}
\end{figure}

\begin{table}[th]
  \centering
  \caption{Grid convergence study for $\Ra = 10^2$.}
  \label{Tab-gridRa2}
  \begin{tabular}{c S[table-format=1.4] S[table-format=1.5] c c}
    \toprule
    Grid & {$\max(\Delta_{x_i})$} & {$p_0$} & {GCI} & {GCI} ratio \\
    \midrule
    $64 \times 64$    & 0.01959   & 0.95752 & --      & --    \\
    $96 \times 96$    & 0.01303   & 0.95744 & 0.009\% & --    \\ 
    $144 \times 144$  & 0.00868  & 0.95740 & 0.004\% & 2.236 \\ 
    $216 \times 216$  & 0.00578 & 0.95738 & 0.002\% & 2.241 \\ 
    \bottomrule
  \end{tabular}
  \begin{tablenotes}
   \item The order of convergence ($a$) is estimate as $a = 1.99$, using values of the 
   most refined grids in Eq. (\ref{Eq4.2}), with refinement ratio $\delta = 1.5$;
   \item GCI is the grid convergence index, as defined in Eq. (\ref{Eq4.3}).
   \item Calculated on non-uniform grid such as $\min(\Delta_{x_i}) = \max(\Delta_{x_i})/2$ for $i=1,2$.
   \end{tablenotes}
\end{table}

\begin{table}[th]
  \centering
  \caption{Grid convergence study for $\Ra = 10^7$.}
  \label{Tab-gridRa7}
  \begin{tabular}{c S[table-format=1.4] S[table-format=1.5] c c}
    \toprule
    Grid & {$\max(\Delta_{x_i})$} & {$p_0$} & {GCI} & {GCI} ratio \\
    \midrule
    $64 \times 64$    & 0.03573   & 0.91990 & --      & --    \\
    $96 \times 96$    & 0.02372   & 0.92138 & 0.170   & --    \\ 
    $144 \times 144$  & 0.01577  & 0.92208 & 0.080   & 2.126 \\ 
    $216 \times 216$  & 0.01049 & 0.92240 & 0.037   & 2.180 \\ 
    \bottomrule
  \end{tabular}
  \begin{tablenotes}
   \item The order of convergence ($a$) is estimated as $a = 1.92$, using values of the 
   most refined grids in Eq. (\ref{Eq4.2}), with refinement ratio $\delta = 1.5$;
   \item GCI is the grid convergence index, as defined in Eq. (\ref{Eq4.3}).
   \item Calculated on non-uniform grid such as $\min(\Delta_{x_i}) = \max(\Delta_{x_i})/21$ for $i=1,2$.
   \end{tablenotes}
\end{table}

Simulations are performed over non-uniform grids, with accumulation of points 
near walls, for numerical efficiency.
Indeed, the resolution at the boundary region can be refined using the hyperbolic tangent transformation, given by 
\begin{equation}\label{Eq4.18}
    \tilde{x}_i 
    = 
    \dfrac{
        1 
        + 
        \tanh\left[ \alpha \left(x_i - 1/2 \right) \right] 
    }
    {2 \tanh\left( \alpha/2 \right)},
\end{equation}
for each direction (i.e., $i = 1,2$).
The real constant $\alpha$ controls the grid clustering intensity. 
As $\alpha \rightarrow 0$, the transformation tends to the identity 
transformation. 
As $\alpha$ increases, the accumulation becomes more pronounced near 
the domain boundary. 
The parameter $\alpha$ is determined numerically to achieve a specified 
spacing ratio (e.g., $10$ for tenfold resolution enhancement 
at boundaries).

This analysis, as justified above, is based on the hydrodynamic pressure 
($p_0$). 
Tables \ref{Tab-gridRa2} and \ref{Tab-gridRa7} synthesise the finds 
of the grid convergence study, for the cases $\Ra = 10^2$ and 
$10^7$, respectively.
The grid partition is the same for both directions ($i= 1,2$), 
with $\max(\Delta_{x_i}) = m \min(\Delta_{x_i})$ in the interior region, 
with $m = 2$ for $\Ra = 10^2$, and $m = 21$ for $\Ra = 10^7$.
Both cases encompass fourth grid resolutions, departing from $64 \times 64$ 
to $216 \times 216$, with a refinement ratio of $\delta = 1.5$.
The accumulation of points in the proximity of boundary is much more pronounced for the case $\Ra = 10^7$ ($21$ times the interior resolution), because it requires more resolution to describe, properly, variations 
inside the thin thermal boundary layer, i.e., to capture strong gradients
near walls.
As evidenced, the order of convergence is compatible with a second-order 
method.
Furthermore, it is possible to infer the importance of a non-uniform 
grid, refined only in the required regions, for efficient 
computation. 
With relatively fewer degrees of freedom, results are very close to those 
of the reference, as will be presented in the next section.

\subsubsection{Results comparison}

\begin{figure}[t]
\centering
     \begin{subfigure}[t]{210pt}
         \centering
         \includegraphics[width=\textwidth]{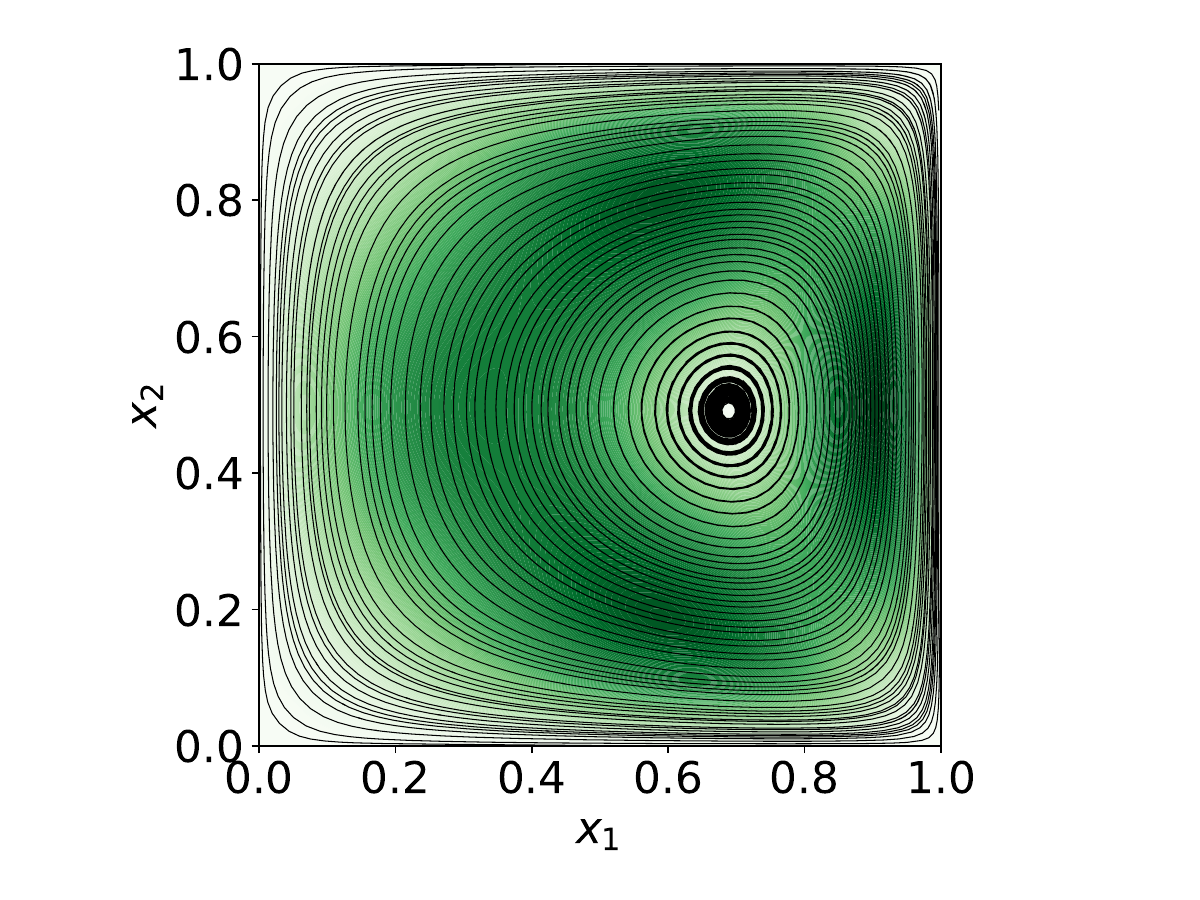}
         \caption{}
         \label{fig-qualiA}
     \end{subfigure}
     \begin{subfigure}[t]{210pt}
         \centering
         \includegraphics[width=\textwidth]{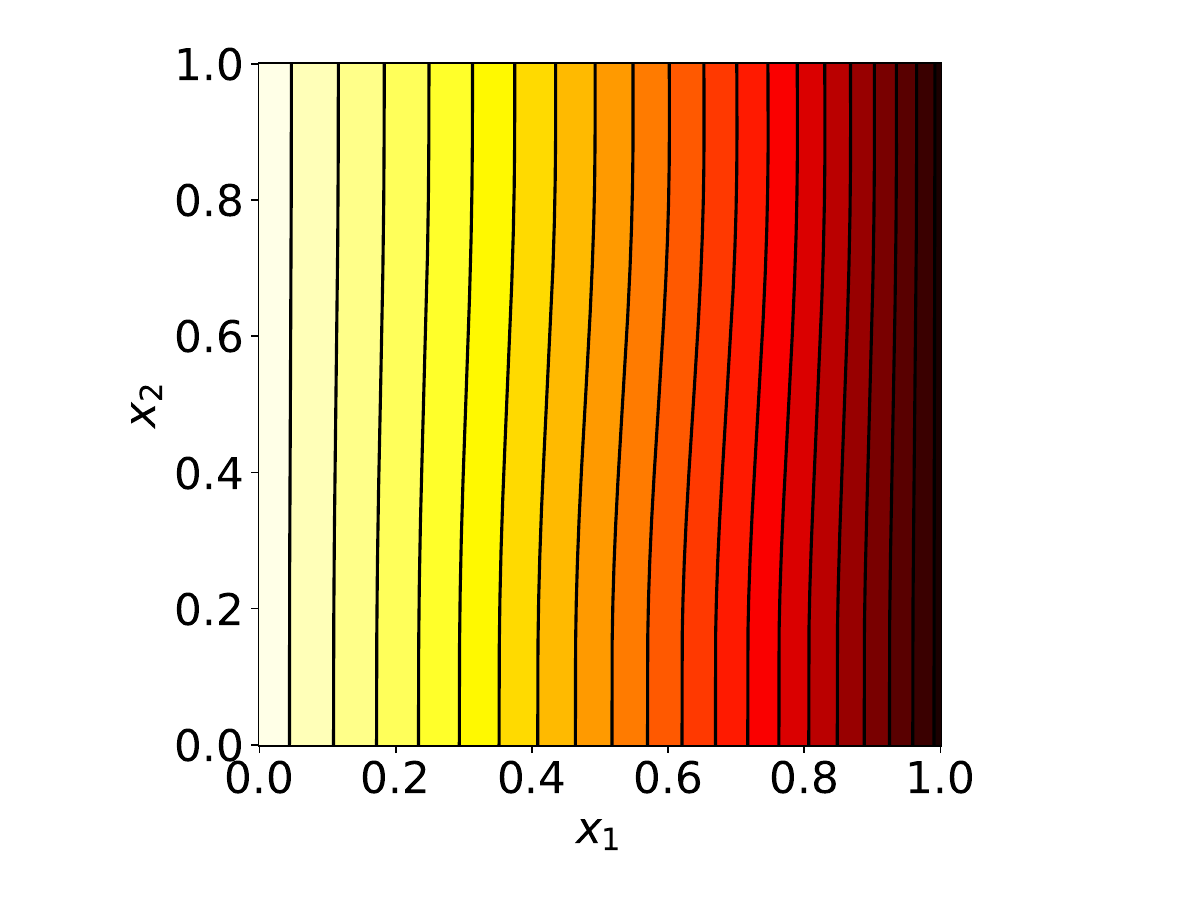}
         \caption{}
         \label{fig-qualiB}
     \end{subfigure}
     \begin{subfigure}[t]{210pt}
         \centering
         \includegraphics[width=\textwidth]{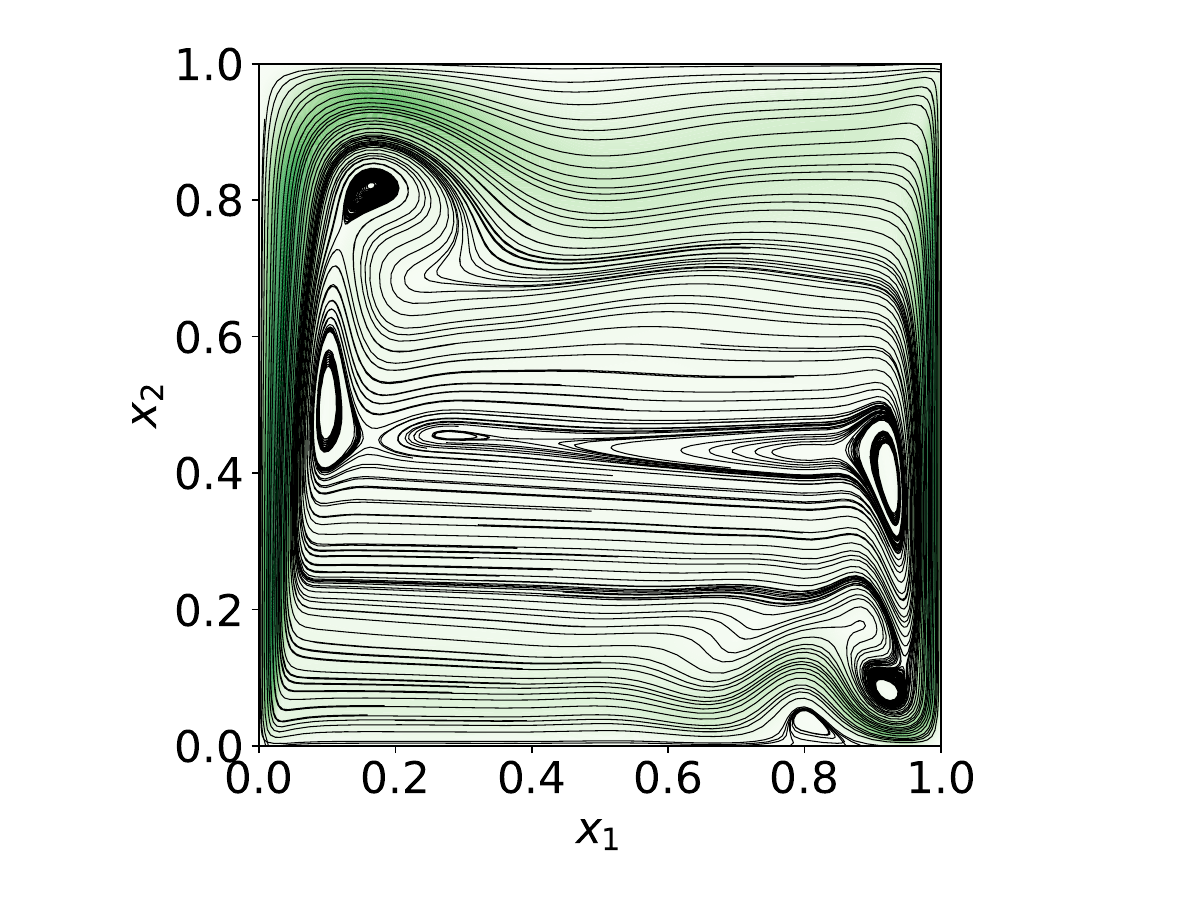}
         \caption{}
         \label{fig-qualiC}
     \end{subfigure}
     \begin{subfigure}[t]{210pt}
         \centering
         \includegraphics[width=\textwidth]{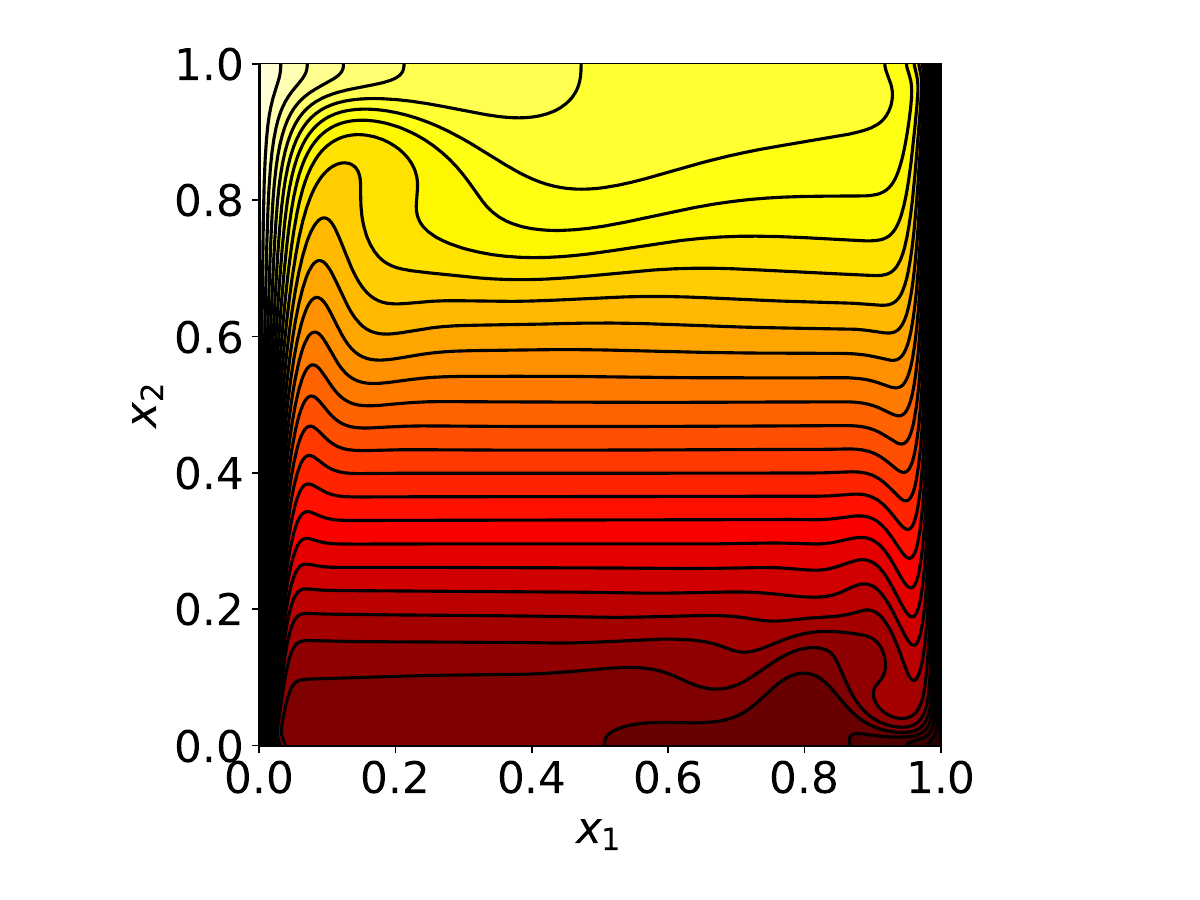}
         \caption{}
         \label{fig-qualiD}
     \end{subfigure}
\caption{
    Streamlines -- (a) and (c) -- and isotherms -- (b) and (d) -- for a qualitative 
    verification of cases $\Ra = 10^2$ and $\Ra = 10^7$, respectively.
    Light (dark) regions indicate areas of low (high) speed 
    or high (low) temperatures.
}
\label{fig-quali}
\end{figure}

\begin{figure}[h!]
\centering
    \begin{subfigure}[t]{210pt}
         \centering
         \includegraphics[width=\textwidth]{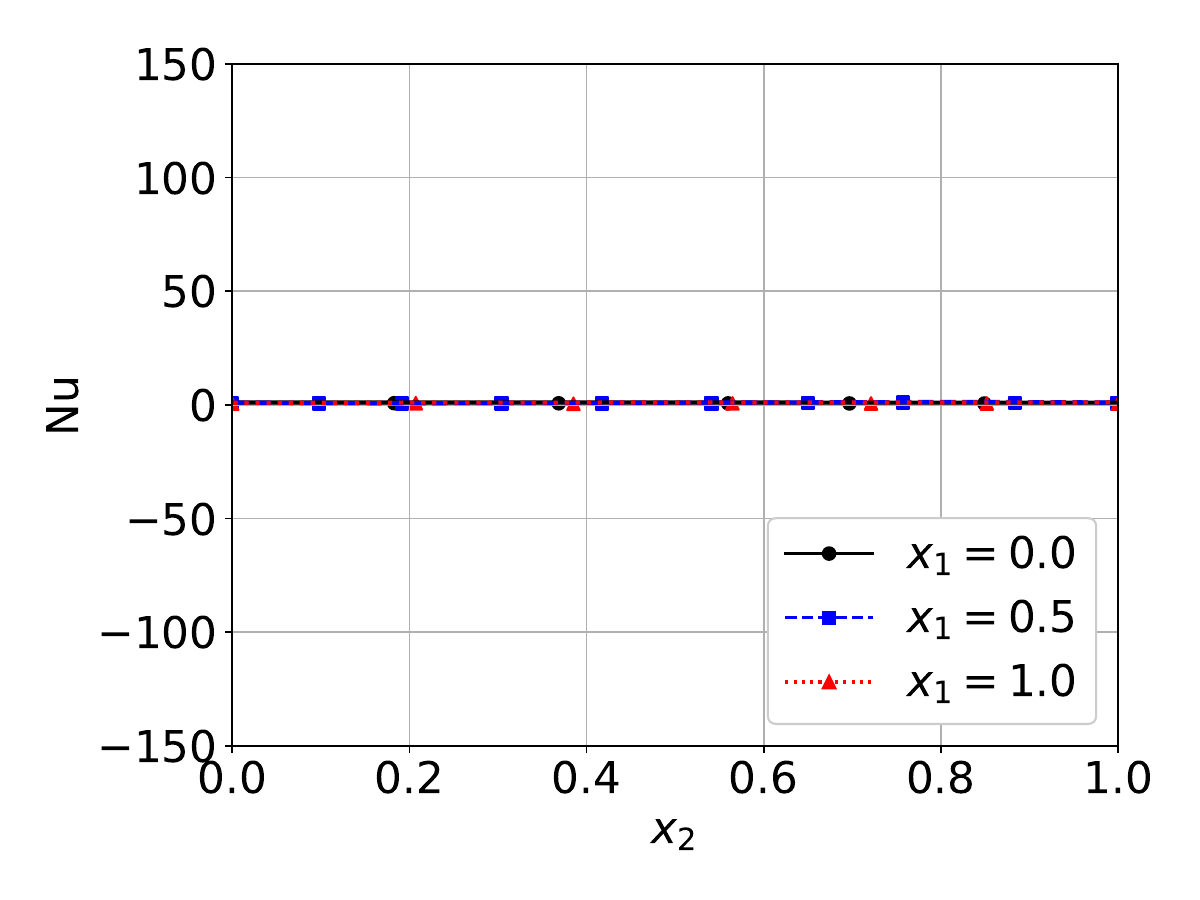}
         \caption{}
         \label{fig-profE2a}
    \end{subfigure}
    \begin{subfigure}[t]{210pt}
         \centering
         \includegraphics[width=\textwidth]{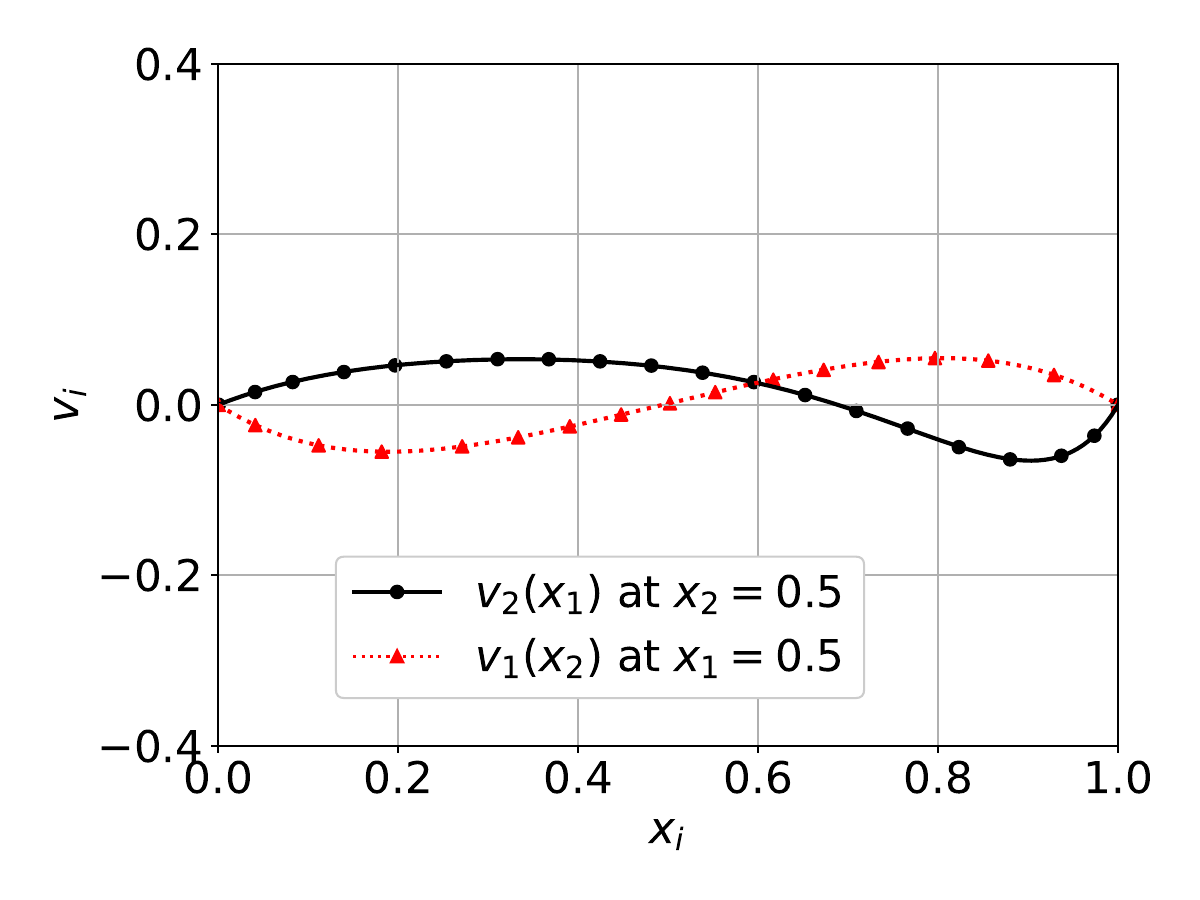}
         \caption{}
         \label{fig-profE2b}
    \end{subfigure}
    \begin{subfigure}[t]{210pt}
         \centering
         \includegraphics[width=\textwidth]{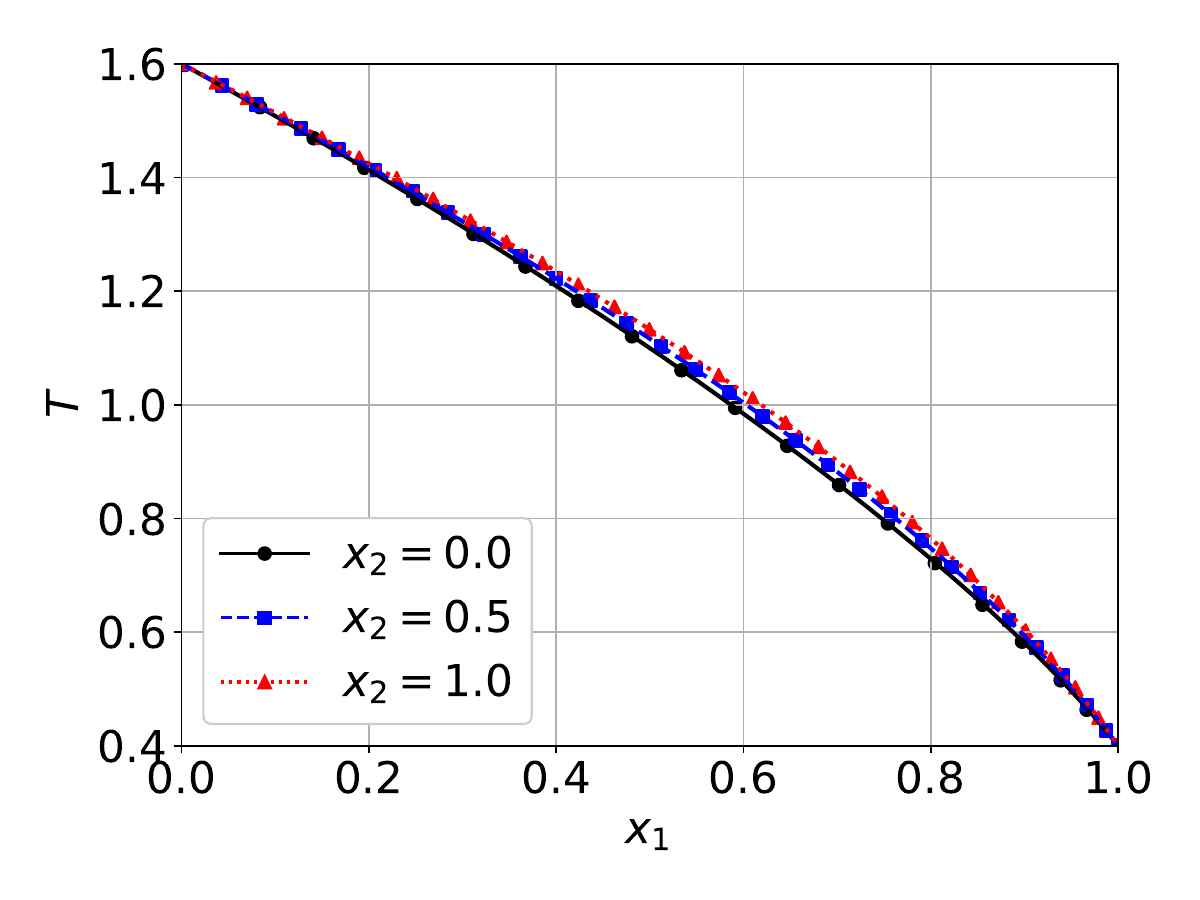}
         \caption{}
         \label{fig-profE2c}
    \end{subfigure}
    \begin{subfigure}[t]{210pt}
         \centering
         \includegraphics[width=\textwidth]{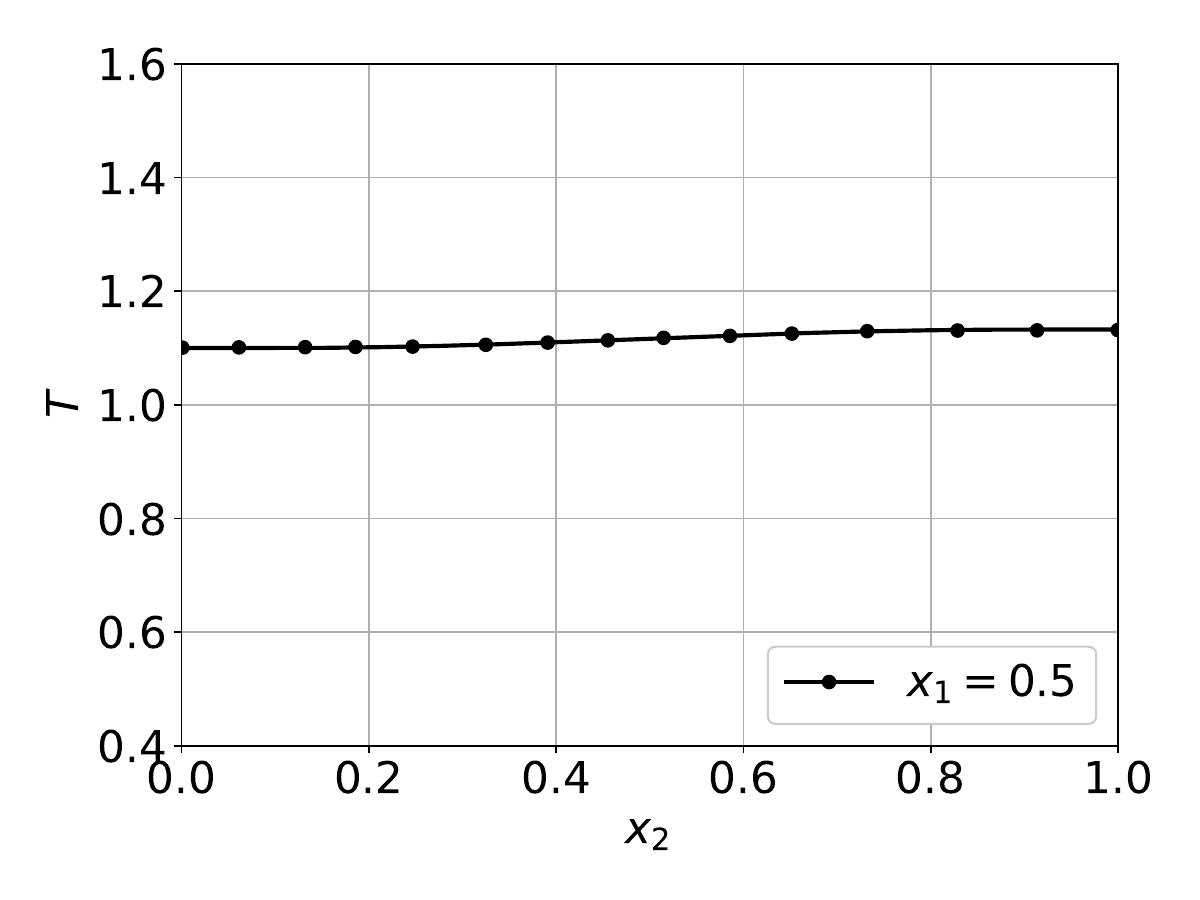}
         \caption{}
         \label{fig-profE2d}
    \end{subfigure}
    \begin{subfigure}[t]{210pt}
         \centering
         \includegraphics[width=\textwidth]{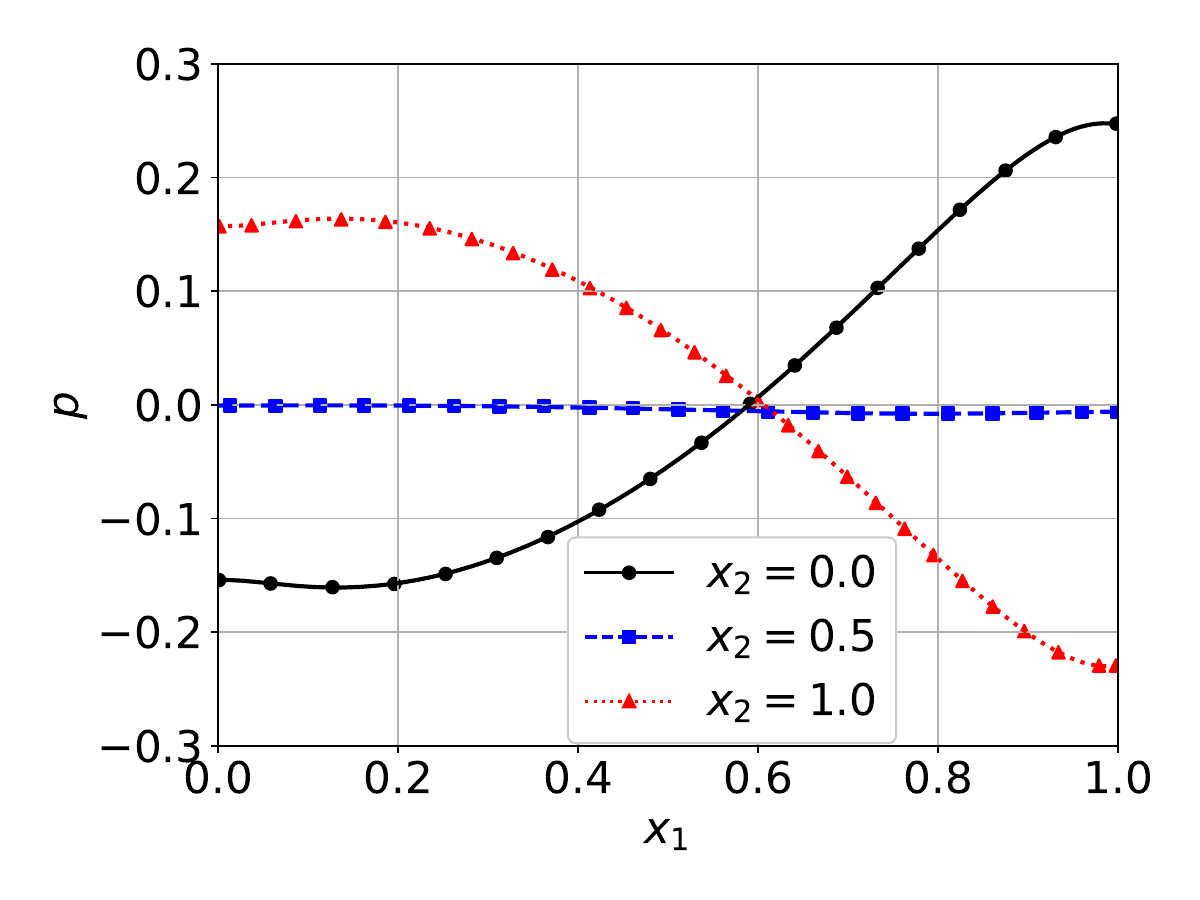}
         \caption{}
         \label{fig-profE2e}
    \end{subfigure}
    \begin{subfigure}[t]{210pt}
         \centering
         \includegraphics[width=\textwidth]{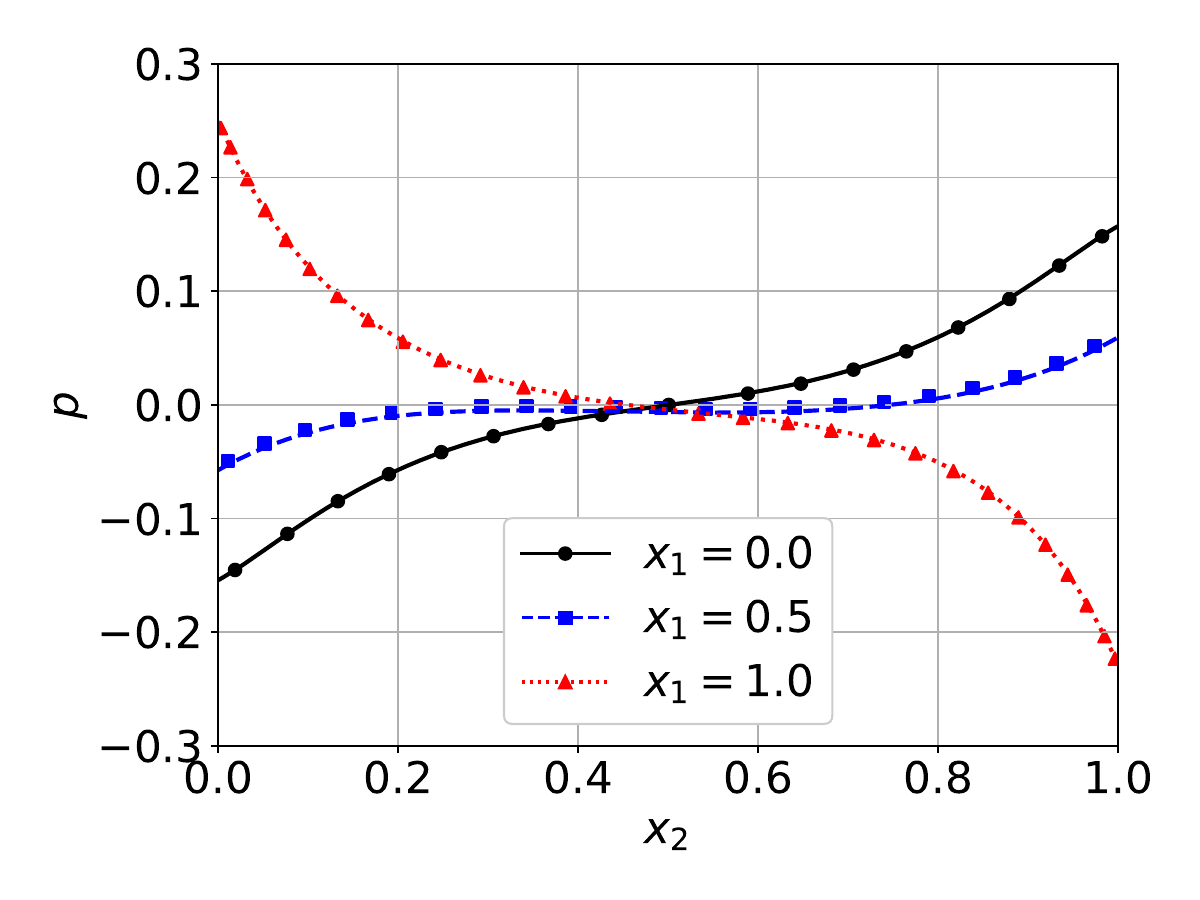}
         \caption{}
         \label{fig-profE2f}
    \end{subfigure}
\caption{
    Profiles of local \Nus[l], $\Nus$ (a), ; vertical, $v_2$, and horizontal, $v_1$, velocity components (b), temperature, $T$, over horizontal (c) and vertical (d) axes; and hydrodynamic pressure, $p$, over horizontal (e) and vertical (f) axes.
    Lines indicate computed profiles over a $216 \times 216$ grid, while markers denote reference profiles, for $\Ra = 10^2$. 
}
\label{fig-profE2}
\end{figure}

\begin{figure}[h!]
\centering
    \begin{subfigure}[t]{210pt}
         \centering
         \includegraphics[width=\textwidth]{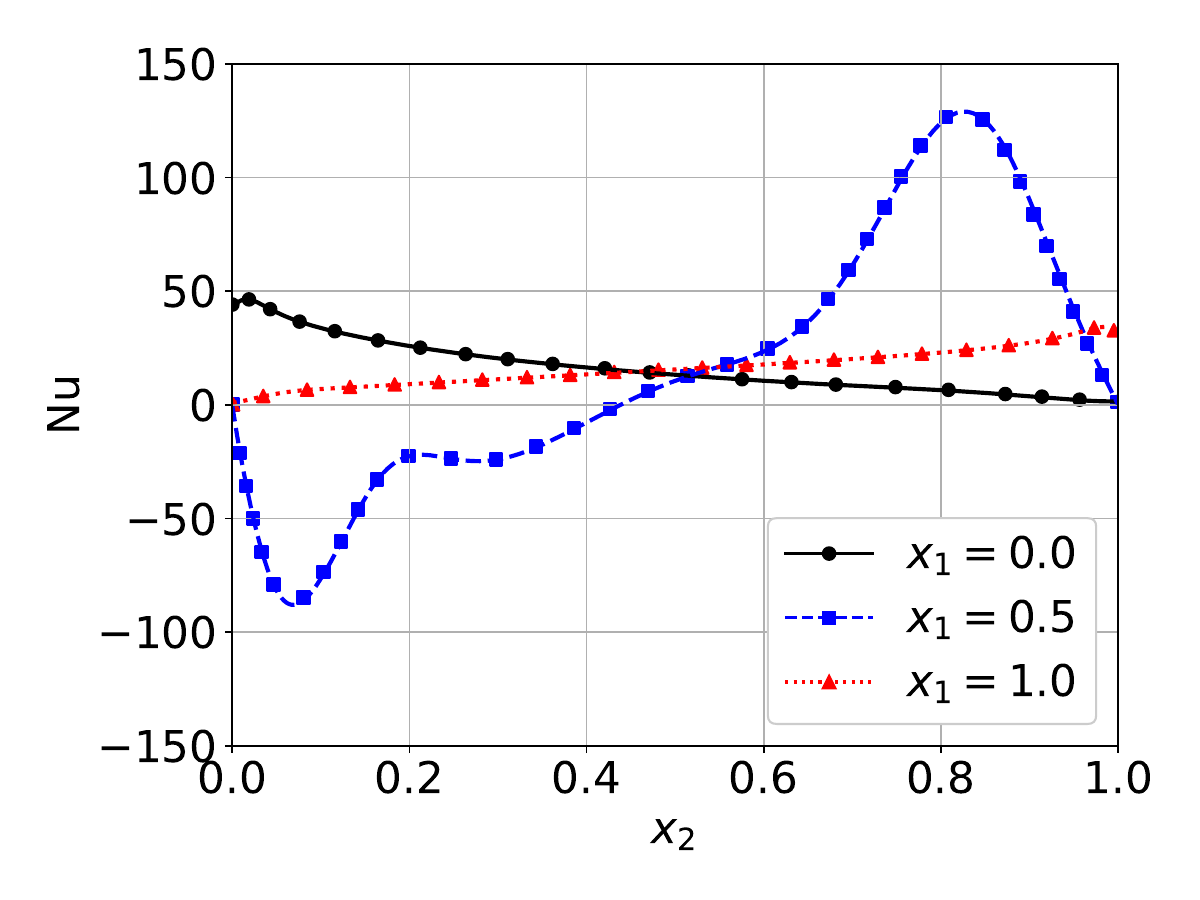}
         \caption{}
         \label{fig-profE7a}
    \end{subfigure}
    \begin{subfigure}[t]{210pt}
         \centering
         \includegraphics[width=\textwidth]{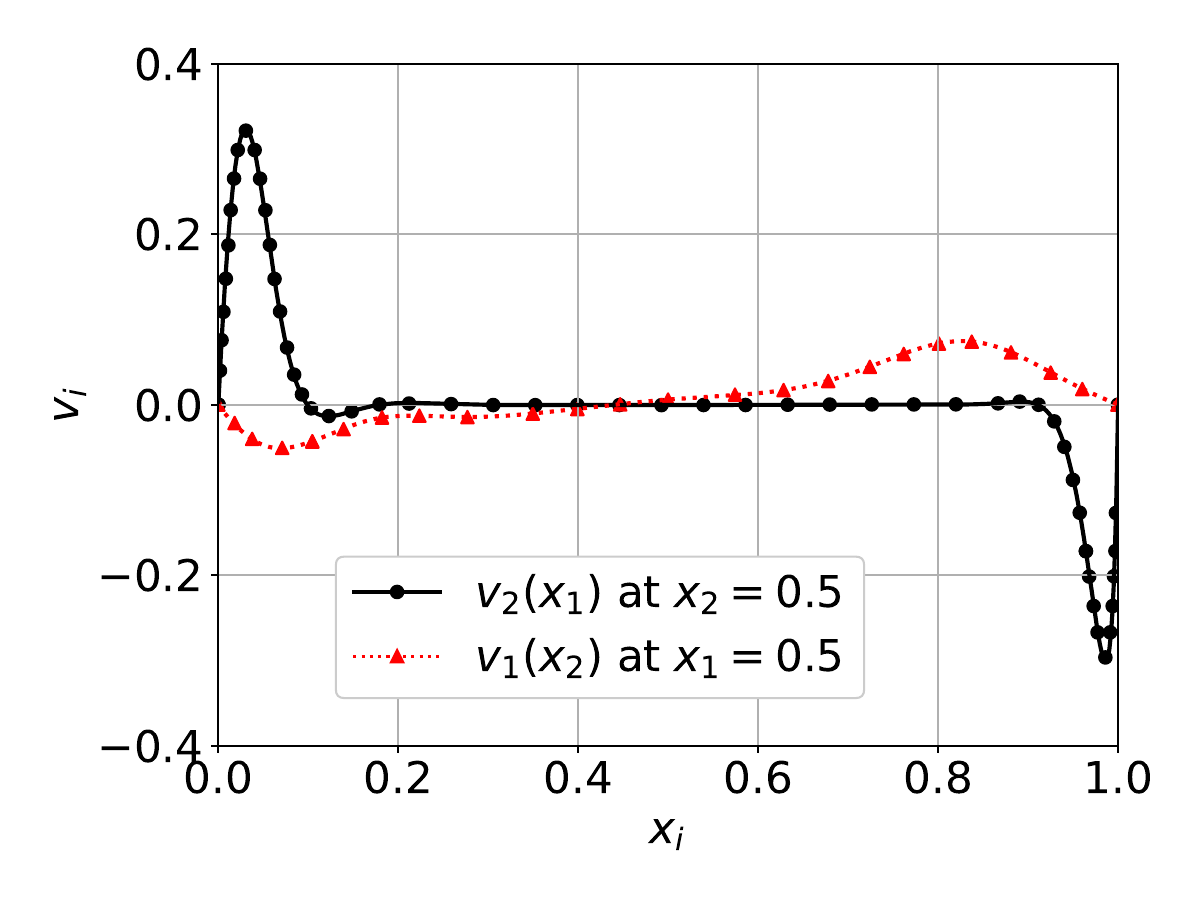}
         \caption{}
         \label{fig-profE7b}
    \end{subfigure}
    \begin{subfigure}[t]{210pt}
         \centering
         \includegraphics[width=\textwidth]{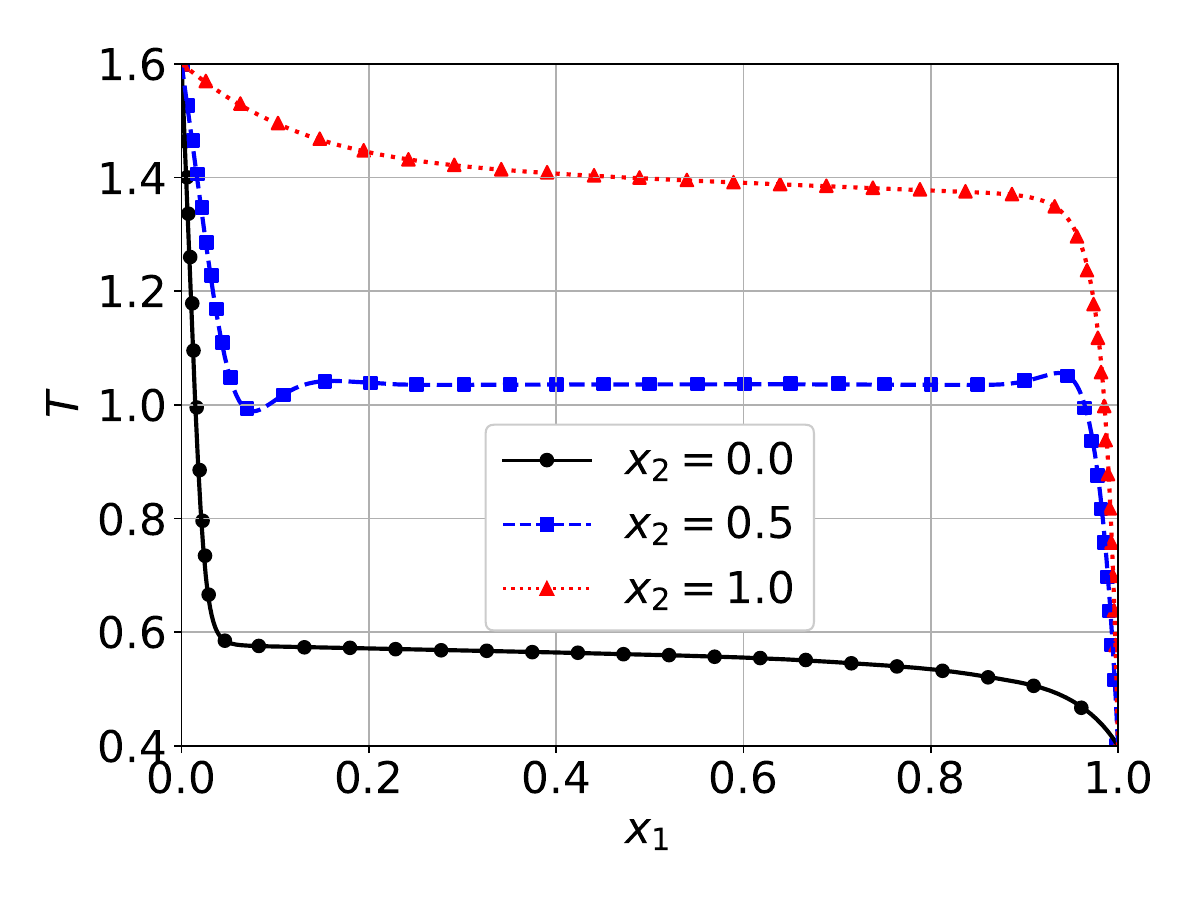}
         \caption{}
         \label{fig-profE7c}
    \end{subfigure}
    \begin{subfigure}[t]{210pt}
         \centering
         \includegraphics[width=\textwidth]{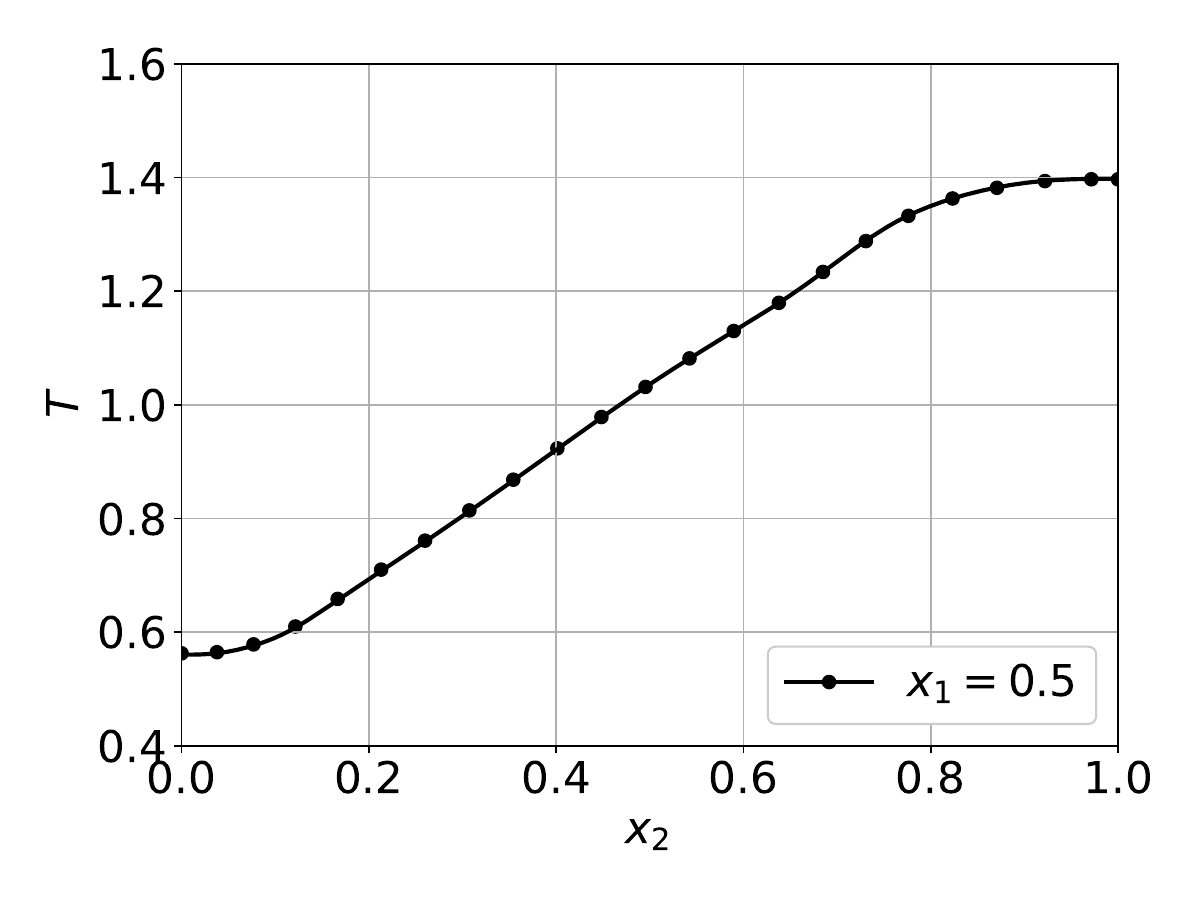}
         \caption{}
         \label{fig-profE7d}
    \end{subfigure}
    \begin{subfigure}[t]{210pt}
         \centering
         \includegraphics[width=\textwidth]{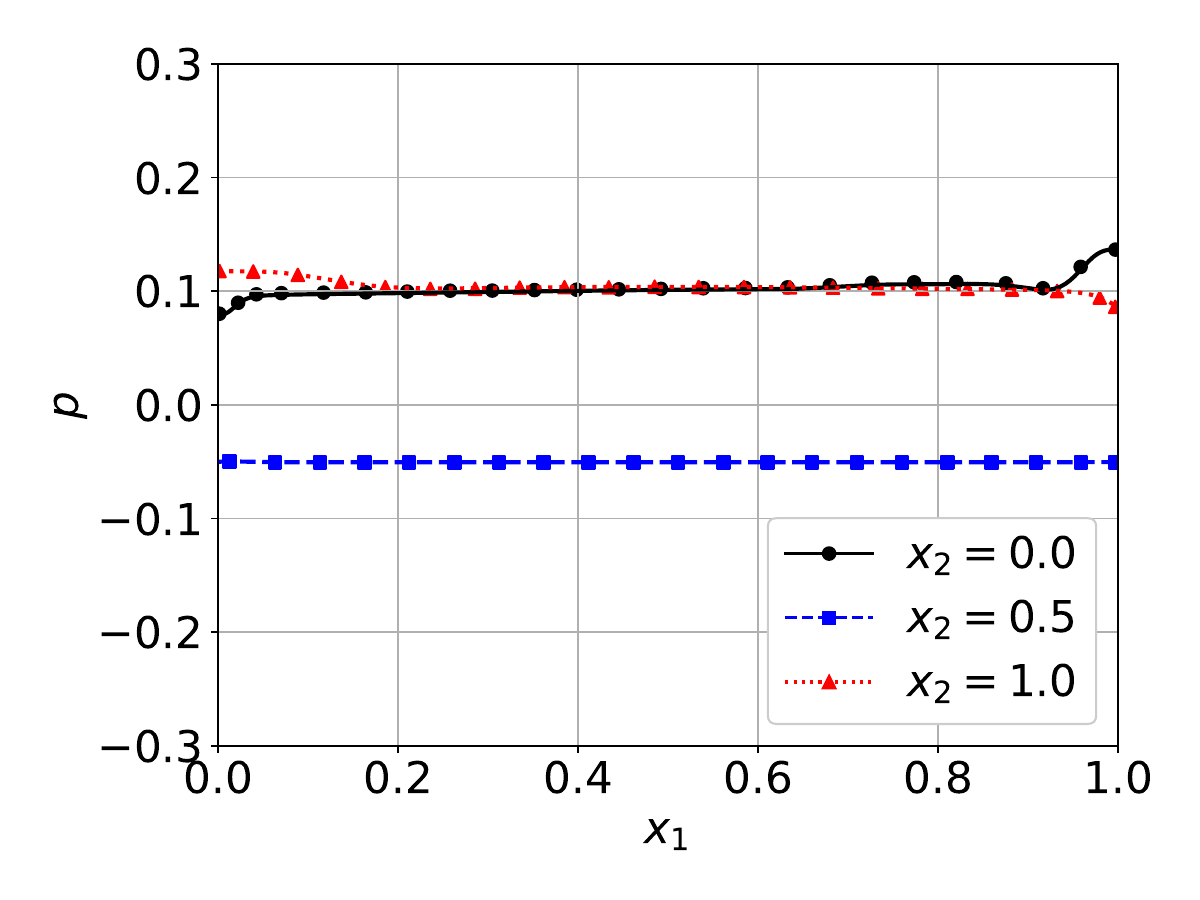}
         \caption{}
         \label{fig-profE7e}
    \end{subfigure}
    \begin{subfigure}[t]{210pt}
         \centering
         \includegraphics[width=\textwidth]{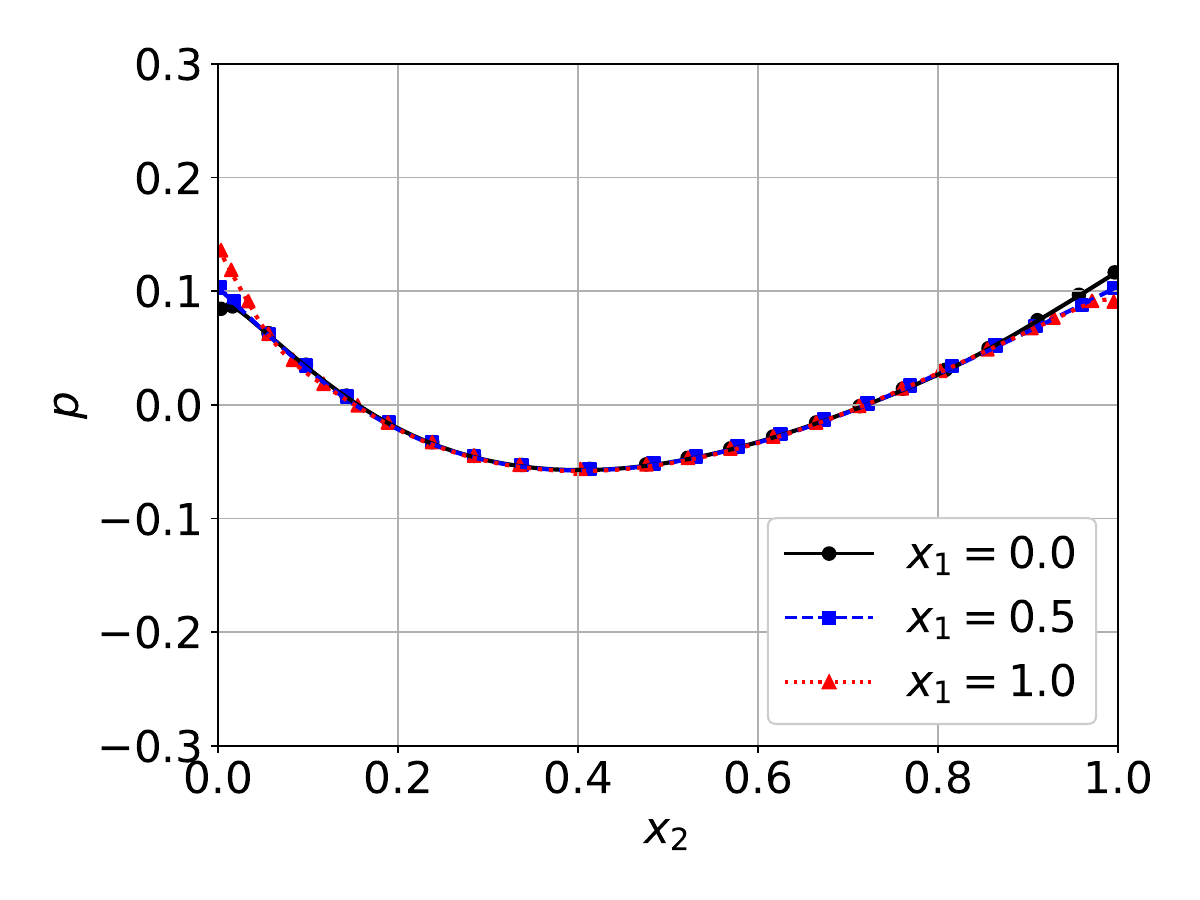}
         \caption{}
         \label{fig-profE7f}
    \end{subfigure}
\caption{
    Profiles of local \Nus[l], $\Nus$ (a), ; vertical, $v_2$, and horizontal, $v_1$, velocity components (b), temperature, $T$, over horizontal (c) and vertical (d) axes; and hydrodynamic pressure, $p$, over horizontal (e) and vertical (f) axes.
    Lines indicate computed profiles over a $216 \times 216$ grid, while markers denote reference profiles, for $\Ra = 10^7$.
}
\label{fig-profE7}
\end{figure}

This verification is organised into three levels.
The first level involves visualising streamlines and isotherms, 
which provides a more qualitative check.
It is presented in Fig. \ref{fig-quali}, 
for both cases ($\Ra = 10^2$ and $10^7$).
The second level addresses the local profiles of relevant 
properties. 
Since these data are only available in graphical form, they 
have been extracted using an appropriate tool 
\cite{WebPlotDigitizer}. 
This process inevitably introduces uncertainty, depending on 
factors such as figure resolution and line width, but it enables 
a systematic comparison of variables behaviour.
Figures \ref{fig-profE2} and \ref{fig-profE7} show the computed profiles 
(represented by lines) in comparison with the reference profiles 
(represented by marks), for the smaller and greater \Ra[l], respectively.
It should also be noted that, at low Mach number limit, the flow 
responds only to pressure differences, meaning that any solution 
with the correct pressure gradient is sufficient. 
However, to enable a direct comparison with the reference,
(hydrodynamic) pressures must be aligned to the same baseline. 
Therefore, data obtained with the present formulation was shifted 
to match the reference pressure baseline, using the average 
between the maximum and minimum profile values as a basis for this 
adjustment.
The third level provides a more rigorous quantitative 
verification by comparing the values of key (global) parameters to 
those tabulated in the reference, as compiled in 
Tab. \ref{Tab-quantComp}.

In short, the calculated and reference solutions show consistent 
agreement across all levels of analysis, verifying the proper 
modelling of energy transfer processes, and the code correctness. 

\begin{table}[th]
  \centering
  \begin{threeparttable}
  \caption{Comparison of global properties.}
  \label{Tab-quantComp}
  \begin{tabular}{c c c c c c}
    \toprule
    $\Ra$ & Data Type & {$p_0$} & {$\overline{\Nus}_L$} & {$\overline{\Nus}_M$} & {$\overline{\Nus}_R$} \\
    \midrule
    \multirow{2}{*}{$10^2$} 
        & Calculated & 0.95738 & 0.9787 & 0.9787 & 0.9786 \\ 
        & Reference  & 0.95736 & 0.9787 & 0.9787 & 0.9787 \\
    \addlinespace
    \multirow{2}{*}{$10^7$}
        & Calculated & 0.92245 & 16.248 & 16.241 & 16.222\\ 
        & Reference  & 0.92263 & 16.241 & 16.241 & 16.240 \\
    \bottomrule
  \end{tabular}
  \begin{tablenotes}
    \small
    \item Simulations performed on non-uniformly spaced grid ($216 \times 216$), with distinct accumulation of points (near walls) for each case: $2:1$ for $\Ra = 10^2$, and $21:1$ for $\Ra = 10^7$.
    \item $\overline{\Nus}_A$ denotes spatially averaged Nusselt number at left ($A = L$), middle ($A = M$), and right ($A = R$) walls.
    \item Reference data from \cite{vierendeels/2003:IJNMHFF}.
  \end{tablenotes}
  \end{threeparttable}
\end{table}


\subsection{Double Tsuji flame}
\label{DTf}

\begin{figure}[t]
\centering
\includegraphics[width=270pt]{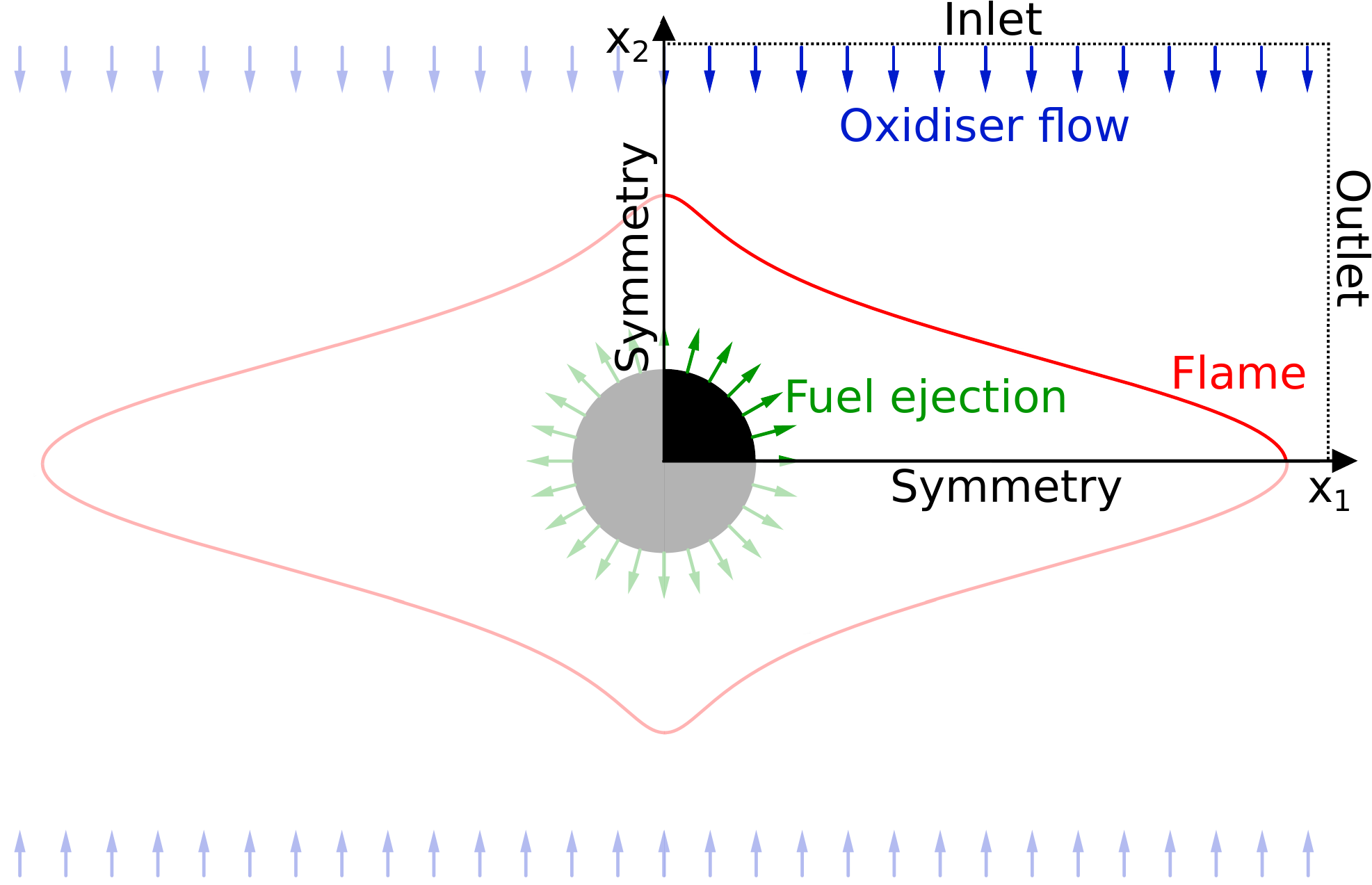}
\caption{
Schematic representation of the double Tsuji burner: 
an infinitely long cylinder (2D represented), which ejects fuel 
radially and evenly, placed in the centre of impinging oxidant 
flows (a planar unconfined counterflow).
By symmetry, the first quadrant represents totally the system, a priori.
}
\label{fig-dTsuji}
\end{figure}

This case is a complete test for the proposed method.
Consider an infinitely long circular cylindrical burner, with 
radius $\hat{R}$, that radially and uniformly ejects fuel, with 
mass fraction $\hat{Y}_{F,b}$ and velocity magnitude $\hat{u}_b$, 
in the centre of an atmosphere generated by planar impinging 
jets, composing a counterflow of oxidiser, with a mass fraction 
$\hat{Y}_{O,\infty}$ and strain rate $\hat{a}$. 
This setup is referred to as the double Tsuji burner, and the 
resulting flame is known as the double Tsuji flame 
\cite{severino/22:AMM,fachini/2025:CST}.

Double Tsuji flames exhibit characteristics of both counterflow 
and coflow diffusion flames, with smooth transitions between these 
behaviours across their spatial distribution \cite{severino/21:CTM, severino/22:AMM}. 
As such, they can be regarded as a ``generalised flame’’, 
encapsulating features of flames under diverse flow conditions, 
within a single configuration. 
This makes them an ideal test case for verifying numerical 
methods for reacting flow simulations. 
Additionally, the cylindrical burner with radial ejection 
introduces a geometric complexity that demands a proper 
technique to represent such configuration on 
orthogonal meshes. 
In fact, to accomplish it, an immersed boundary method (IBM) with 
mass flux, as described in Sec. \ref{sub-IBM}, is employed.

To consider the radial fuel ejection from the burner, a linear 
source of mass ($S_m$) is added to the continuity 
equation, consistently with the fuel ejection velocity.
More specifically,
\begin{equation}\label{Eq4.19}
    S_m(x_1,x_2) \coloneqq \chi \dfrac{\rho_b U_b}{(x_1^2 + x_2^2)^{1/2}},
\end{equation}
together with a radial ejection velocity
\begin{equation}\label{Eq4.20}
    v_{{ib}_j}(x_1,x_2) \coloneqq \dfrac{U_b}{(x_1^2 + x_2^2)^{1/2}} x_j
\end{equation}
for $j = 1, 2$.
Mixture fraction ($Z_{ib}$) and excess enthalpy ($H_{ib}$), at the burner, 
are given by
\begin{equation}\label{Eq4.21}
    Z_{ib} = S + 1 \quad \text{,} \quad H_{ib} = ( S + 1 ) / Q + 1
\end{equation}

For next discussions, in the absence of an explicit definition, 
the following parameters are considered: $\Pr = 0.71$, $\Pe = 71$ 
(or $\Rey = 100$), $U_b = 0.1$,  $S = 14.89$, $Q = 4.17$, 
$\Le_F = 0.97$, $\Le_O = 1.11$, with $\Dar_{ib} = 10^{-4}$ and 
$\varepsilon = 0.5$.

\paragraph{Boundary conditions}

\begin{subequations}\label{Eq4.22}
    \begin{gather}
        v_1 = \partial_{x_1} v_2 = \partial_{x_1} Z = \partial_{x_1} H = 0
        \quad
        ,
        \quad
        \text{on the vertical symmetry axis ($x_1 = 0$)};
       \label{Eq4.22a} \\ 
        \partial_{x_2} v_1 = v_2 = \partial_{x_2} Z = \partial_{x_2} H = 0
        \quad
        ,
        \quad
        \text{on the horizontal symmetry axis ($x_2 = 0$)};
        \label{Eq4.22b} \\
        v_1 - x_1 = v_2 + x_2 = Z - Z_\infty = H - H_\infty
        = 0
        \quad
        ,
        \quad
        \text{at the inlet ($x_2 \rightarrow \infty$)} \\
        \partial_{x_1} v_1 = \partial_{x_1} v_2 = \partial_{x_1} Z = \partial_{x_1} H
        = 0
        \quad
        ,
        \quad
        \text{at the outlet ($x_1 \rightarrow \infty$)}.
        \label{Eq4.22c}
    \end{gather}
\end{subequations}
in which $Z_\infty \coloneqq 0$ and $H_\infty \coloneqq ( S + 1 ) / Q + 1$.

For an open domain, such as in the present case, global mass conversation
(throughout the computational domain) can be challenging.
Since the scope of this study is not directly related to boundary conditions,
a simple outflow extrapolation is imposed, as defined above.
Global mass conservation is, then, monitored, with the outlet flux being 
uniformly modified, as necessary.

\paragraph{Initial condition}

Temporal integration initiates from a potential flow 
\cite{severino/21:CTM,severino/22:AMM}:

\begin{subequations}\label{Eq4.23}
    \begin{gather}
        v_1(x_1,x_2)
        =
        \ \ 
        x_1
        \left[
            1       
            -
            \dfrac{ 
                \left(
                    x_1^2 - 3x_2^2 
                \right)
            }  
            {
                \left( 
                    x_1^2+x_2^2 
                \right)^3
            }
            +
            \dfrac{ 
                U_b 
            }  
            {   
                x_1^2 + x_2^2
            }
        \right];
    \label{Eq4.23a}
       \\ 
       v_2(x_1,x_2)
       = 
       - x_2 
       \left[ 
            1
            +
            \dfrac{ 
                \left(
                    3 x_1^2 - x_2^2
                \right) 
            }  
            {
                \left(
                    x_1^2 + x_2^2
                \right)^3
            }
            -
            \dfrac{ 
                U_b 
            }  
            {   
                x_1^2 + x_2^2
            }
        \right],
    \label{Eq4.23b}
    \end{gather}    
\end{subequations}
and asymptotic solutions for $Z$ and $H$ \cite{severino/22:AMM}, modified to
also include the ejection effect:
\begin{subequations}\label{Eq4.24}
    \begin{gather}
       Z(x_1,x_2)
       = 
       \Bigg[ \dfrac{ 1 + U_b/w_F^2}{ (x_1/w_F)^2 + U_b/w_F^2 } \Bigg]^{1/2} 
         \exp (- x_2^2 / 2)
       \label{Eq4.24a}\\
       H(x_1,x_2)
        =
         \left[
            \dfrac{S+1}{Q}(T_F - 1) - 1
         \right]
         Z
         +
        \dfrac{S+1}{Q}
        +
        1
    \label{Eq4.24b}
    \end{gather}
\end{subequations}
in which $w_F$ and $T_F$ are the width (location on the $x_1$-axis) and
temperature of the flame, estimated in the reference.

\subsubsection{Steady-state solution}
\label{SSfields}

\begin{figure}[t]
\centering
    \begin{subfigure}[t]{231pt}
         \centering
         \includegraphics[width=\textwidth]{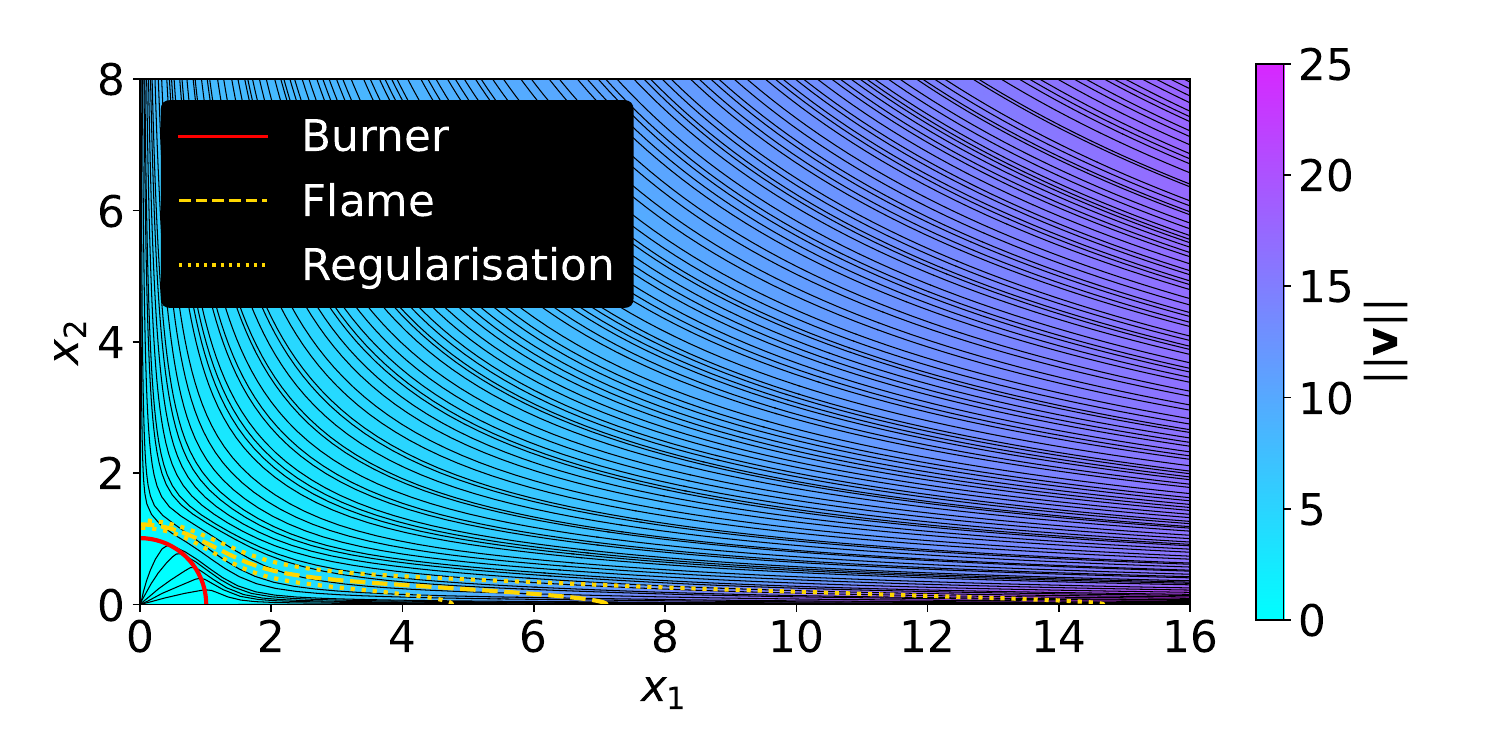}
         \caption{}
         \label{fig-DTfieldsA}
    \end{subfigure}
    \begin{subfigure}[t]{231pt}
         \centering
         \includegraphics[width=\textwidth]{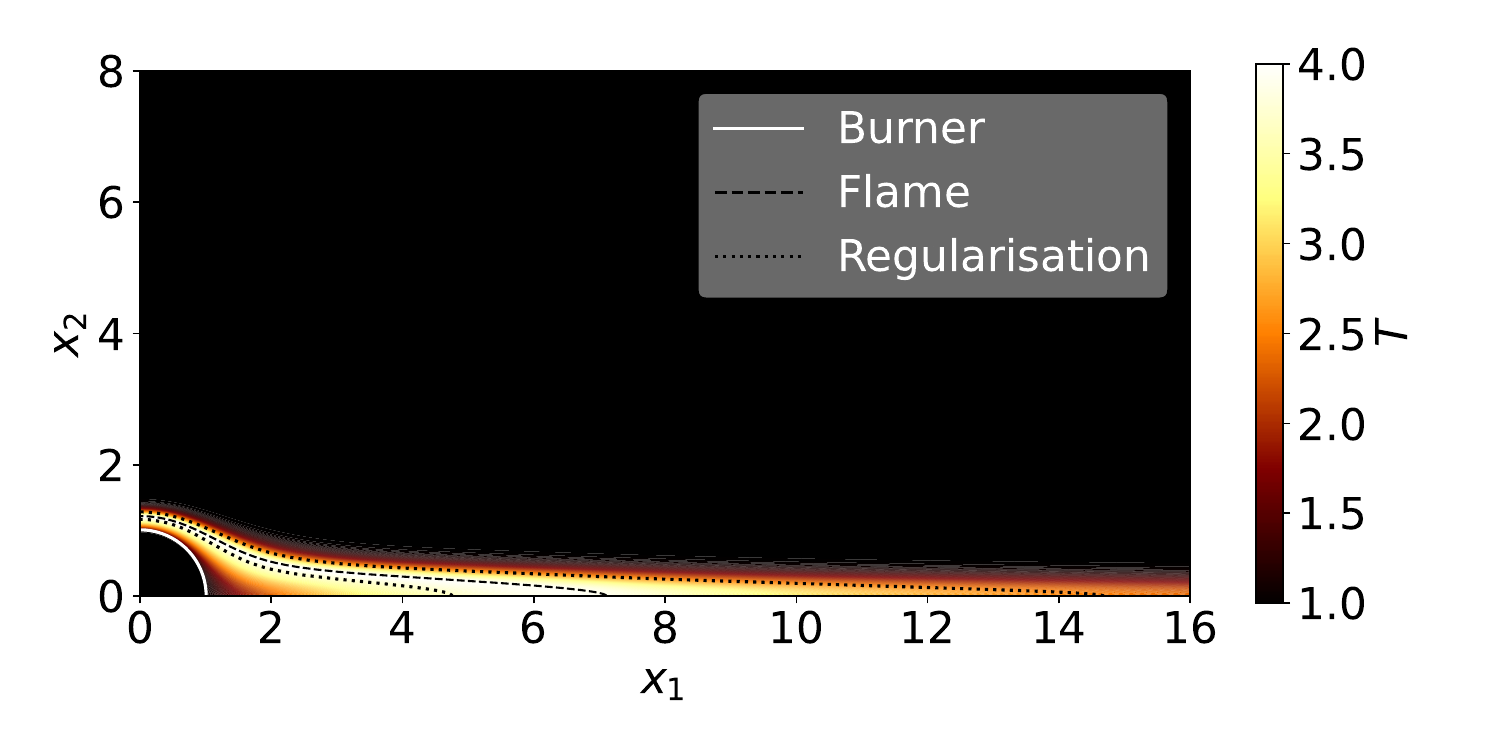}
         \caption{}
         \label{fig-DTfieldsB}
    \end{subfigure}
\caption{
    Streamlines with velocity magnitude indication (a) and temperature field 
    for the double Tsuji burner.
    Light (dark) portions indicate low (high) speed or high (low) temperatures.
}
\label{fig-DTfields}
\end{figure}

Flow topology, velocity and temperature are presented in Fig. 
\ref{fig-DTfields}.
As the effect of impinging flows, strong variations of velocity and 
temperature are confined to the stagnation region (over $x_2$-axis),
and around the $x_1$-axis.
Additional lines illustrate the cylinder surface (solid line), 
flame position (dashed line), and flame regularisation thickness 
(between dotted lines).

\subsubsection{Domain size dependence}
\label{sizeDep}

\begin{table}[th]
  \centering
  \caption{Domain convergence study.}
  \label{Tab-domainConv}
  \begin{threeparttable}
    \begin{tabular}{c c S[table-format=1.6] S[table-format=-1.6] S[table-format=1.6] S[table-format=-1.6]}
      \toprule
      {Parameter} & {Value} & {$h_F$} & {$\Delta h_F (\%)$} & {$w_F$} & {$\Delta w_F (\%)$} \\
      \midrule
      $L_{x_1}$ & 10 & 1.21969 & \multicolumn{1}{c}{--} & 7.13264 & \multicolumn{1}{c}{--} \\
                & 15 & 1.21904 & -0.053 & 7.10868 & -0.337 \\
                & 20 & 1.21900 & -0.004 & 7.10706 & -0.023 \\
      \multicolumn{6}{c}{\textit{Horizontal Convergence ($L_{x_2} = 10$)}} \\
      \midrule
      $L_{x_2}$ & 10 & 1.21900 & \multicolumn{1}{c}{--} & 7.10706 & \multicolumn{1}{c}{--} \\
                & 15 & 1.22072 & 0.141 & 7.12096 & 0.195 \\
                & 20 & 1.22156 & 0.069 & 7.12879 & 0.110 \\
      \multicolumn{6}{c}{\textit{Vertical Convergence ($L_{x_1} = 20$)}} \\
      \bottomrule
    \end{tabular}
    \begin{tablenotes}
      \item $\Delta$: Relative change from previous domain size.
    \end{tablenotes}
  \end{threeparttable}
\end{table}

Consider $L_{x_1}$ and $L_{x_2}$ as the numerical domain size in the 
(horizontal) $x_1$-direction and the (vertical) $x_2$-direction, 
respectively.
A non-uniform grid is generated for the domain 
$[0,L_{x_1}] \times [0,L_{x_2}]$, with $L_{x_i} = 20$ ($i=1,2$) and  resolution $700 \times 700$.
The most refined regions are around symmetry axes, with grid spacing
as small as $\Delta x_i = 5\times 10^{-3}$ for both
directions (i.e., $i=1,2$).
That is, the burner regions is the most refined one.
The coarsest region is in the outlet vicinity, for the horizontal 
spacing ($\Delta x_1$), and around the inlet, for the vertical 
spacing ($\Delta x_2$), with 
$\Delta x_i \approx 0.1$ for $i=1,2$.
Taking subsets of this initial grid, the following subdomains length
are considered for the domain size dependence test: 
$L_{x_1} = 10, 15$ and $20$ with $L_{x_2} = 10$, 
for the horizontal direction; 
$L_{x_2} = 10, 15$ and $20$ with $L_{x_1} = 20$, 
for the vertical direction.
The flame positions on the horizontal ($w_F$) and vertical 
($h_F$) axes are selected as the monitoring parameters, 
denoted as the flame width and height, in this order.
As shown in Tab. \ref{Tab-domainConv}, increasing the horizontal domain, from 
$L_{x_1} = 15$ to $20$, resulted in a maximum relative variation of 
$\Delta w_F \approx -0.02\%$ in the flame width, and 
$\Delta h_F \approx -0.004\%$ in its height.
For an equivalent decrease in the vertical domain, from 
$L_{x_2} = 15$ to $10$, a maximum relative variation of 
$\Delta w_F = -0.19\%$ in the flame width, and 
$\Delta h_F = -0.14\%$ in its height, are also observed in Tab. 
\ref{Tab-domainConv}. 
Given the observed small deviations, and the convergence trend, a 
domain size of $L_{x_1} = 20$ and $L_{x_2} = 10$ ensures that boundary 
effects arising from domain truncation alter relevant parameters by 
no more than $0.2\%$.
This meets a rigid domain size convergence criterion, and 
consequently, this domain will be employed in all subsequent 
simulations.


\subsubsection{Parameters dependence}
\label{regPar}

\begin{table}[th]
  \centering
  \caption{Parameters convergence study.}
  \label{Tab-parConv}
  \begin{threeparttable}
    \begin{tabular}{c c S[table-format=1.6] S[table-format=-1.6] S[table-format=1.6] S[table-format=-1.6]}
      \toprule
      {Parameter} & {Value} & {$h_F$} & {$\Delta h_F (\%)$} & {$w_F$} & {$\Delta w_F (\%)$} \\
      \midrule
      $\Dar_{ib}$ & $1.6 \cdot 10^{-4}$ & 1.21261 & \multicolumn{1}{c}{--} & 7.21502 & \multicolumn{1}{c}{--} \\
                &  $8 \cdot 10^{-5}$ & 1.22125 & 0.707 & 7.07152 & -2.029 \\
                &  $4 \cdot 10^{-5}$ & 1.22575 & 0.367 & 6.99839 & -1.045 \\
      \midrule
      $\varepsilon$ & 0.9 & 1.21878 & \multicolumn{1}{c}{--} & 7.12582 & \multicolumn{1}{c}{--} \\
                & 0.5 & 1.21901 & -0.019 & 7.10695 & 0.265 \\
                & 0.1 & 1.21913 & -0.010 & 7.08865 & 0.258 \\
      \bottomrule
    \end{tabular}
    \begin{tablenotes}
      \item $\Delta$: Relative change from previous parameter value.
    \end{tablenotes}
  \end{threeparttable}
\end{table}

The effects of the IB Darcy number ($\Dar_{ib}$) and the
flame regularisation thickness ($\varepsilon$) are accessed here.

\paragraph{IB Darcy number}
As inferred from the discussion for the Taylor--Couette flow 
(refer to Sec. \ref{TCflow}), $\Dar_{ib} = 10^{-4}$ is a good starting 
point.
In addition, the values $\Dar_{ib} = 1.6 \cdot 10^{-4}, 8 \cdot 10^{-5}$ and
$4 \cdot 10^{-5}$ are tested.
The first part of Tab. \ref{Tab-parConv} compiles the results.
Solutions are little sensitive to the variations of $\Dar_{ib}$ in that
interval, with maximum variations of $2\%$.

\paragraph{Flame regularisation}
The regularisation of the flame is conveniently done around the 
stoichiometric level of the mixture fraction (i.e. $Z=1$), 
rather than directly around the spatial position of the flame
(refer to Sec. \ref{flameReg}).
This approach simplifies mathematical modelling and numerical 
implementation, but does not allow precise spatial control of 
the regularisation thickness, a priori, as illustrated by the
region delimited by dotted lines in Fig. \ref{fig-DTfields}.
Nonetheless, this approach proved to be efficient and sufficiently 
accurate, with little sensitivity to the regularisation 
thickness ($\varepsilon$), according to the results 
for $\varepsilon = 0.1, 0.5$ and $0.9$, in the second part of Tab. 
\ref{Tab-parConv}.

\subsubsection{Grid resolution dependence}
\label{gridStudy}

\begin{figure}[t]
\centering
\includegraphics[width=250pt]{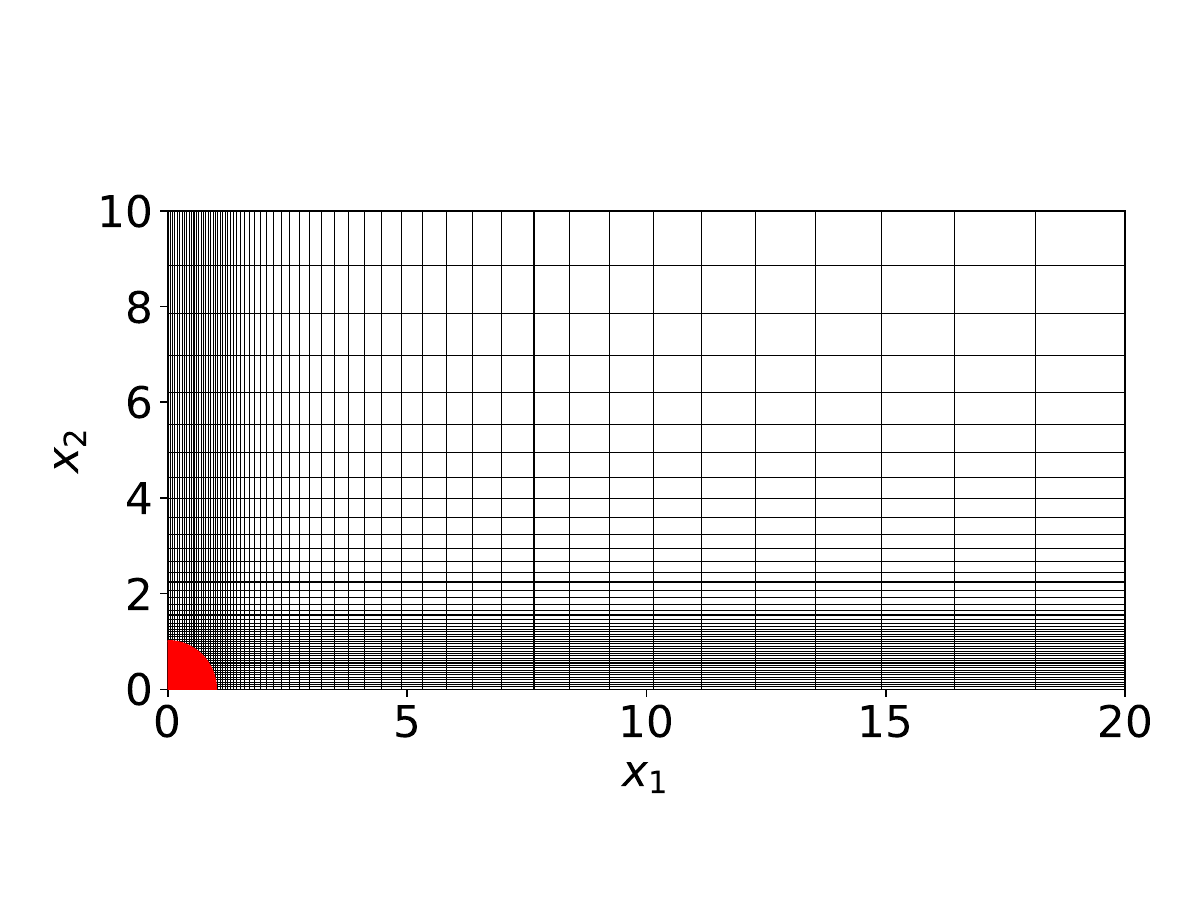}
\caption{
Illustration of the grid topology used for double-Tsuji-burner simulations.
The circular cylindrical burner is represented by a filled circle at the origin.
}
\label{fig-DTgrid}
\end{figure}

\begin{table}[th]
  \centering
  \caption{Grid convergence study.}
  \label{Tab-gridConv}
  \begin{threeparttable}
    \begin{tabular}{c S[table-format=1.6] S[table-format=-1.6] S[table-format=1.6] S[table-format=-1.6]}
      \toprule
      {Resolution} & {$h_F$} & {$\Delta h_F (\%)$} & {$w_F$} & {$\Delta w_F (\%)$} \\
      \midrule
      $600 \times 480$ & 1.21901 & \multicolumn{1}{c}{--} & 7.10695 & \multicolumn{1}{c}{--} \\
      $900 \times 720$ & 1.21897 & -0.003 & 7.10683 & -0.002 \\
      \bottomrule
    \end{tabular}
    \begin{tablenotes}
      \item $\Delta$: Relative change from previous grid resolution.
    \end{tablenotes}
  \end{threeparttable}
\end{table}

In this case, strongest gradients are concentrated in the stagnation region, upstream the cylinder, and around the flame sheet.
Thus, a grid with variable refinement is used, with augmented resolution in those zones, as depicted in Fig. \ref{fig-DTgrid}.
Results are synthesised in Tab. 
\ref{Tab-gridConv}. 
The relative variations from $600 \times 480$ to $900 \times 720$ are very small, attesting that the results are, asymptotically, converged.

\subsubsection{OpenFOAM comparison}
\label{OpenComp}

Results obtained from the present approach are compared with that 
from OpenFOAM 8.
The same hypotheses admitted in Sec. \ref{Formulation} are imposed 
in OpenFOAM simulations.
Since this functionality was unavailable in standard libraries, 
the power law for the transport coefficients was implemented. 
Additionally, constant Lewis numbers ($Le_i$) for each species ($i$) 
of a complete methane reaction were specified using the 
reactingLeFoam solver \cite{reactingLeFoam}, with values specified in \cite{smoke/1991}.

\begin{figure}[h!]
\centering
    \begin{subfigure}[t]{230pt}
         \centering
         \includegraphics[width=\textwidth]{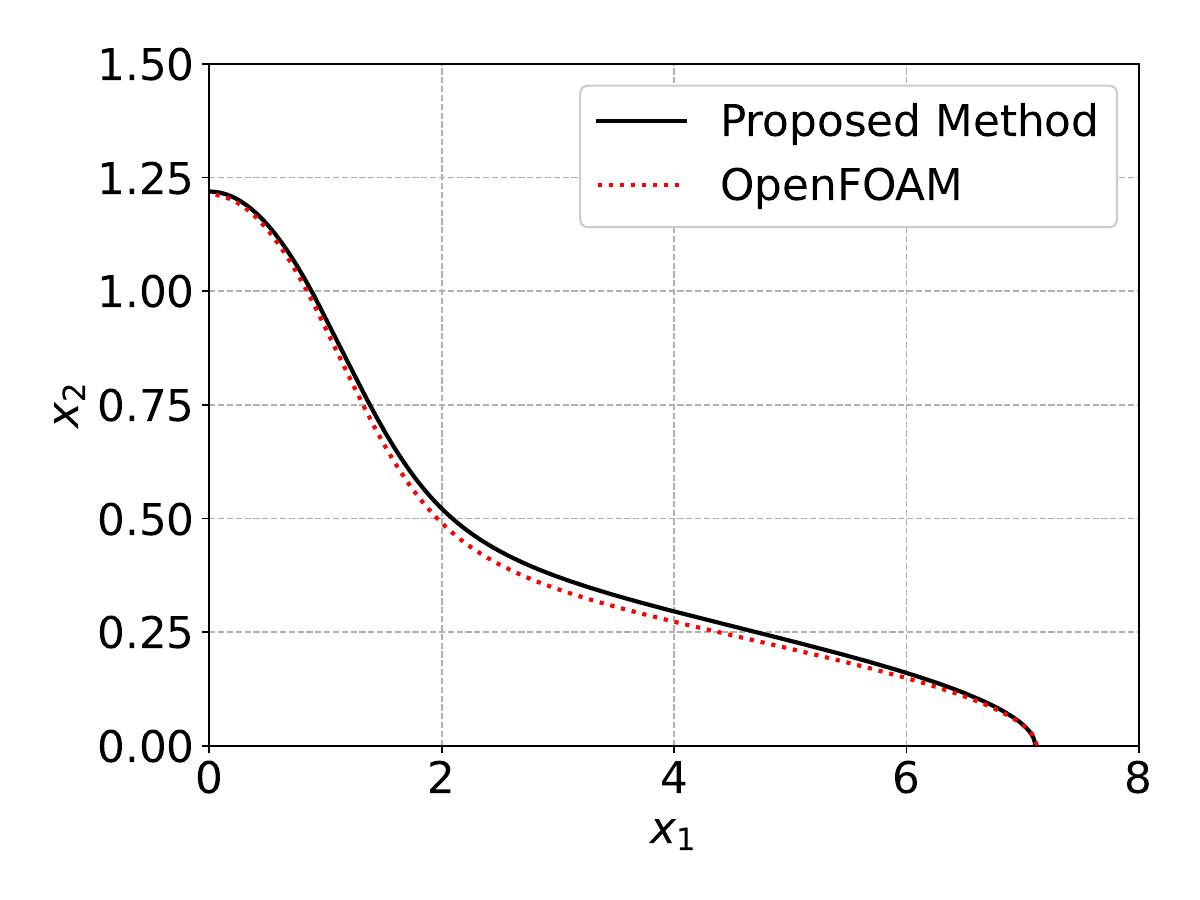}
         \caption{}
         \label{fig-DTFcompA}
    \end{subfigure}
    \begin{subfigure}[t]{230pt}
         \centering
         \includegraphics[width=\textwidth]{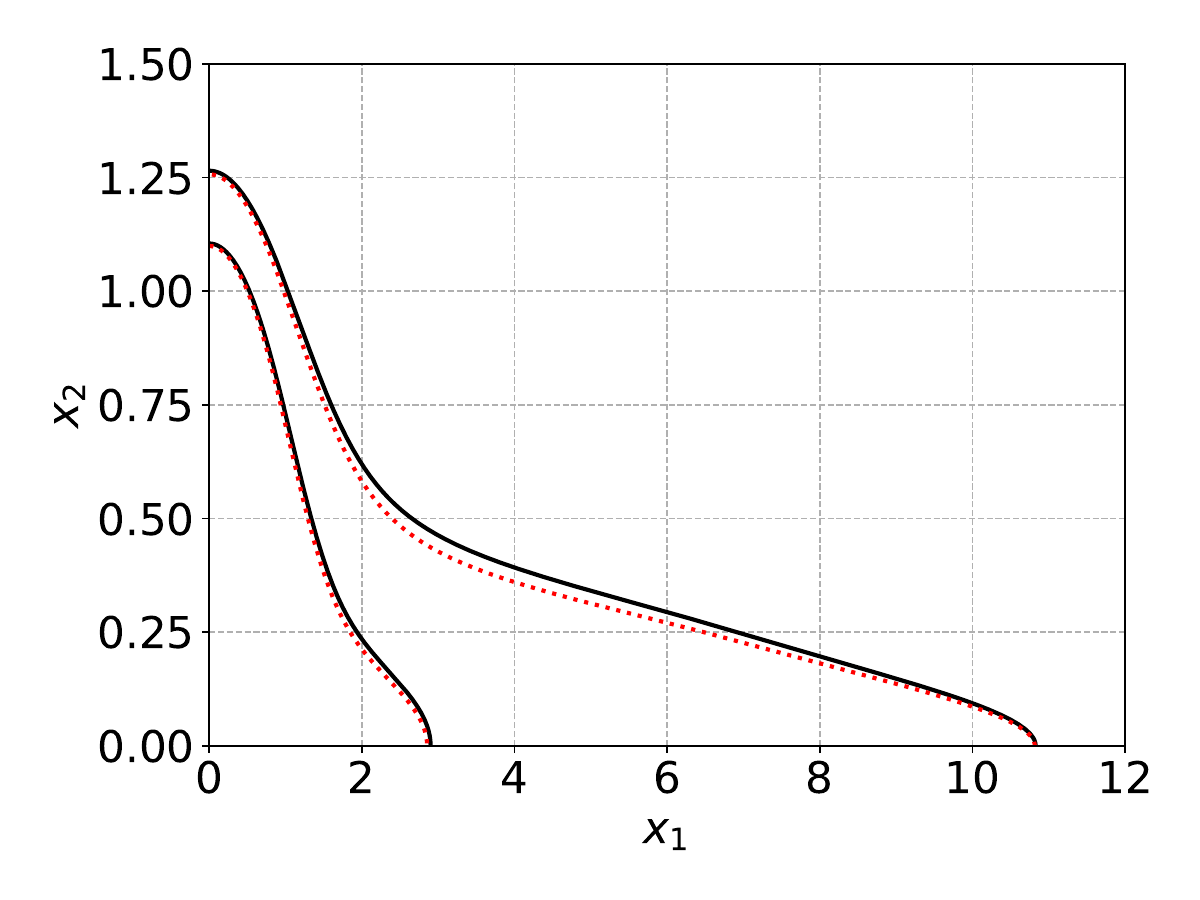}
         \caption{}
         \label{fig-DTFcompB}
    \end{subfigure}
\caption{
    Comparison between flame shape (a) and the isotherm $T = 3$ (b) from simulations using the proposed method and OpenFOAM.
}
\label{fig-DTFcomp}
\end{figure}

As exhibited in Fig. \ref{fig-DTFcomp}, a very good agreement of flame shapes and isotherms are obtained.
Minor deviations were expected and can be attributed to the inherent distinctions in the numerical methods employed by the respective solver (e.g., spatial and operator discretisations, flame regularisation procedures,
and the circular cylinder representations).
Therefore, the properness and accuracy of the proposed methods are established.


\section{Conclusion}
\label{Conclusion}
\addvspace{10pt}

This work proposes a robust numerical method for simulations of low-Mach-number 
flows, ranging from incompressible to strongly-variable density ones, without and with
combustion.
The numerical approach is based on the transformation of the original set of partial
differential equations, using spatial discretisation, into a set of ordinary differential
equations (method of lines).
Thus, a predictor-corrector scheme could be applied to them, using a fractional time-step 
(or projection) method to couple pressure and velocity.
It consists in the solution of a Poisson equation for pressure, whose boundary conditions 
are inherited from the interfaces fluxes, when based on staggered or collocated grids with 
the definition of auxiliary fluxes.
A second-order Adams--Bashforth (Adams--Moulton) temporal scheme is used for the predictor 
(corrector) stage.
Equations are spatially discretised into a collocated grid using centred finite differences, 
specially designed to produce strong ``odd-even coupling'', via interpolation of auxiliary
fluxes.
A regularisation technique for flame properties is proposed to mitigate the discontinuity issue,
induced by the flame sheet approximation.
A new feature of it is the regularisation based on the temperature derivative, from which the 
singularity effectively emerges, rather than directly on temperature, as classically performed. 
Another relevant contribution is the extension of the immersed boundary method (IBM) to consider
mass fluxes from the immersed body surfaces, which naturally demands a mass source addition 
to the continuity equation.
Selected test cases demonstrated the capability and limitations of the method under different
conditions, focusing on distinct aspects.
Namely, the correctness and accuracy of the projection method are tested using Taylor--Green
vortices.
IBM is analysed using a Taylor--Couette flow, providing insightful conclusions about the
penalisation method and related parameters selection.
The capability to handle strong density gradients and thermal-energy transport is verified in a 
differentially heated cavity.
Finally, the composition of these various methods, including IBM for velocity, mixture 
fraction and excess enthalpy, species mass transport, differential-preferential diffusion, and 
flame regularisation are checked using the double Tsuji flame.
After the complete discussion of this work, it is possible to state that the proposed numerical 
method is stable, efficient and accurate.

  
\section*{Declaration of Competing Interest}
\label{DCI}
No potential conflict of interest was reported by the authors.

\section*{Acknowledgments}
\label{Acknowledgments}

This article is dedicated to the memory of Professor Amable Liñán. \\

This study was financed, in part, by the São Paulo Research Foundation 
(FAPESP), Brasil. Process Numbers 2021/09246-9, 2021/10689-2 and 
2022/14361-4.
This study was financed, in part, the Conselho Nacional de 
Desenvolvimento  Científico e Tecnológico (CNPq) -- under grants 
307922/2019-7, 161887/2021-0, and 301607/2025-7;
This study was financed, in part, by the Coordenação de Aperfeiçoamento 
de Pessoal de Nível Superior -- Brasil (CAPES) -- Finance Code 001.
D.R. acknowledges funding by MCIN/AEI/10.13039/501100011033 and the European Union’s FEDER (Funder ID: 10.13039/501100011033), under grant PID2024-157642MB-I00. \\
Research carried out using the computational resources of the Center 
for Mathematical Sciences Applied to Industry (CeMEAI) funded by 
FAPESP (grant 2013/ 07375-0).

\bibliography{usedRef}

\end{document}